\documentclass{aa}

\usepackage[varg]{txfonts}
\usepackage{amsmath}
\usepackage{enumitem}
\usepackage{graphicx}
\usepackage{wrapfig}
\usepackage{adjustbox}
\usepackage[colorinlistoftodos]{todonotes}
\usepackage{subcaption}
\usepackage{placeins}
\usepackage{caption}
\usepackage{natbib,twoopt}
\usepackage{array}
\usepackage{multirow}
\usepackage[]{hyperref} 
\bibpunct{(}{)}{;}{a}{}{,}             

\makeatletter
  \newcommandtwoopt{\citeads}[3][][]{\href{http://adsabs.harvard.edu/abs/#3}%
    {\def\hyper@linkstart##1##2{}%
     \let\hyper@linkend\@empty\citealp[#1][#2]{#3}}}
  \newcommandtwoopt{\citepads}[3][][]{\href{http://adsabs.harvard.edu/abs/#3}%
    {\def\hyper@linkstart##1##2{}%
     \let\hyper@linkend\@empty\citep[#1][#2]{#3}}}
  \newcommandtwoopt{\citetads}[3][][]{\href{http://adsabs.harvard.edu/abs/#3}%
    {\def\hyper@linkstart##1##2{}%
     \let\hyper@linkend\@empty\citet[#1][#2]{#3}}}
  \newcommandtwoopt{\citeyearads}[3][][]%
    {\href{http://adsabs.harvard.edu/abs/#3}
    {\def\hyper@linkstart##1##2{}%
     \let\hyper@linkend\@empty\citeyear[#1][#2]{#3}}}

\newcommand{\lcol}[1]{\multicolumn{2}{l}{#1}}
\newcommand{\rcol}[1]{\multicolumn{2}{r}{#1}}
\makeatother

\usepackage{tikz}
\usetikzlibrary{shapes, arrows, positioning}

\definecolor{dgreen}{rgb}{0, 0.5, 0}

\begin{document}

\title{The ALMA-PILS survey: Inventory of complex organic molecules towards IRAS~16293--2422~A}

\author{        S. Manigand$^{\ref{inst-NBI}}$\and 
                                J. K. J\o rgensen$^{\ref{inst-NBI}}$\and 
                                H. Calcutt$^{\ref{inst-Sweden}}$\and
                                H. S. P. M{\"u}ller$^{\ref{inst-Koln}}$\and 
                                N. F. W. Ligterink$^{\ref{inst-Bern}}$\and 
                                A. Coutens$^{\ref{inst-LAB}}$\and 
                                M. N. Drozdovskaya$^{\ref{inst-Bern}}$\and
                                E. F. van Dishoeck$^{\ref{inst-Leiden},\ \ref{inst-MPE}}$ \and 
                                S. F. Wampfler$^{\ref{inst-Bern}}$ 
                                }
                                
\titlerunning{Inventory of complex organic molecules towards IRAS~16293--2422~A}
\authorrunning{S. Manigand et al.}

\institute{
Niels Bohr Institute \& Centre for Star and Planet Formation, University of Copenhagen, \O ster Voldgade 5--7, DK-1350 Copenhagen K., Denmark\label{inst-NBI}\and
Department of Space, Earth and Environment, Chalmers University of Technology, 41296, Gothenburg, Sweden\label{inst-Sweden} \and
I. Physikalisches Institut, Universit{\"a}t zu K{\"o}ln, Z{\"u}lpicher Str. 77, 50937 K{\"o}ln, Germany\label{inst-Koln} \and
Center for Space and Habitability (CSH), University of Bern, Gesellschaftsstrasse 6, 3012 Bern, Switzerland \label{inst-Bern} \and
Laboratoire d'Astrophysique de Bordeaux, Univ. Bordeaux, CNRS, B18N, all{\'e}e Geoffroy Saint-Hilaire, 33615 Pessac, France \label{inst-LAB} \and
Leiden Observatory, Leiden University, PO Box 9513, NL-2300 RA Leiden, The Netherlands\label{inst-Leiden}\and
Max-Planck Institut f{\"u}r Extraterrestrische Physik (MPE), Giessenbachstr. 1, 85748 Garching, Germany\label{inst-MPE}
}

\date{Received ??? / Accepted ???}

\abstract{Complex organic molecules are detected in many sources in the warm inner regions of envelopes surrounding deeply embedded protostars. Exactly how these species form remains an open question.}
{This study aims to constrain the formation of complex organic molecules through comparisons of their abundances towards the Class 0 protostellar binary IRAS~16293--2422.}
{We utilised observations from the Atacama Large Millimetre/submillimetre Array (ALMA) Protostellar Interferometric Line Survey (PILS) of IRAS 16293--2422. The species identification and the rotational temperature and column density estimation were derived by fitting the extracted spectra towards IRAS~16293--2422~A and IRAS~16293--2422~B with synthetic spectra. The majority of the work in this paper pertains to the analysis of IRAS~16293--2422~A for a comparison with the results from the other binary component, which have already been published.}
{We detect 15 different complex species, as well as 16 isotopologues towards the most luminous companion protostar IRAS~16293--2422~A. Tentative detections of an additional 11 isotopologues are reported. We also searched for and report on the first detections of methoxymethanol (CH$_3$OCH$_2$OH) and \textit{trans}-ethyl methyl ether (t-C$_2$H$_5$OCH$_3$) towards IRAS~16293--2422~B and the follow-up detection of deuterated isotopologues of acetaldehyde (CH$_2$DCHO and CH$_3$CDO). Twenty-four lines of doubly-deuterated methanol (CHD$_2$OH) are also identified. }
{The comparison between the two protostars of the binary system shows significant differences in abundance for some of the species, which are partially correlated to their spatial distribution. The spatial distribution is consistent with the sublimation temperature of the species; those with higher expected sublimation temperatures are located in the most compact region of the hot corino towards IRAS~16293--2422~A. 
This spatial differentiation is not resolved in IRAS~16293--2422~B and will require observations at a higher angular resolution. In parallel, the list of identified CHD$_2$OH lines shows the need of accurate spectroscopic data including their line strength.}


\maketitle

\section{Introduction}

Complex organic molecules \citep[COMs, i.e. molecules containing six or more atoms with at least one carbon atom;][]{Herbst-2009} are observed in various interstellar environments, ranging from very cold, dense collapsing cloud cores \citep[e.g.][]{Bacmann-2012, Taquet-2017} to protoplanetary discs \citep[e.g.][]{Oberg-2015, Walsh-2016, Favre-2018}, as well as in low- and high-mass star-forming regions \citep[e.g.][]{Blake-1987, Fayolle-2015, Bergner-2017, Ceccarelli-2017, Ospina-Zamudio-2018, Bogelund-2019}.
Meteorite measurements \citep{Ehrenfreund-2001, Botta-2002}, comet coma observations \citep{BockeleeMorvan-2000, Crovisier-2004EG, Biver-2014}, and in situ \textit{Rosetta} mission measurements \citep{LeRoy-2015, Altwegg-2017} have revealed complex organic molecules, such as glycolaldehyde (CH$_2$OHCHO), ethylene glycol ((CH$_2$OH)$_2$), and glycine.
The presence of complex and even pre-biotic species suggests that some of these molecules that formed during the earliest stage of stellar formation are preserved until the formation of small bodies. The extent to which the most complex species were formed at the earliest evolutionary stage of the protosolar nebula and how they were preserved or transformed during the formation of planetesimals remain open and complex questions. 

In this study, we focus  on the formation of COMs during the deeply embedded Class 0 stage.
Class 0 protostars are characterised by an envelope that is rich in diverse molecules, where most of the mass of the system is gathered \citep{Andre-1994}. COMs may be observed in the inner regions of the envelope, which is close to their host protostar, where the temperature exceeds the water ice desorption temperature of $\sim$100~K. These molecule-rich regions are called hot cores for high-mass protostars and hot corinos for low-mass counterparts. At sub-millimetre wavelengths, most of the lines originate from complex species. The measured COM abundances are typically between $10^{-7}$ and $10^{-11}$ with respect to H$_2$ \citep[e.g.][]{Herbst-2009}. In general, the formation of COMs is thought to occur on ice surfaces through a succession of different processes that are initiated by efficient hydrogenation of simple ice species, for example, CO, which leads to the efficient production of methanol \citep[CH$_3$OH;][]{Garrod-2006, Garrod-2008, Fuchs-2009, Chuang-2018}. During the collapse of the pre-stellar core, the temperature starts to increase during the so-called warm-up phase, and more complex species form and desorb into the gas phase in the hot corino. The resulting abundances of these species in the gas depend on the initial abundances of the precursors in the parent cloud, but they are also very sensitive to the physical conditions of the pre-stellar and protostellar phases and may thus vary significantly for different protostellar sources \citep{Garrod-2013, Drozdovskaya-2016}. While the gas-phase formation pathways were not efficient enough to explain such high abundances of the COMs observed in the dense warm envelope, the recent progress in quantum chemical modelling brought attention back to the gas-phase chemistry \citep[e.g.][]{Balucani-2015, Skouteris-2018, Skouteris-2019}. Similar progress has been made recently in the study of the deuteration of COMs through gas-phase reactions \citep{Skouteris-2017}. Careful comparisons between observations and predictions from grain surface and gas-phase reaction experiments and calculations are needed to address the relative importance for the species that are present in their different protostellar environments.

Close protostellar multiple sources, that is, systems with multiple components separated by $\lesssim1000$ au, are particularly interesting laboratories for studying the effect of the physical conditions with time. Indeed, the chemical evolution of such sources is expected to stem from the same initial compositions and physical conditions (gas and dust temperatures as well as ultra-violet irradiation), which are inherited from the common parental cloud.
The Class~0 low-mass protostellar binary IRAS 16293--2422 (IRAS 16293 hereafter), which is located in the $\rho$ Ophiuchus cloud complex at a distance of $\sim$140 pc \citep{Dzib-2018}, is one of the best sources to investigate for that purpose. The components of the protostellar binary are known to be rich in COMs with abundances that are comparable to those seen towards high-mass protostars \citep{VanDishoeck-1995, Cazaux-2003, Caux-2011}. 
Previous interferometric observations revealed abundance differences of specific molecules between the two protostars \citep{Bisschop-2008, Jorgensen-2011, Jorgensen-2012}; however, the origin of the differentiation is still debated. 
The ALMA Protostellar Interferometric Line Survey\footnote{The reduced dataset and all the publications are available on the website \href{http://youngstars.nbi.dk/PILS/}{http://youngstars.nbi.dk/PILS/}} \citep[PILS,][]{Jorgensen-2016} performed an unbiased spectral survey of the system in Band~7, demonstrating the strength of ALMA for detecting new complex species towards the most compact component IRAS 16293B of the system, including CH$_3$Cl \citep{Fayolle-2017}, CH$_3$NC \citep{Calcutt-2018a}, HONO \citep{Coutens-2019}, NH$_2$CN \citep{Coutens-2018}, and CH$_3$NCO \citep{Ligterink-2017}. In addition, many deuterated and less abundant isotopologues have been found for the first time in the interstellar medium (ISM) towards this source, including NHDCHO, NH$_2$CDO \citep{Coutens-2016}, NHDCN \citep{Coutens-2018}, CHD$_2$CN \citep{Calcutt-2018b}, deuterated and $^{13}$C isotopologues of CH$_2$OHCHO, deuterated ethanol (C$_2$H$_5$OH), ketene (CH$_2$CO), acetaldehyde (CH$_3$CHO) and formic acid (HCOOH) \citep{Jorgensen-2016, Jorgensen-2018}, OC$^{33}$S \citep{Drozdovskaya-2018}, and doubly deuterated methyl formate \citep[CHD$_2$OCHO;][]{Manigand-2018}. This paper aims to study the O-bearing COMs in the other protostar IRAS 16293A of the binary system and to compare their abundances with IRAS 16293B. Because of the proximity of the two sources (5"; $\sim$720 au), the two sources are expected to have the same molecular heritage as their parent cloud. Therefore, any abundance difference between the sources could reveal the chemistry that occurred during the earliest evolutionary phases of this low-mass protostellar binary.

Similar comparative studies have been carried out towards the Class 0 binary protostar NGC 1333 IRAS 4A. \cite{LopezSepulcre-2017} observed this source at high angular resolution with ALMA and measured the abundances of eight COMs towards the two components of the binary. The authors notice a large difference in abundance for all the species and suggest that the lack of COMs towards one of the source could be due to a lower mass, with a lower accretion rate, compared to the molecule rich counterpart. This scenario could alternatively suggest that the COM-poor source is not yet in the collapse phase resulting in the formation of the hot corino, which is consistent with the lack of evidence for the presence of a hot corino in this component \citep{Persson-2012}. In contrast, both components in IRAS 16293 are known to harbour hot corinos.

In this study, we present an analysis of the PILS data taken towards IRAS 16293A, which focus on the content of the O-bearing molecules in the hot corino region. This work aims to compare the abundances found for both components of the binary system to better understand the structure and/or the chemistry of this particular binary source. 
This paper is organised as follows. In
Section~2, we describe the observations and spectroscopic data used in this study. 
In Section 3, we present the results of the observations and analyse the spectrum. 
In Section~4, we compare the abundances with those in IRAS 16293B and discuss the different aspects of the two sources. Finally, we summarise the key points of the analysis and the discussion in Section 5.

\section{Data analysis}

In this section, we present the results of the oxygen-bearing COMs identification using the local thermodynamic equilibrium (LTE) analysis. In Section \ref{sec-obs} we present the observations used in this study and the species identification is described in Section 2.2. The LTE model and the $\chi^2$-fitting are detailed in Section \ref{sec-LTE} and \ref{sec-chi2}, respectively.  Finally, the treatment of the uncertainties and the continuum correction are addressed in Section \ref{sec-unc}.

\subsection{\label{sec-obs}Observations}

\begin{figure*}[t]
\centering
\includegraphics[scale=0.7]{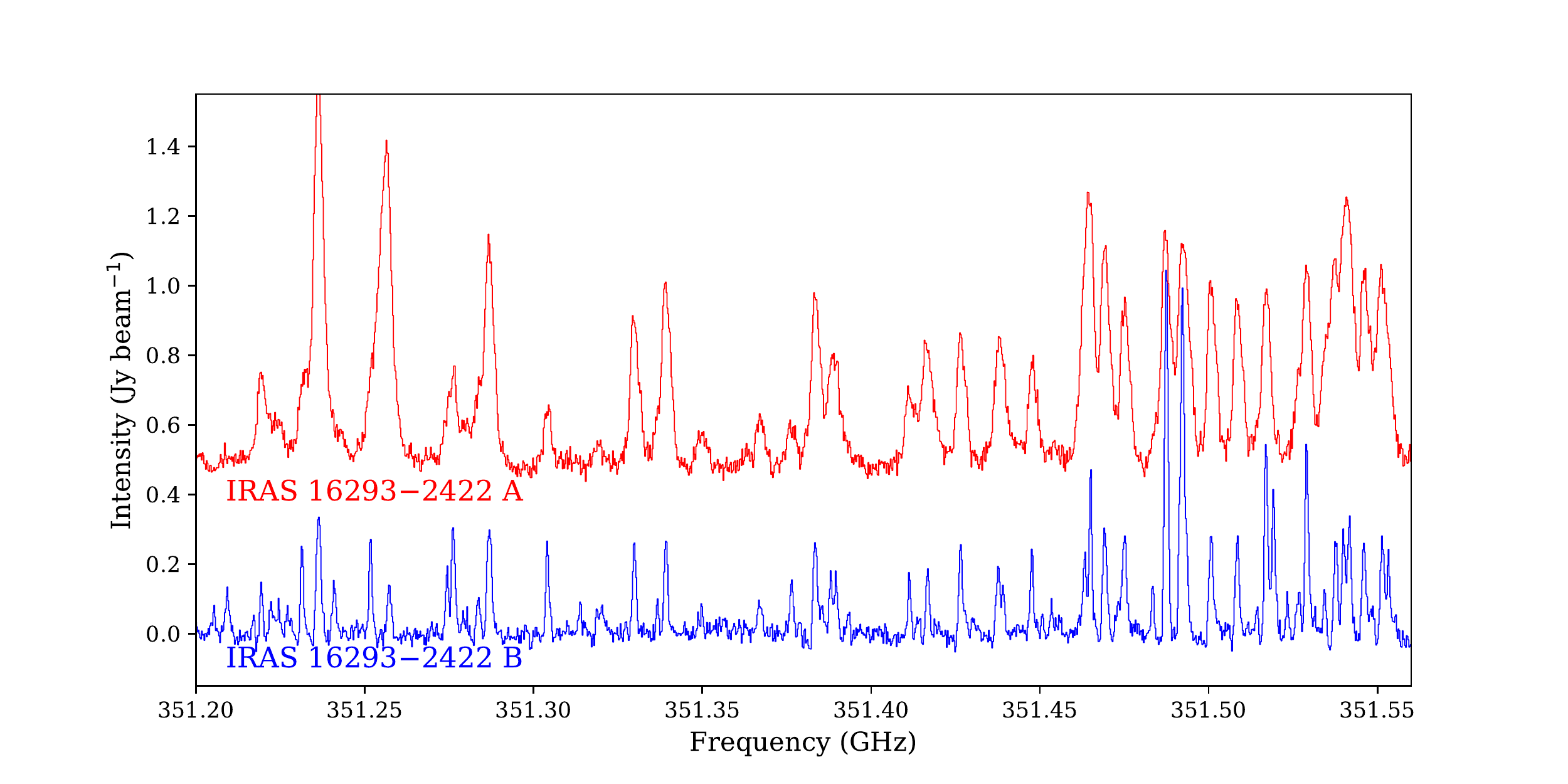}
\caption{\label{fig-exemple-spec} Portion of the spectrum towards IRAS 16293A and IRAS 16293B in red and blue, respectively, extracted on their corresponding offset positions. IRAS 16293A spectrum is offset by 0.5~Jy\,beam$^{-1}$ for illustrative purposes.}
\end{figure*}

We used ALMA data from the PILS survey of the low-mass protostellar binary IRAS 16293 \citep{Jorgensen-2016}.  The data are a combination of the 12~m dishes and the Atacama Compact Array (ACA). The observations cover the frequency range of 329.1 to 362.9~GHz with a spectral resolution of 0.244~MHz, corresponding to $\sim$0.2 km~s$^{-1}$, and a restoring beam of 0\farcs 5. The survey reaches a sensitivity of 4--5 mJy\,beam$^{-1}$\,km\,s$^{-1}$ across the entire frequency range. The ALMA software suite CASA \citep{McMullin-2007} was used to reduce the data, except for the continuum subtraction, which was performed after the data reduction \citep[for more details, see][]{Jorgensen-2016}. The flux calibration across the spectral range is consistent with the 5\% level.
The spectroscopic data of all the species that are analysed in this paper are detailed in Appendix~\ref{sec-spectro} as well as the vibrational correction factors used, in the case of rotational temperatures that are higher than 100~K.

\subsection{\label{sec-results}Species identification}

In this study, oxygen-bearing COMs have been investigated towards IRAS 16293A and IRAS 16293B through the analysis of the spectral emission at one position towards each source. The continuum peak positions, the strong continuum emission, the dynamics of the gas, and the spectral line density make it difficult to analyse the molecular emission. Therefore, the spectra were extracted towards offset positions
at 0\farcs6 from the continuum peak position in the north-east direction
and 0\farcs5 in the south-west direction for IRAS 16293A and B, respectively. This is in agreement with the choice of the extracted spectrum position in the previous PILS studies \citep[e.g.][]{Calcutt-2018b, Jorgensen-2018, Lykke-2017}. Figure~\ref{fig-exemple-spec} shows the spectra towards IRAS 16293A and B.  Most of the lines identified towards IRAS 16293B also appear in IRAS 16293A, albeit at different intensities, which indicates that IRAS 16293A is as rich in molecules as its companion. The line widths are typically two to three times larger towards IRAS 16293A than towards IRAS 16293B, thereby increasing the confusion due to blending effects.

The complexity of the molecular emission towards IRAS16293A and B makes the line assignment to the species difficult when based on the peak frequency alone. In some cases, a species that was not considered to be present in the source might have a significant number of transitions with frequencies coinciding with bright transitions of other species. In order to deal with such line confusion, we coupled the identification of possible O-bearing COMs present in the gas to the LTE analysis. 
All the species detected in the counterpart protostar IRAS16293B and their fainter isotopologues were considered as potential candidate species that compose the envelope of IRAS16293A. Additionally, related and more complex O-bearing COMs have also been added to the list of candidates, such as trans-ethyl methyl ether (\emph{t}-C$_2$H$_5$OCH$_3$) and methoxymethanol (CH$_3$OCH$_2$OH).

\subsection{\label{sec-LTE}LTE model and detection criteria}

We compared the extracted spectrum with a synthetic model in order to identify and derive the abundances of species that are present in the gas. The model applies the radiative transfer equations to a homogeneous column of gas along the line of sight under the LTE assumption. Only the optically thin lines, which have a line opacity lower than 0.1, are considered in the comparison of the modelled spectrum. For each species, a single excitation temperature was assumed to describe the level populations, which we refer to as rotational temperature.

The brightness temperature $T_\mathrm{b}(v)$ depends on the line opacity $\tau(v)$, through the relation:
\begin{equation}\label{eq:Tb}
T_\mathrm{b}(v) = T_\mathrm{0} \left( \Omega\frac{1}{\mathrm{e}^{T_\mathrm{0}/T_\mathrm{rot}}-1} - \frac{1}{\mathrm{e}^{T_\mathrm{0}/T_\mathrm{CMB}}-1} \right) \left( 1-\mathrm{e}^{-\tau(v)} \right),
\end{equation}
with $T_\mathrm{0} = \mathrm{h}\nu/\mathrm{k_B}$, $T_\mathrm{rot}$, the rotational temperature of the gas and the cosmic microwave background temperature $T_\mathrm{CMB}=2.73$~K. The filling factors of the gas emission, $\Omega$, is equal to 0.5 \citep{Jorgensen-2016}, corresponding to a source size of 0{\farcs}5. The source size is the same as what was used in the previous studies of IRAS 16293B based on the same dataset.

Under the LTE assumption and assuming a Gaussian profile, the line opacity is expressed as:
\begin{equation}
\tau(v) = \tau_\mathrm{0} \exp{\frac{-(v - v_\mathrm{LSR})^2}{2}\left(\frac{2\sqrt{2\mathrm{ln}2}}{\Delta\nu_\mathrm{FWHM}}\right)^2}
\end{equation}
\begin{equation}
\text{and } \tau_0 = \frac{2\sqrt{2\mathrm{ln}2}}{\Delta v_\mathrm{FWHM}\sqrt{2\mathrm{\pi}}} \ \frac{N_\mathrm{tot} g_\mathrm{u} A_\mathrm{ul} c^3}{8\mathrm{\pi} Q(T_\mathrm{rot}) \nu_\mathrm{0}^3} \left(\mathrm{e}^{T_\mathrm{0}/T_\mathrm{rot}} - 1\right)  \mathrm{e}^{-\frac{E_\mathrm{u}}{\mathrm{k_B} T_\mathrm{rot}}} 
\end{equation}
where $g_\mathrm{u}$ is the upper state degeneracy, $E_\mathrm{u}$ is the upper state energy, $A_\mathrm{ul}$ is the Einstein coefficient of the transition, $N_mathrm{tot}$ is the column density, and $Q(T_\mathrm{rot})$ is the partition function evaluated at $T_\mathrm{rot}$. We note that the brightness temperature and the line opacity are functions of the velocity $v$, representing the velocity shift with respect to the peak velocity, $v_\text{LSR}$, which was set at the rest frequency of the modelled line. The parameters $T_\mathrm{ex}$, $N_\mathrm{tot}$, $v_\mathrm{LSR}$ and $\Delta v_\mathrm{FWHM}$, the full line width at half-maximum in velocity units, are free parameters in the model. 

The detection criteria for a species in the PILS data is mostly affected by the high density of the line present in the spectrum, which is more problematic towards IRAS16293A due to the broader lines. A line is considered to be unblended, and thus usable for the fit, in the following cases: if there is no line from another species in the $\nu_0\pm2$FWHM frequency range (i.e. Rayleigh criteria); and if there is a line in the $\nu_0\pm2$FWHM frequency range but the peak intensity of this line is lower than 10\% the peak intensity of the line of interest or less than 1$\sigma$.

\noindent The frequency range considered in the Rayleigh criteria is extended to $\pm 4$FWHM at the proximity of a very optically thick blending line, for example, the low-E$_{up}$ CH$_3$OH lines.

\subsection{\label{sec-chi2}$\chi^2$-fitting}

Towards the analysed position, 0\farcs 6 north-east of IRAS 16293A, a line width FWHM of 2.2--2.4~km\,s$^{-1}$, and a peak velocity $v_\text{lsr}$ of 0.8~km\,s$^{-1}$ reproduce the emission line shapes of all the species in this study, suggesting that they are indeed located in the same region around the protostar. The line width and the peak velocity towards the offset position from IRAS 16293B, which are 1.0 and 2.7~km\,s$^{-1}$, respectively, are consistent with the range of values from the previous PILS studies \citep[0.8--1.0~km\,s$^{-1}$ and 2.5--2.7~km\,s$^{-1}$, respectively,][]{Coutens-2016, Lykke-2017, Calcutt-2018a, Persson-2018, Ligterink-2017}.

The species that are considered have been fitted on a grid in column density and rotational temperature, ranging from 10$^{14}$ to 10$^{18}$~cm$^{-2}$ in logarithmic scale and from 75 to 300~K by increments of 5~K, respectively. Each of them were iteratively fitted from the most to least abundant and put in a reference spectrum that was taken into account for the fit of the next species. 

The comparison between the observed spectrum and the model is quantified by the $\chi^2$ minimisation, which is defined as
\begin{equation}\label{eq:chi2}
\chi^2 = \sum_\text{lines}{\int_\text{line profile} \omega(v)\left(\frac{I_\text{b}(v) - T_\text{b}(v)}{\sigma_\text{b}(v)}\right)^2 \text{d}v},
\end{equation}
where $I_b(v)$ and $\sigma_b(v)$ are the line intensity and the intensity uncertainty of the extracted spectrum, respectively. We note that the intensity uncertainty includes the contribution from the calibration uncertainty $\sigma_\text{cal}$, the root mean square (RMS) $\sigma_\text{RMS}$, and the following: $\sigma_\text{b}(v) = \sqrt{\sigma^2_\text{RMS}(v) + \sigma^2_\text{cal}(v)}$.

The weighting factor $\omega(v)$ is equal to three if $T_b(v) > I_b(v)$, and it is one otherwise. In the case of this study, this weighting factor provides a better constraint on possible blending effects, which could bias the minimisation. In other words, the weighting factor increases the $\chi^2$ value when the model is above the spectrum. We chose to break the symmetry of the model that way to allow the fitting procedure to converge to lower column density values compared to those derived with a symmetric $\chi^2$, which could include possible contributions from species that are close enough for blending. The choice of the weighting factor value of three is arbitrary and depends on the degree of line confusion in the data.

\subsection{\label{sec-unc}Uncertainties and continuum corrections}

The uncertainties are usually given by the covariance matrix, which is estimated in the standard $\chi^2$ minimisation methods. However, the modified $\chi^2$ used in this study no longer allows for the use of the covariance matrix to get the uncertainties. Instead, the uncertainties have been estimated using a simple Monte Carlo simulation, where the spectrum as well as an additional Gaussian noise of $\sigma_b(v)$ amplitude were fitted to the synthetic spectrum several times. This gives a distribution of each parameter; here, the excitation temperature and the column density are called the posterior probability distributions. This versatile method has also been successfully used to estimate the uncertainties on the 3~$\mu$m OH band model parameters as well as in reflectance spectroscopy of meteorites and asteroids \citep{Potin-subm}.
In summary, the posterior probability distribution of the parameters represents the distribution of the possible values that can take each parameter, given the uncertainties of the data. The formalism of this Monte Carlo simulation is detailed in Appendix \ref{sec-MCerror}. 

This statistical estimation of the uncertainties gives much lower relative errors on the excitation temperature and column density compared to the actual data relative uncertainties. This suggests that the proper statistical estimate of the errors is not conservative considering the underlying hypotheses of the LTE model and how they are handled in the minimisation. 
Indeed, the $\chi^2$ minimisation assumes that the model is perfect and determines the best set of parameters to fit the observations. However, the observations include many effects that are neglected by the LTE analysis but which appear in the spectrum. These effects can be, for example, non-uniform source coverage, beam dilution, self-absorption effects, or abundance gradients of the species in the line of sight (with the associated velocity shift). 
Therefore, we chose to take conservative values of $\sim$20 and $\sim$30\% for the rotational temperatures and the column densities relative uncertainties, respectively, for IRAS 16293A. For consistency, the relative uncertainty values used in the study of \cite{Jorgensen-2018} are assumed for IRAS 16293B, that is, $\sim$20 and $\sim$10\% for the rotational temperatures and the column densities, respectively.

The dust emission is treated as a continuum brightness temperature, which is directly derived from the continuum emission in the same frequency range. The high density of both sources makes it so the dust emission is partially coupled to the gas emission. In order to take into account the continuum contribution to the line emission, the correction factor $A_\text{corr}$ was applied to the column density
\begin{equation}
A_\text{corr} = \frac{\Omega\frac{1}{\mathrm{e}^{T_\mathrm{0}/T_\mathrm{rot}}-1} - \frac{1}{\mathrm{e}^{T_\mathrm{0}/T_\mathrm{CMB}}-1}}{\Omega\frac{1}{\mathrm{e}^{T_\mathrm{0}/T_\mathrm{rot}}-1} - \Omega_\text{dust}\frac{1}{\mathrm{e}^{T_\mathrm{0}/T_\mathrm{d}}-1}}
\end{equation} 
in which $T_\text{d}$ is the dust brightness temperature and $\Omega_\text{dust}$ is the dust emission filling factor, which is assumed to be equal to $\Omega$.
Under the LTE assumption, the use of a correction factor is equivalent to replacing $T_\text{CMB}$ with the dust brightness temperature $T_\text{d}$ in Equation \ref{eq:Tb}. The dust brightness temperatures used in this paper are 30.1~K and 21.2~K towards the offset position from continuum peak positions towards IRAS16293A and B, respectively.
The value of the continuum correction factor ranges from $\sim$1.1 at 300~K to $\sim$1.4 at 90~K and it does not strongly depend  on the frequency across the spectral range of the PILS data.

\begin{table*}[t]
\caption{\label{tab-fitresults} Best-fit results of all the detected and tentatively detected species towards IRAS 16293A.}
\centering
\begin{tabular}[height=0.99\textheight]{lccccccc}
\hline\hline
Species & \#$_\text{lines}$ & $E_\text{up}$ range & $T_\text{rot}$ & $N_\text{tot}$ & \multicolumn{2}{c}{$N_\text{tot}$/$N_\mathrm{CH_3OH}$ \hspace{0.5cm} $N_\text{tot}$/$N_\text{main}$} & isotopic ratio\tablefootmark{a} \\
 & & (K) & (K) & (cm$^{-2}$) & & & \\ 
\hline
CH$_3$OH & 18  & $61-1201$ & $130\pm 26$ & $1.3\pm0.4\times10^{19}$ & \lcol{ } & \Large \textcolor{white}{I} \\ 
CH$_3^{18}$OH & 7 & $35-338$ & $130\pm 26$\tablefootmark{c} & $2.3\pm0.6\times10^{16}$ & \rcol{$1.8\pm0.8 \times10^{-3}$} & $565\pm237$\\ 
$^{13}$CH$_3$OH & 8 & $329-653$ & $130\pm 26$\tablefootmark{d} & $2.0\pm0.7\times10^{17}$ & \rcol{$1.5\pm0.6\times10^{-2}$} & $65\pm27$ \\ 
CH$_2$DOH & 24 & $61-896$ & $130\pm 26$ & $1.1\pm0.3\times10^{18}$ & \rcol{$8.3\pm3.6\times10^{-2}$} & $2.8\pm0.4$ \% \\ 
CH$_3$OD & 6& $104-288$ & $130\pm 26$\tablefootmark{e} & $2.8\pm0.8\times10^{17}$ & \rcol{$2.2\pm0.9\times10^{-2}$} & $2.2\pm0.9$ \% \\ 
CH$_3$OCHO\tablefootmark{$\dagger$} & 159 & $86-693$ & $115\pm 23$ & $2.7\pm0.8\times10^{17}$ & \lcol{$2.1\pm0.9\times10^{-2}$} & \Large \textcolor{white}{I} \\ 
CH$_3$O$^{13}$CHO & 6 & $255-308$ & $115\pm 23$\tablefootmark{$\star$} & $3.6\pm1.1\times10^{15}$ & \rcol{$1.3\pm0.5\times10^{-2}$} & $75\pm32$ \\ 
CH$_2$DOCHO\tablefootmark{$\dagger$} & 82 & $62-417$ & $115\pm 23$ & $2.3\pm0.7\times10^{16}$ & \rcol{$8.5\pm3.6\times10^{-2}$} & $2.8\pm0.4$ \% \\ 
CH$_3$OCDO\tablefootmark{$\dagger$} & 3 & $282-369$ & $115\pm 23$ & $4.5\pm1.3\times10^{16}$ & \rcol{$1.7\pm0.7\times10^{-2}$} & $1.7\pm0.7$ \% \\ 
CHD$_2$OCHO\tablefootmark{$\dagger$} & 5 & $266-319$ & $115\pm 23$ & $5.3\pm1.6\times10^{15}$ & \rcol{$2.0\pm0.8\times10^{-2}$} & $8.2\pm0.6$ \% \\ 
CH$_3$OCH$_3$ & 7 & $122-662$ & $100\pm 20$ & $5.2\pm1.6\times10^{17}$ & \lcol{$4.0\pm1.7\times10^{-2}$} & \Large \textcolor{white}{I} \\ 
CH$_3$O$^{13}$CH$_3$ & 11 & $47-234$ & $100\pm 20$ & $1.2\pm0.4\times10^{16}$ & \rcol{$2.3\pm1.0\times10^{-2}$} & $86\pm18$\\ 
s-CH$_2$DOCH$_3$ & 10 & $46-337$ & $100\pm 20$ & $2.4\pm0.7\times10^{16}$ & \rcol{$4.6\pm1.9\times10^{-2}$} & $2.3\pm0.5$ \%\\
a-CH$_2$DOCH$_3$ & 20 & $68-243$ & $100\pm 20$\tablefootmark{f} & $6.4\pm2.0\times10^{16}$ & \rcol{$1.2\pm0.5\times10^{-1}$} & $3.1\pm0.3$ \%\\
H$_2$CO & 2 & $173-(1542)$ & $155\pm 31$\tablefootmark{$\star$} & $1.4\pm0.4\times10^{17}$ & \lcol{$1.1\pm0.5\times10^{-2}$} & \Large \textcolor{white}{I} \\ 
H$_2$C$^{18}$O & 3 & $97-279$ & $155\pm 31$\tablefootmark{$\star$} & $2.1\pm1.0\times10^{14}$ & \rcol{$1.5\pm0.6\times10^{-3}$} & $667\pm280$ \\ 
H$_2^{13}$CO & 4 & $98-240$ & $155\pm 31$ & $1.5\pm0.5\times10^{15}$ & \rcol{$1.1\pm0.5\times10^{-2}$} & $93\pm39$ \\ 
H$_2$C$^{17}$O & 4 & $61-157$ & $155\pm 31$\tablefootmark{$\star$} & $<6.0\times10^{13}$ & \rcol{$<4.3\times10^{-4}$} & $>2333$ \\ 
HDCO & 4 & $26-219$ & $155\pm 31$\tablefootmark{$\star$} & $6.9\pm2.1\times10^{15}$ & \rcol{$4.9\pm2.1\times10^{-2}$} & $2.5\pm0.5$ \%\\ 
D$_2$CO & 3 & $127-370$ & $155\pm 31$\tablefootmark{$\star$} & $5.8\pm1.7\times10^{15}$ & \rcol{$4.1\pm1.7\times10^{-2}$} & $20.2\pm4.2$ \%\\ 
C$_2$H$_5$OH & 59 & $72-453$ & $135\pm 27$ & $8.0\pm2.4\times10^{16}$ & \lcol{$6.1\pm2.6\times10^{-3}$} & \Large \textcolor{white}{I} \\ 
CH$_3$CHDOH\tablefootmark{b} & 10 & $55-198$ & $135\pm 27$\tablefootmark{g} & $7.0\pm2.1\times10^{15}$ & \rcol{$8.8\pm3.7\times10^{-2}$} & $4.4\pm0.9$ \% \\ 
CH$_3$CH$_2$OD\tablefootmark{b} & 9 & $48-192$ & $135\pm 27$\tablefootmark{$\star$} & $<3.5\times10^{15}$ & \rcol{$<4.4\times10^{-2}$} & $<4.4$ \% \\ 
a-CH$_2$DCH$_2$OH\tablefootmark{b} & 9 & $62-205$ & $135\pm 27$\tablefootmark{$\star$} & $<3.5\times10^{15}$ & \rcol{$<4.4\times10^{-2}$} & $<4.4$ \% \\ 
s-CH$_2$DCH$_2$OH\tablefootmark{b} & 7 & $48-199$ & $135\pm 27$\tablefootmark{$\star$} & $<5.2\times10^{15}$ & \rcol{$<6.5\times10^{-2}$} & $<3.3$ \% \\ 
t-HCOOH & 8 & $35-530$ & $90\pm 18$ & $1.3\pm0.4\times10^{16}$ & \lcol{$1.0\pm0.4\times10^{-3}$} & \Large \textcolor{white}{I} \\ 
t-H$^{13}$COOH & 6 & $141-160$ & $90\pm 18$\tablefootmark{$\star$} & $<3.3\times10^{14}$ & \rcol{$<2.5\times10^{-2}$} & $>39$ \\ 
t-DCOOH & 7 & $147-225$ & $90\pm 18$\tablefootmark{$\star$} & $<5.1\times10^{14}$ & \rcol{$<3.9\times10^{-2}$} & $<3.9$ \% \\ 
t-HCOOD & 6 & $131-208$ & $90\pm 18$\tablefootmark{$\star$} & $<2.5\times10^{14}$ & \rcol{$<1.9\times10^{-2}$} & $<1.9$ \% \\ 
CH$_2$CO & 5 & $357-996$ & $135\pm 27$ & $9.1\pm2.7\times10^{15}$ & \lcol{$7.0\pm2.9\times10^{-4}$} & \Large \textcolor{white}{I} \\ 
CHDCO & 8 & $161-313$ & $135\pm 27$\tablefootmark{$\star$} & $<3.6\times10^{14}$ & \rcol{$<4.0\times10^{-2}$} & $<2.0$ \% \\ 
HNCO & 3 & $333-794$ & $180\pm 36$ & $1.5\pm0.5\times10^{16}$ & \lcol{$1.2\pm0.5\times10^{-3}$} & \Large \textcolor{white}{I} \\
DNCO & 5 & $51-366$ & $180\pm 36$\tablefootmark{$\star$} & $<2.5\times10^{14}$ & \rcol{$<1.7\times10^{-2}$} & $<1.7$ \% \\
CH$_3$CHO & 20 & $153-385$ & $140\pm 28$ & $3.5\pm1.1\times10^{15}$ & \lcol{$2.7\pm1.1\times10^{-4}$} & \Large \textcolor{white}{I} \\ 
c-C$_2$H$_4$O & 7 & $43-376$ & $95\pm 19$ & $6.9\pm2.1\times10^{15}$ & \lcol{$5.3\pm2.2\times10^{-4}$} & \\ 
CH$_3$COCH$_3$ & 85 & $80-322$ & $125\pm 25$ & $2.4\pm0.7\times10^{16}$ & \lcol{$1.8\pm0.8\times10^{-3}$} & \\ 
CH$_3$COOH & 17 & $129-407$ & $110\pm 22$ & $4.5\pm1.4\times10^{15}$ & \lcol{$3.5\pm1.5\times10^{-4}$} & \\ 
CH$_2$(OH)CHO & 9 & $80-323$ & $155\pm 31$ & $1.3\pm0.4\times10^{15}$ & \lcol{$1.0\pm 0.4\times10^{-4}$} & \\ 
NH$_2$CHO & 8 & $151-453$ & $145\pm 29$ & $1.9\pm0.6\times10^{15}$ & \lcol{$1.5\pm0.6\times10^{-4}$} & \\ 
aGg'-(CH$_2$OH)$_2$ & 20 & $108-363$ & $145\pm 29$ & $4.4\pm1.3\times10^{15}$ & \lcol{$3.4\pm1.4\times10^{-4}$} & \\ 
gGg'-(CH$_2$OH)$_2$ & 12 & $97-115$ & $145\pm 29$\tablefootmark{$\star$} & $<4.8\times10^{15}$ & \lcol{$<3.7\times10^{-4}$} & \\
C$_2$H$_5$CHO & 3 & 297 -- 330 & $120\pm 24$ & $<1.2\times10^{15}$ & \lcol{$<9.2\times10^{-5}$} & \\
\hline 
\end{tabular}
\tablefoot{
\tablefoottext{a}{The D/H ratio, including statistics correction, is expressed in \%, whereas other isotopic ratios correspond to $^{12}$C/$^{13}$C or $^{16}$O/$^{18}$O.}
\tablefoottext{b}{Abundances of C$_2$H$_5$OH isotopologues were derived using the anti conformer entries and corrected to take the gauche conformer into account (see spectroscopic data section in Appendix \ref{sec-spectro}).}
\tablefoottext{c}{T$_\text{rot}$ converged to 105~K and N$_\text{tot}$ = $2.2\pm0.7\times10^{16}$ cm$^{-2}$.}
\tablefoottext{d}{T$_\text{rot}$ converged to 120~K and N$_\text{tot}$ = $2.3\pm0.7\times10^{17}$ cm$^{-2}$.}
\tablefoottext{e}{T$_\text{rot}$ converged to 105~K and N$_\text{tot}$ = $3.4\pm0.7\times10^{17}$ cm$^{-2}$.}
\tablefoottext{f}{T$_\text{rot}$ converged to 115~K and N$_\text{tot}$ = $6.6\pm2.0\times10^{16}$ cm$^{-2}$.}
\tablefoottext{g}{T$_\text{rot}$ converged to 125~K and N$_\text{tot}$ = $6.6\pm2.0\times10^{15}$ cm$^{-2}$.}
\tablefoottext{c-g}{The rotational temperature of the main isotopologue was  used to derive the column density, despite the convergence to a slightly different $T_{rot}$.}
\tablefoottext{$\star$}{The rotational temperature did not converge; the main isotopologue value is assumed.}
\tablefoottext{$\dagger$}{\cite{Manigand-2018}.}}
\end{table*}

\section{Results}

In total, 15 different species as well as 16 isotopologues, have been securely detected, while 11 additional species and isotopologues are tentatively identified towards IRAS 16293A. Table \ref{tab-fitresults} shows the result of the LTE analysis with the number of lines taken into account in the minimisation, the best-fit parameters (i.e. T$_\text{rot}$, N$_\text{tot}$, and $\Delta v_\text{FWHM}$), the abundance with respect to the main isotopologue for the same species, and the corresponding isotopic ratio.

\subsection{Abundances and rotational temperatures}

The relative abundances of the oxygen species are determined from their column density relative to the column density of CH$_3$OH. We note that H$_2$ was not used as a reference for the estimation of the abundances because only the lower limit of H$_2$ column density can be determined from the optically thick dust emission. \cite{Calcutt-2018b} and \cite{Jorgensen-2016} estimated the lower limit of H$_2$ column density towards IRAS16293A and B, respectively. The D/H ratio is based on the column density ratio and is corrected from the statistics of H atoms in the deuterated chemical group of the molecule. For example, the column density ratio of CH$_2$DOH is three times higher than its D/H ratio. In addition, the statistical correction includes the symmetries in the sense of the rotational spectroscopy. For example, the asymmetric deuterated dimethyl ether (a-CH$_2$DOCH$_3$) has four possible sites where the D atom could be placed, that is, out of the C--O--C plane, regardless of the rotational emission of the molecule. Thus, its column density ratio is four times higher than the D/H ratio. Similarly, the symmetric s-CH$_2$DOCH$_3$ conformer has two in-plane possible sites for the D atom, which leads to a factor of two between the column density ratio and the D/H ratio. The rotational temperature of the isotopologues was fixed to the value derived for their main isotopologue, except for formaldehyde (H$_2$CO) where only the H$_2^{13}$CO rotational temperature could be constrained. For several isotopologues, the fit converged to a different rotational temperature as the main isotopologue. The difference is lower than the uncertainty for the rotational temperature and does not significantly affect the column density. The results of the minimisation for these species are indicated in the notes of Table \ref{tab-fitresults}.
Despite the large frequency range of the survey, H$_2$CO isotopologues only have a few lines in the range, and most of them are either optically thick or the transition levels are not populated at all (E$_\text{up}>1000$~K). The rotational temperature was derived using two optically thin lines of H$_2^{13}$CO, with upper state energies of 98 and 240~K. The optically thin lines, which were used to derive the column density of the other isotopologue, are the same lines used by \cite{Persson-2018} in the analysis of IRAS 16293B.

Similar to the study of IRAS 16293B \citep{Jorgensen-2018}, a large number of rarer isotopologues were also identified towards IRAS 16293A. We note that CH$_3^{18}$OH and H$_2$C$^{18}$O show a $^{16}$O/$^{18}$O ratio close to the local interstellar medium value of 557$\pm$30 \citep{Wilson-1999}. The upper limit column density of H$_2$C$^{17}$O is consistent with the canonical $^{16}$O/$^{17}$O ratio of $2005\pm155$ \citep{Penzias-1981, Wilson-1999}. The $^{13}$C-isotopologues that were detected do not show any significant deviation from the local ISM $^{12}$C/$^{13}$C ratio of 68$\pm$15 \citep{13-C-ratio}. Lower $^{12}$C/$^{13}$C ratios have been reported towards IRAS 16293B \citep{Jorgensen-2018} and were interpreted by a chemical process that is similar to the deuteration enhancement on the ice grain surface, which occurs during the pre-stellar formation stage. However, the mean $^{12}$C/$^{13}$C ratio of 76$\pm$12 towards IRAS 16293A is consistent with the local ISM value, which indicates that the isotopic enhancement mechanism observed in IRAS 16293B does not occur in IRAS 16293A.
Deuterated isotopologues of CH$_3$OH, H$_2$CO, C$_2$H$_5$OH, dimethyl ether (CH$_3$OCH$_3$), CH$_3$OCHO, CH$_2$CO, HCOOH, and isocyanic acid (HNCO) are present in the gas with D/H ratios of 2--5\%. As it was observed towards IRAS 16293B \citep{Persson-2018, Manigand-2018}, the doubly-deuterated isotopologues D$_2$CO and CHD$_2$OCHO are significantly enhanced compared to the singly-deuterated isotopologues. The implications are discussed in the next section of the paper.

New spectroscopic data for deuterated CH$_3$CHO isotopologues were recently reported by \cite{Coudert-2019} who add information about CH$_2$DCHO to that of CH$_3$CDO from \cite{Elkeurti-2010}, which was utilised for the first detection reported by \cite{Jorgensen-2018}. To test their experimental results, \citeauthor{Coudert-2019} used the new spectroscopic data to search for CH$_2$DCHO and CH$_3$CDO in the PILS dataset. They report the presence of 93 and 43 transitions of CH$_2$DCHO and CH$_3$CDO across a 10~GHz frequency range, respectively, and they made rough estimates of the column densities based on the archival 12m array PILS data only. In order to carry out a consistent comparison of the PILS results in terms of assumptions concerning analysed positions, source sizes, velocity shift, and FWHM as well as using the combined 12m+ACA dataset, we analysed the data for those isotopologues with the same methodology in this paper as the other PILS papers. The derived column density and rotational temperature are reported in Table \ref{tab:newdetect}. The new column density reported for CH$_3$CDO is consistent with the value that was derived in the previous study of \cite{Jorgensen-2018}.

\subsection{New detections}

In addition to the 31 identified isotopologues towards IRAS 16293A, two new species, CH$_3$OCH$_2$OH and \textit{t}-C$_2$H$_5$OCH$_3$, were detected towards IRAS 16293B, exclusively. Only an upper limit column density was derived for the two species towards IRAS 16293A, assuming the same rotational temperature as for IRAS 16293B. Column densities, relative abundances to CH$_3$OH, and upper limits are reported in Table \ref{tab:newdetect}. Both species have already been detected in the ISM before; however, this is the first time they were detected in a low mass star-forming region. 
\cite{Mcguire-2017} report the detection of CH$_3$OCH$_2$OH towards the high-mass protostellar core NGC 6334I MM1, with a relative abundance of $\sim$0.03 with respect to CH$_3$OH. This abundance is significantly higher than those towards IRAS 16293A and IRAS 16293B and this is discussed in the next section along with the other O-bearing species. We note that 
\textit{t}-C$_2$H$_5$OCH$_3$ was tentatively detected in several high-mass sources, such as W51 e1/e2 \citep{Fuchs-2005}, which was refuted by \cite{Carroll-2015} and Orion KL \citep{Tercero-2015}. Later, \cite{Tercero-2018} confirmed the detection towards the Orion KL compact bridge region. The last study reported a rotational temperature of 150$\pm$20~K and a column density of $3.0\pm0.9\times 10^{15}$ cm$^{-2}$, leading to an abundance of $3.7\times10^{-9}$ with respect to H$_2$. This estimation is consistent with the abundance derived towards IRAS 16293A and B.

\begin{table}[t]
\centering
\caption{\label{tab:newdetect}\small Rotational temperature, column density, and relative abundance with respect to CH$_3$OH for the newly detected species.}
\begin{tabular}{lccc}
\hline\hline
\multirow{2}{*}{Species} & \multicolumn{3}{c}{IRAS 16293B} \\
 & $T_\text{rot}$ (K) & $N_\text{tot}$ (cm$^{-2}$) & $N_\text{tot}$ / $N_\mathrm{CH_3OH}$ \\
\hline
\textit{t}-C$_2$H$_5$OCH$_3$ & 100 & $1.8\pm0.2\times10^{16}$ &  $1.8\times10^{-3}$  \\
CH$_3$OCH$_2$OH & 130 & $1.4\pm0.2\times10^{17}$ &  $1.4\times10^{-2}$  \\
CH$_3$CDO\tablefootmark{$\dagger$} & 125\tablefootmark{$\star$} & $7.4\pm0.7\times10^{15}$ & $7.4\times10^{-4}$ \\
CH$_2$DCHO\tablefootmark{$\dagger$} & 125\tablefootmark{$\star$} & $6.2\pm0.6\times10^{15}$ & $6.2\times10^{-4}$ \\
\hline
\multirow{2}{*}{Species} & \multicolumn{3}{c}{IRAS 16293A} \\
 & $T_\text{rot}$ (K) & $N_\text{tot}$ (cm$^{-2}$) & $N_\text{tot}$ / $N_\mathrm{CH_3OH}$ \\
\hline
\textit{t}-C$_2$H$_5$OCH$_3$ & 100\tablefootmark{$\star\star$} & $<2.7\times10^{16}$ &  $<2.1\times10^{-3}$  \\
CH$_3$OCH$_2$OH & 130\tablefootmark{$\star\star$} & $<2.7\times10^{17}$ &  $<2.1\times10^{-2}$  \\
CH$_3$CDO & 140\tablefootmark{$\star$} & $<6.0\times10^{14}$ & $<4.6\times10^{-5}$ \\
CH$_2$DCHO & 140\tablefootmark{$\star$} & $<6.0\times10^{14}$ & $<4.6\times10^{-5}$ \\
\hline
\end{tabular}
\tablefoot{
\tablefoottext{$\dagger$}{CH$_2$DCHO and CH$_3$CDO were initially reported in \cite{Coudert-2019}.}
\tablefoottext{$\star$}{The rotational temperature is assumed to be the same as the main isotopologue, which is 100 and 125~K for IRAS 16293A and IRAS 16293B, respectively \citep{Lykke-2017}.}
\tablefoottext{$\star\star$}{The rotational temperature did not converge; the same value as IRAS 16293B is assumed.}
}
\end{table}

\subsection{Identification of CHD$_2$OH lines}

Thirty-one lines of CHD$_2$OH were found in the range of the observations.
CHD$_2$OH has already been detected at lower frequencies towards IRAS 16293 \citep{Parise-2002} using the IRAM 30-metre telescope. At the present day, only a small number of lines are publicly available in terms of spectroscopic data, including the line strength values, although, more extended frequency lists can be found in the literature \citep[e.g.][]{Ndao-2016, Mukhopadhyay-2016}. Therefore, this species requires special treatment.

To confirm the identification CHD$_2$OH in spectrum, each line was independently fitted to a simple
Gaussian line profile in terms of amplitude and by using the same velocity peak $v_\text{lsr}$ and FWHM as CH$_3$OH, that is, 0.8 and 2.2 km~s$^{-1}$  as well as 2.7 and 1.0 km s$^{-1}$  for IRAS 16293A and IRAS 16293B, respectively. 
The line frequencies match 24 and 16 lines of the 31 candidates, for IRAS 16293A and B, respectively, without any hint of blending effect with other species. Figures \ref{App-CHD2OH-idA-fig} and \ref{App-CHD2OH-idB-fig} show the Gaussian fits of all the lines. The integrated intensity over $\pm$FWHM around the rest frequency was derived for each unblended lines and is noted in Table \ref{App-CHD2OH-id-tab}.
The average integrated intensity of the lines assigned to CHD$_2$OH is higher than 0.3 Jy~beam$^{-1}$~km~s$^{-1}$ for IRAS 16293B. Also, several lines towards IRAS 16293B show a small blue-shifted absorption counterpart, which is a characteristic of infalling motions \citep{Pineda-2012}, which are associated with optically thick lines. In comparison to the other detected species, especially to CH$_2$DOH and CH$_3$OH, these intensities correspond to high line opacities due to a high column density. This suggests a high abundance of CHD$_2$OH as well, and it is a key factor motivating further searches for CD$_3$OH, as was already detected by \cite{Parise-2004}, and CD$_3$OD. Despite the substantial data available on CD$_3$OH and CD$_3$OD \citep{Mollabashi-1993, Walsh-1998, Xu-2004, Muller-2006}, we did not find any line that corresponds to those of these two species in the present observations.

\begin{figure*}[h!]
\centering
  \includegraphics[width=17cm]{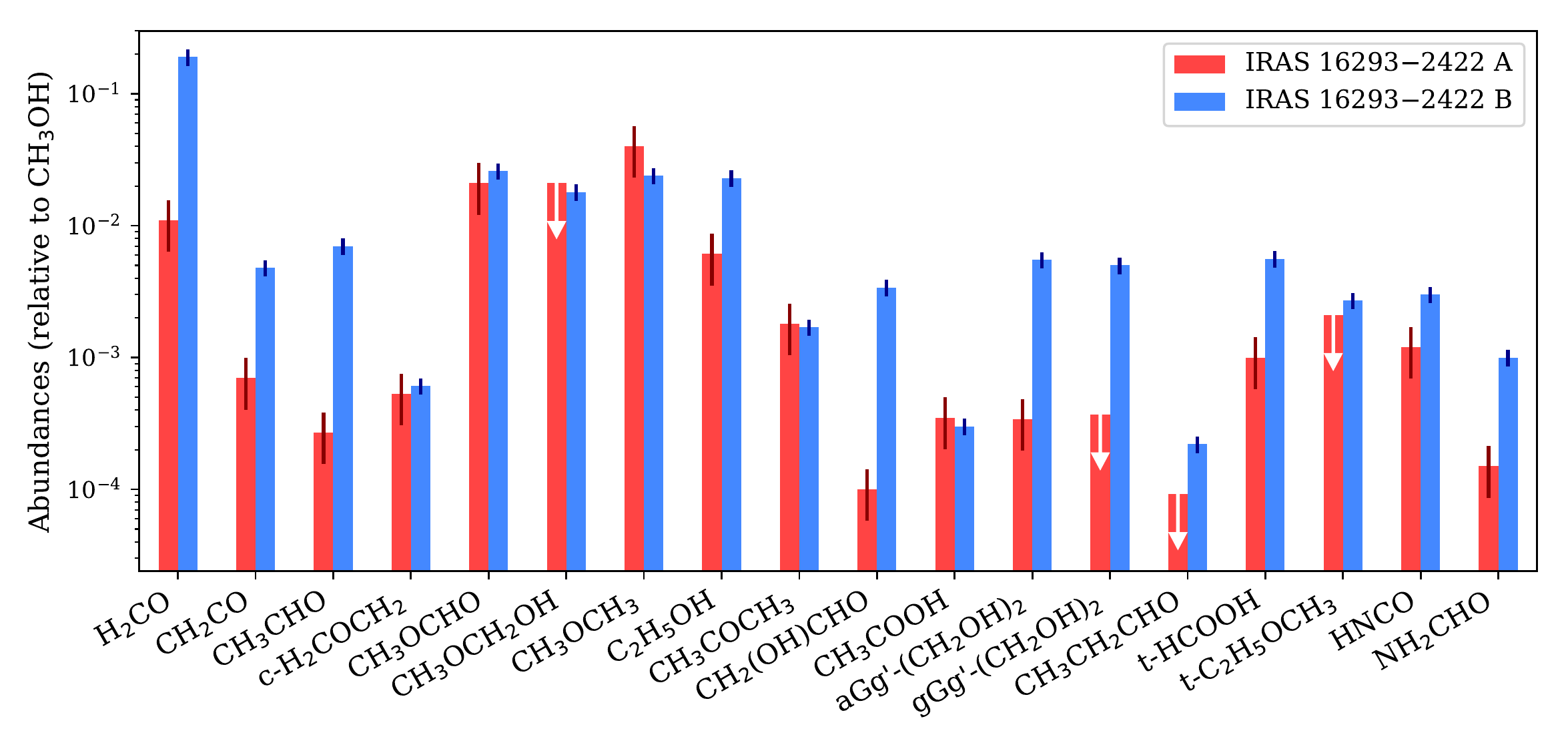}
    \caption{\small Relative abundances of the main isotopologues with respect to CH$_3$OH, towards IRAS 16293A in red and IRAS 16293B in blue. The abundances towards IRAS 16293B are based on \cite{Persson-2018} for H$_2$CO, \cite{Lykke-2017} for CH$_3$CHO, c-H$_2$COCH$_2,$ and CH$_3$COCH$_3$,  \cite{Jorgensen-2016} for CH$_2$(OH)CHO and CH$_3$COOH, and \cite{Jorgensen-2018} for CH$_2$CO, CH$_3$OCH$_3$, C$_2$H$_5$OH, CH$_3$OCHO, (CH$_2$OH)$_2,$ and t-HCOOH. We note that 30\% and 10\% of the uncertainties on the column density are considered for IRAS 16293A and IRAS 16293B, respectively.}
    \label{fig-abundances}
\end{figure*}

\section{Discussion}

\subsection{Comparison between IRAS 16293 A and B}

Most of the species and their isotopologues reported in this study are present towards both IRAS 16293A and B; however, their abundances are different in a few cases.

\begin{figure*}[h!]
\begin{subfigure}[c]{0.57\textwidth}
\includegraphics[width=\textwidth]{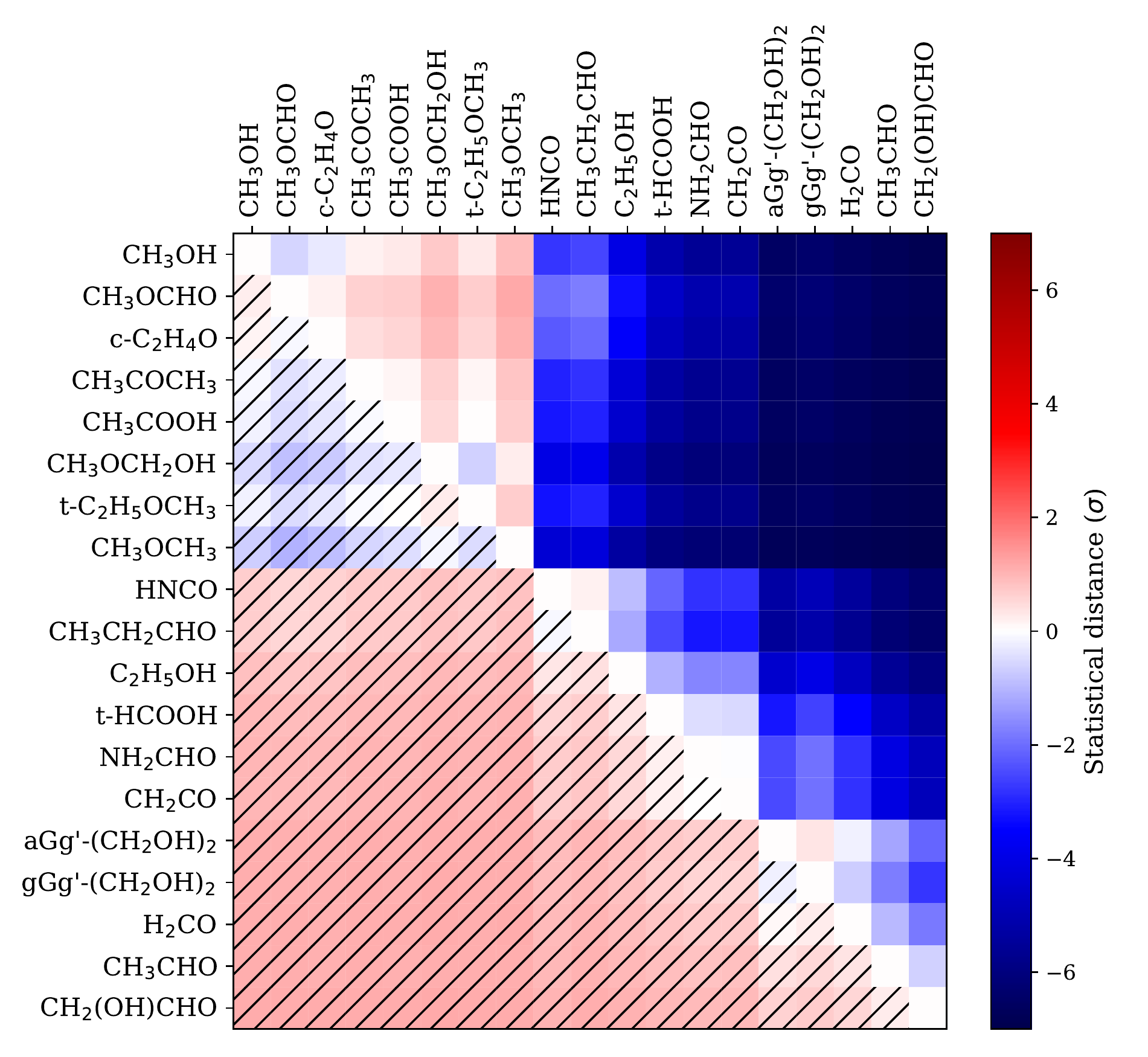}
\end{subfigure}
\begin{subfigure}[c]{0.43\textwidth}
\adjustbox{trim ={0.0\width} {0.37\height} {0.0\width} {0.0\height}, clip=true}{
    \includegraphics[width=\textwidth]{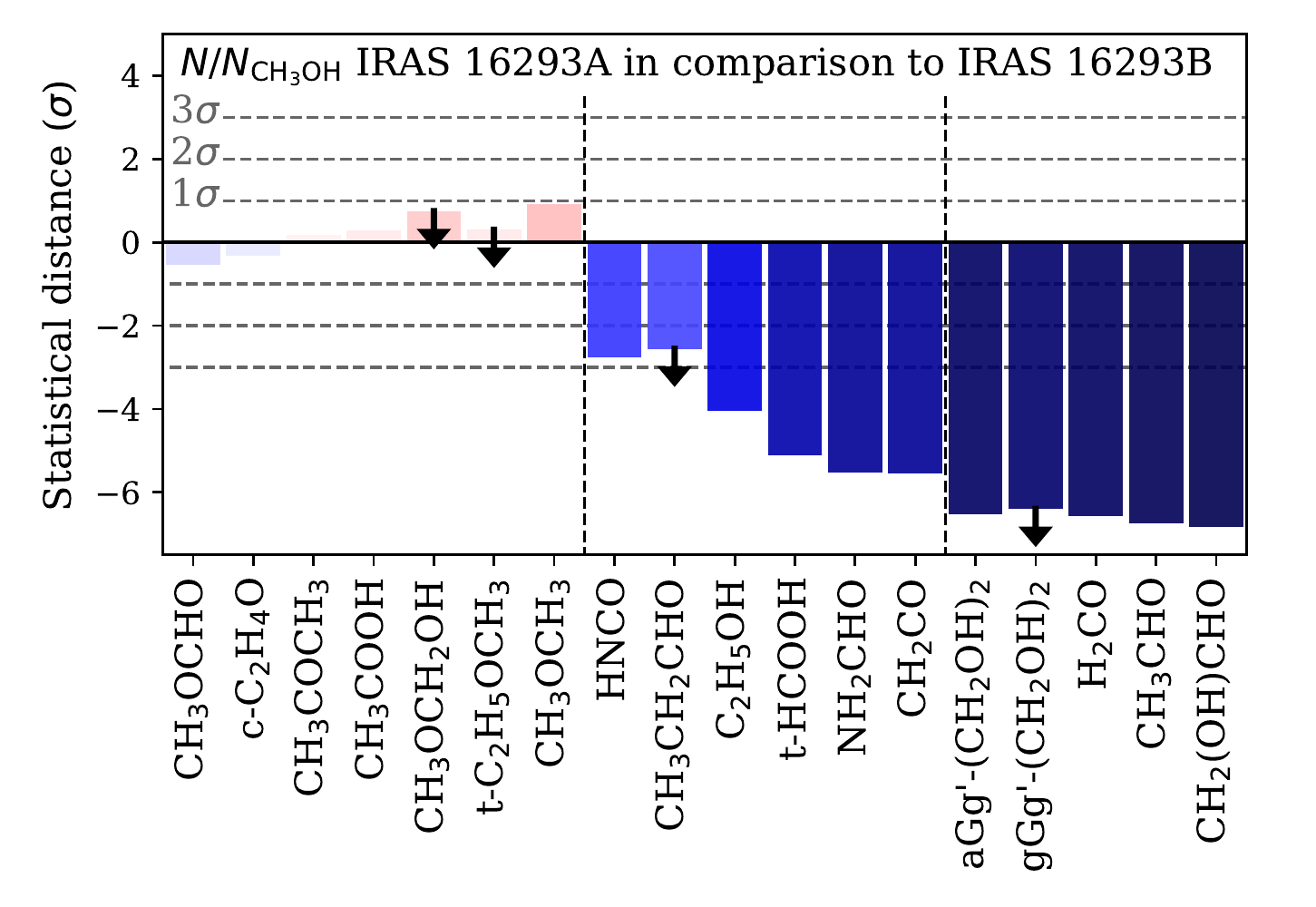}}\\
\adjustbox{trim ={0.0\width} {0.0\height} {0.0\width} {0.04\height}, clip=true}{
    \includegraphics[width=\textwidth]{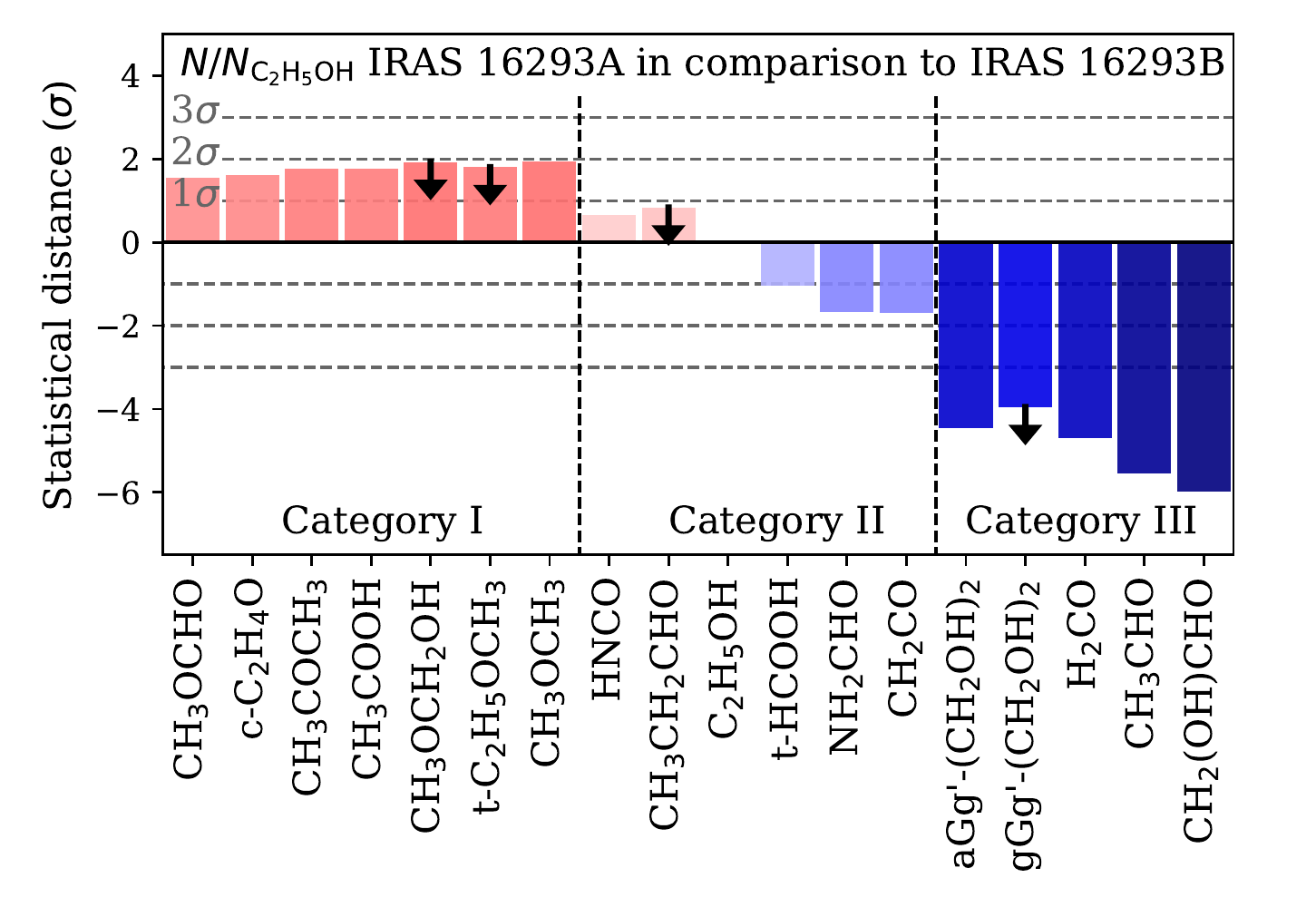}}
\end{subfigure}
\caption{\label{fig-crosscomp}\small (\emph{left}) Cross-comparison of species from IRAS 16293A in comparison to those from IRAS 16293B. The statistical distance $S$ is represented in the colour scale for each pair of species. One cell with a given value $S$ of the plot should be read as 'The abundance of the species $X$ with respect to $Y$ towards IRAS 16293A is $S$ sigma higher or lower compared to IRAS 16293B', with $X$ and $Y$ species taken from the X-axis and Y-axis, respectively. A positive value of $S$ corresponds to $N_\text{X}/N_\text{Y}$ higher in IRAS 16293A in comparison to IRAS 16293B. (\emph{right}) 'CH$_3$OH' and 'C$_2$H$_5$OH' rows of the cross-comparison map. The two plots show the comparison of abundances with respect to CH$_3$OH (upper panel) and C$_2$H$_5$OH (lower panel) towards IRAS 16293A and IRAS 16293B in terms of statistical distance.}
\end{figure*}

Figure \ref{fig-abundances} shows the abundances of the main isotopologues with respect to CH$_3$OH towards  IRAS 16293 A and B. The column densities, from which the abundances were derived, were determined by using the spectrum extracted from the offset positions, as described in the previous section for IRAS 16293A, which were used in previous studies of IRAS 16293B \citep{Coutens-2016, Lykke-2017, Jorgensen-2016}. Towards IRAS 16293A, the abundances relative to CH$_3$OH of CH$_3$OCHO, ethylene oxide (c-C$_2$H$_4$O), CH$_3$OCH$_3$, CH$_3$COCH$_3$, acetic acid (CH$_3$COOH), and HNCO are similar to the abundance observed towards IRAS 16293B. On the other hand, the abundances of CH$_2$CO, H$_2$CO, CH$_3$CHO C$_2$H$_5$OH, CH$_2$(OH)CHO, (CH$_2$OH)$_2$, \emph{t}-HCOOH, and formamide (NH$_2$CHO) are significantly lower towards IRAS 16293A than those towards IRAS 16293B. This selective differentiation could reflect the differentiation of the species across the two protostars. However, it is not possible to rule out the effects of the CH$_3$OH abundance with respect to H$_2$, for which the column density cannot be properly estimated. In addition, considering a single species as a reference for the abundance of all the species can interfere with the relation between the abundance of several species. For example, in the case of IRAS16293 and the comet 67P/Churyumov-Gerasimenko, \cite{Drozdovskaya-2019} show that it is more consistent to discuss the abundance of N-bearing and S-bearing species with respect to CH$_3$CN and CH$_3$SH, respectively.

In order to complete the picture, we estimated the statistical distance $S$ of abundance ratios of each pair of species between IRAS 16293 A and B, as detailed in Appendix \ref{sec-stat}. Figure \ref{fig-crosscomp} shows the colour map of this cross-comparison; each cell corresponds to the statistical distance of the abundance ratio of two species towards IRAS 16293A as compared to IRAS 16293B. The advantage of this comparative analysis is that the abundances do not rely on a single species. This technique offers a global view of the comparison in terms of molecular abundance between the offset position from the two sources. The first and the tenth rows of the map are shown in the right panels of the figure and correspond to the comparison of abundances relative to CH$_3$OH and C$_2$H$_5$OH, respectively, towards IRAS 16293A in comparison to IRAS 16293B. 
The upper right panel shows a clear distinction between the similar abundance of species with respect to CH$_3$OH towards IRAS 16293A and B as well as those with a higher abundance towards IRAS 16293B in comparison to those towards IRAS 16293A. This separation is represented with a vertical dashed line, between CH$_3$OCH$_3$ and HNCO. A similar separation, which is located between CH$_2$CO and H$_2$CO, is shown on the lower right panel where the abundances are plotted with respect to C$_2$H$_5$OH towards IRAS 16293A in comparison to the abundances towards IRAS16293B. These separations represent a factor of two difference in the relative abundance ratio between IRAS16293A and B. This corresponds to a significance of 2$\sigma$, given the column density uncertainties.

The cross-comparison of all the main isotopologues between IRAS 16293 A and B suggests three distinct categories of species.
The first category (noted Category I in Fig. \ref{fig-crosscomp}) includes the species, such as CH$_3$OH, CH$_3$OCHO, c-C$_2$H$_4$O, CH$_3$COCH$_3$, CH$_3$COOH, CH$_3$OCH$_2$OH, \emph{t}-C$_2$H$_5$OCH$_3,$ and CH$_3$OCH$_3$, which have similar abundances between IRAS 16293A and B, or a low absolute statistical distance. H$_2$CO, CH$_3$CHO, CH$_2$(OH)CHO, and (CH$_2$OH)$_2$ show a much lower abundance with respect to all the other species. This corresponds to the righter most part of the distance map  (Category III in Fig. \ref{fig-crosscomp}). The remaining species are in an intermediate category  (Category II in Fig. \ref{fig-crosscomp}) and are characterised by a lower abundance with respect to the first category species and a higher abundance with respect to the most depleted species for IRAS 16293A in comparison to IRAS 16293B.
This comparative analysis suggests that there are at least two processes that deplete some of the species towards the offset position of IRAS 16293A compared to the offset position of IRAS 16293B.
The possible causes of these depletions are discussed in the following sections.

\subsection{Spatial extent}

\begin{figure*}[h]
\adjustbox{trim ={0.0\width} {0.13\height} {0.24\width} {0.04\height}, clip=true}{
	\includegraphics[scale=0.47]{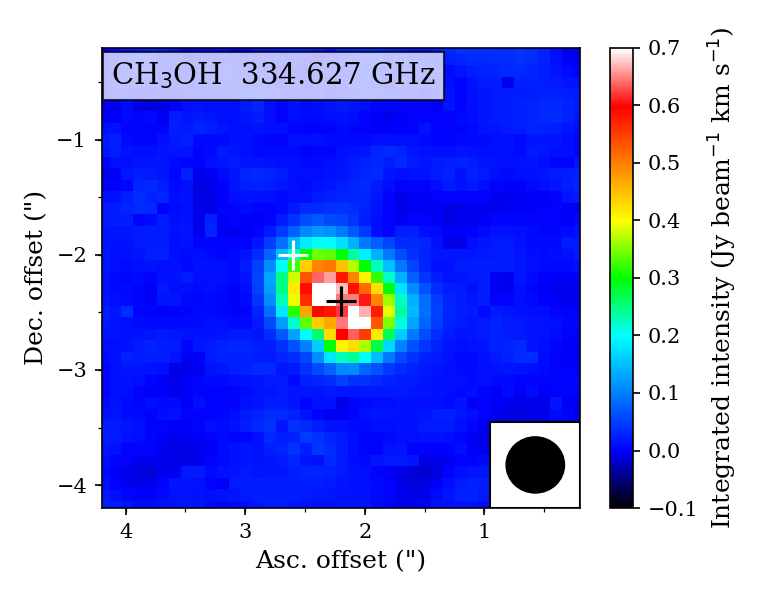} 
	}
\adjustbox{trim ={0.125\width} {0.13\height} {0.24\width} {0.04\height}, clip=true}{
	\includegraphics[scale=0.47]{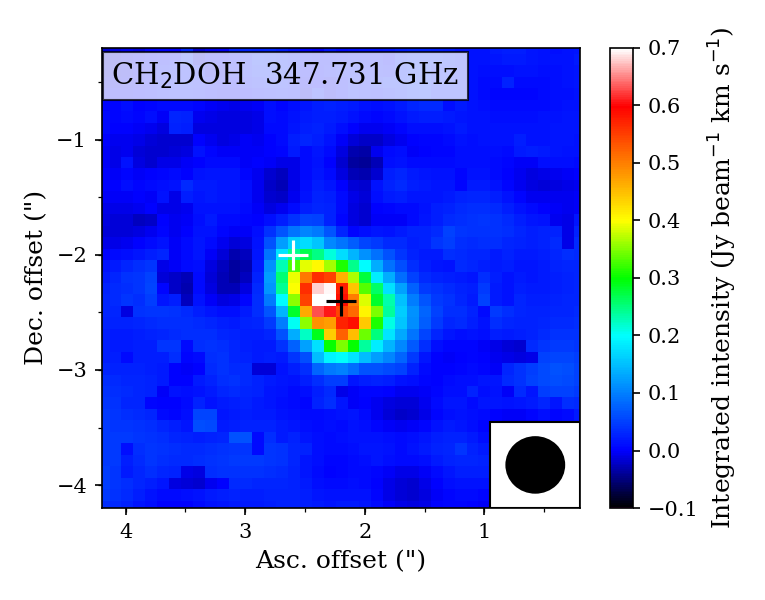} 
	}
\adjustbox{trim ={0.125\width} {0.13\height} {0.24\width} {0.04\height}, clip=true}{
	\includegraphics[scale=0.47]{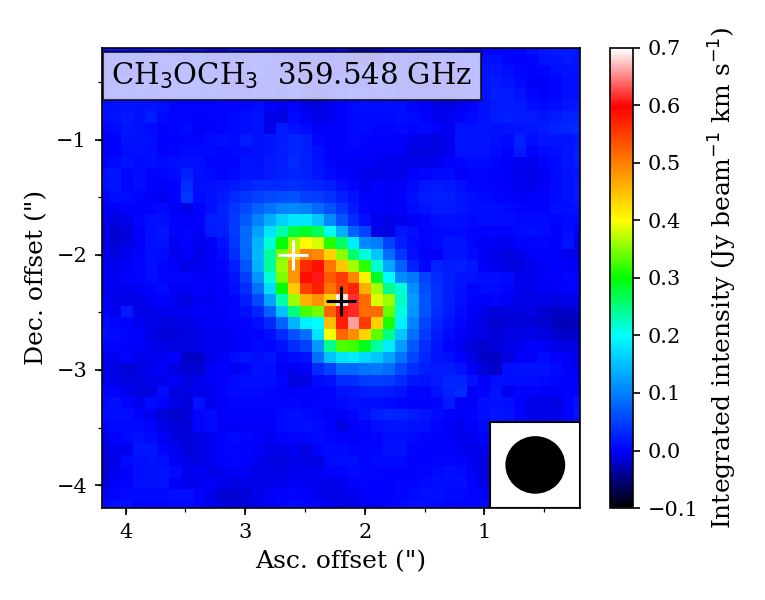} 
	}
\adjustbox{trim ={0.125\width} {0.13\height} {0.0\width} {0.04\height}, clip=true}{
	\includegraphics[scale=0.47]{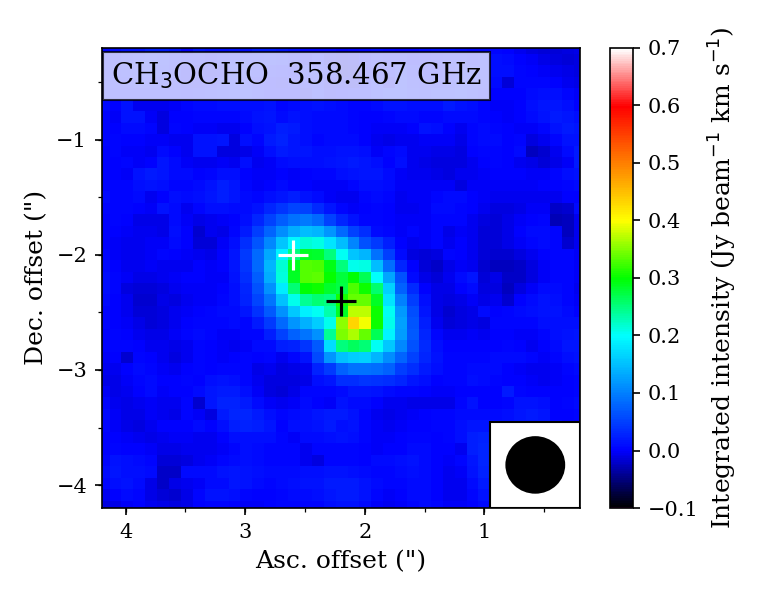} 
	} 
\\
\adjustbox{trim ={0.0\width} {0.13\height} {0.24\width} {0.04\height}, clip=true}{
	\includegraphics[scale=0.47]{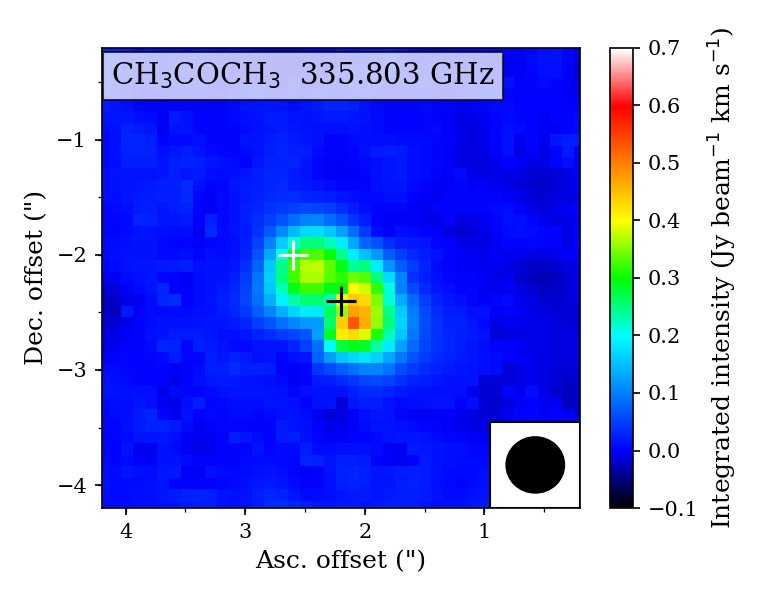} 
	}
\adjustbox{trim ={0.125\width} {0.13\height} {0.24\width} {0.04\height}, clip=true}{
	\includegraphics[scale=0.47]{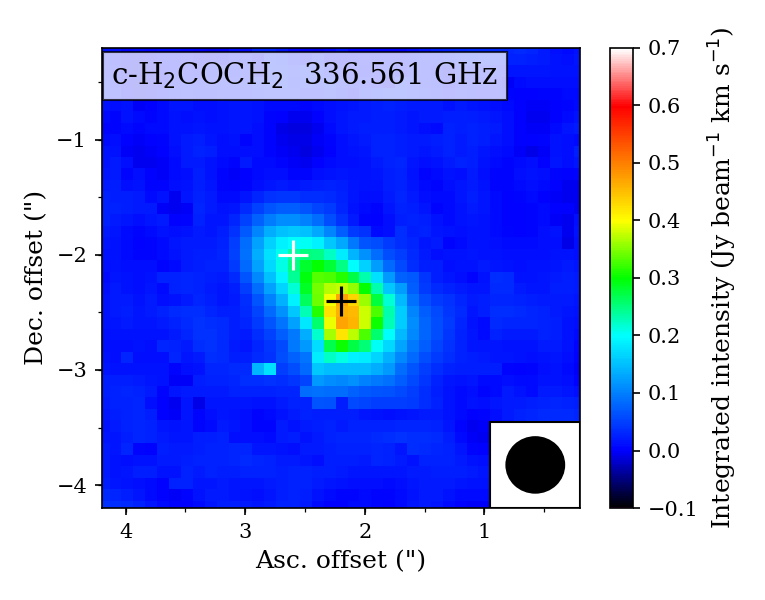} 
	}
\adjustbox{trim ={0.125\width} {0.13\height} {0.24\width} {0.04\height}, clip=true}{
	\includegraphics[scale=0.47]{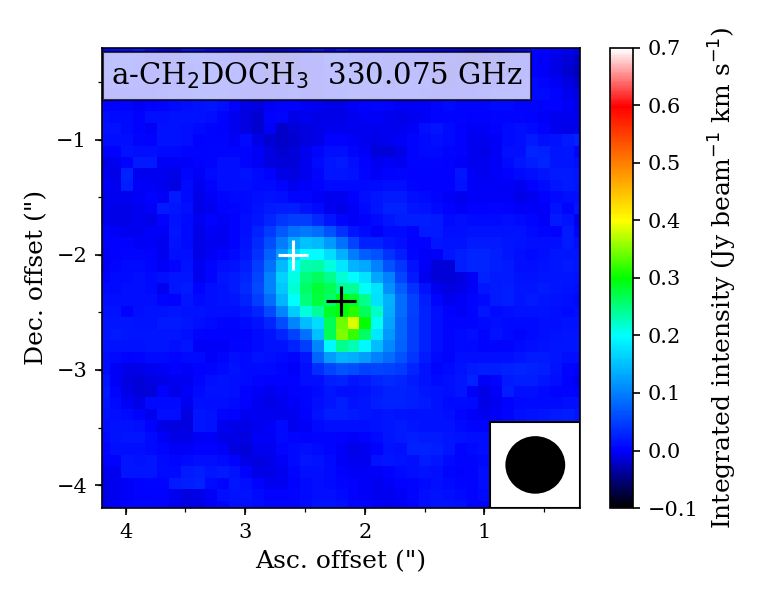} 
	}
\adjustbox{trim ={0.125\width} {0.13\height} {0.0\width} {0.04\height}, clip=true}{
	\includegraphics[scale=0.47]{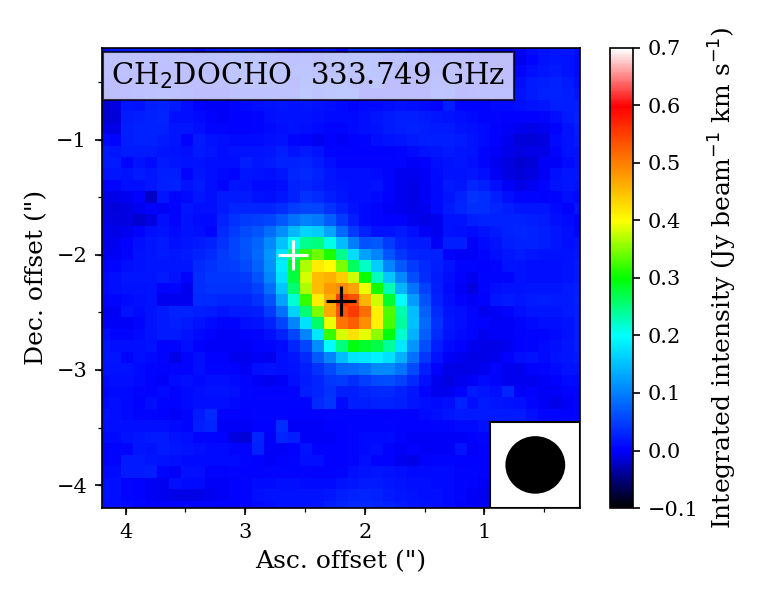} 
	}
\\
\adjustbox{trim ={0.0\width} {0.13\height} {0.24\width} {0.04\height}, clip=true}{
	\includegraphics[scale=0.47]{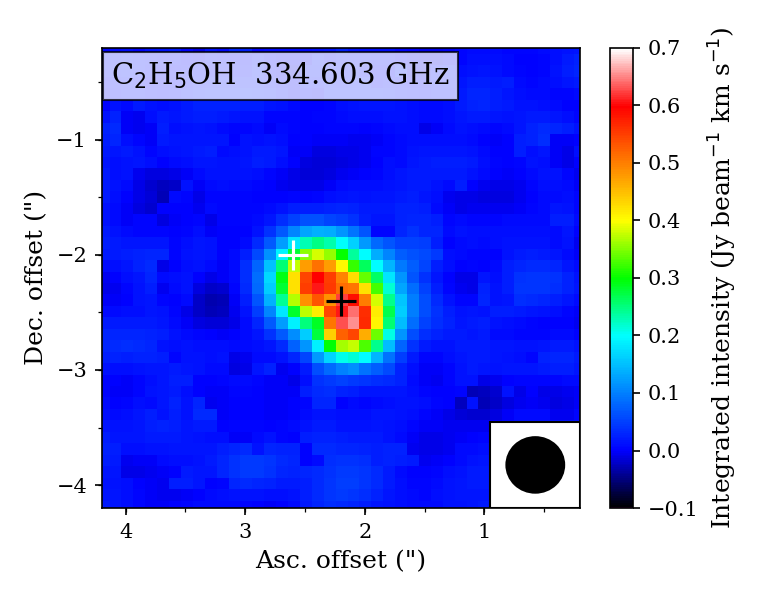} 
	}
\adjustbox{trim ={0.125\width} {0.13\height} {0.24\width} {0.04\height}, clip=true}{
	\includegraphics[scale=0.47]{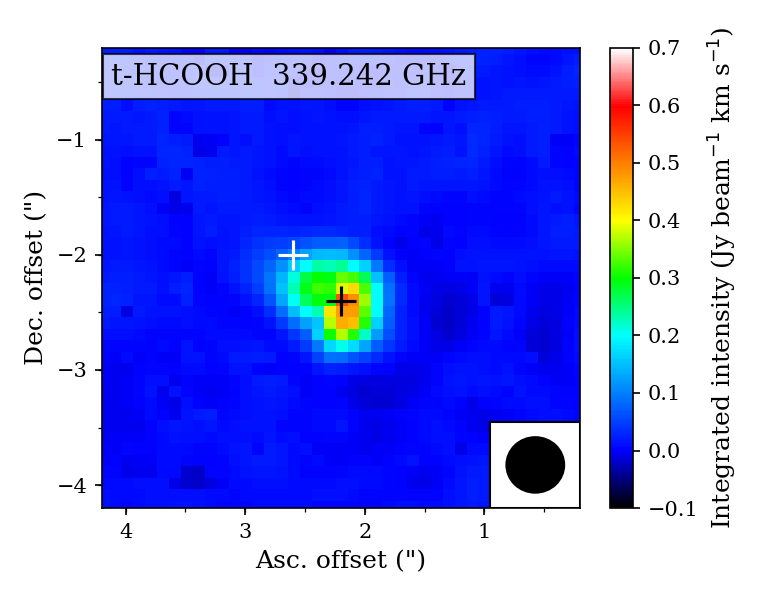} 
	}
\adjustbox{trim ={0.125\width} {0.13\height} {0.24\width} {0.04\height}, clip=true}{
	\includegraphics[scale=0.47]{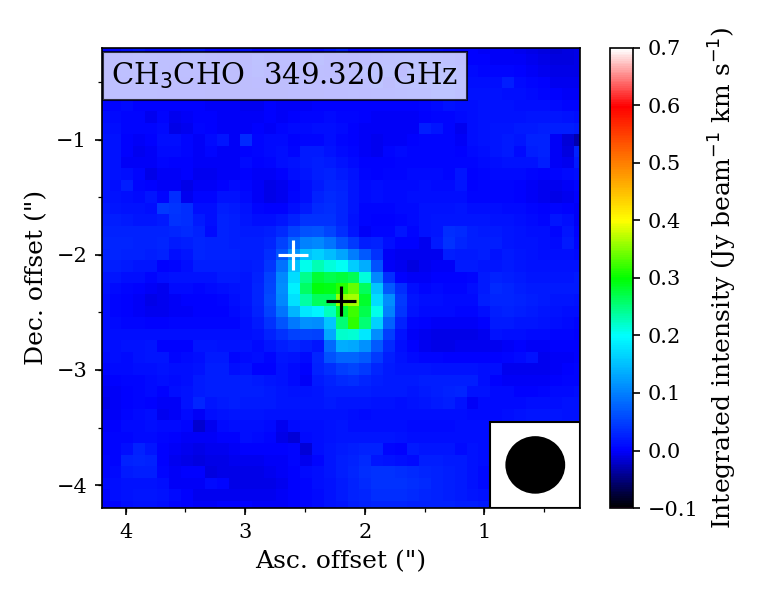} 
	}
\adjustbox{trim ={0.125\width} {0.13\height} {0.0\width} {0.04\height}, clip=true}{
	\includegraphics[scale=0.47]{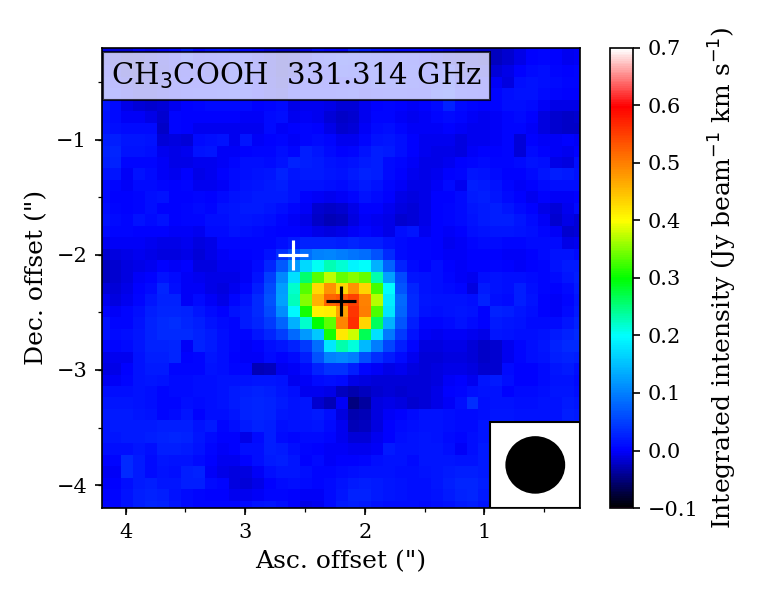} 
	}
\\
\adjustbox{trim ={0.0\width} {0.13\height} {0.24\width} {0.04\height}, clip=true}{
	\includegraphics[scale=0.47]{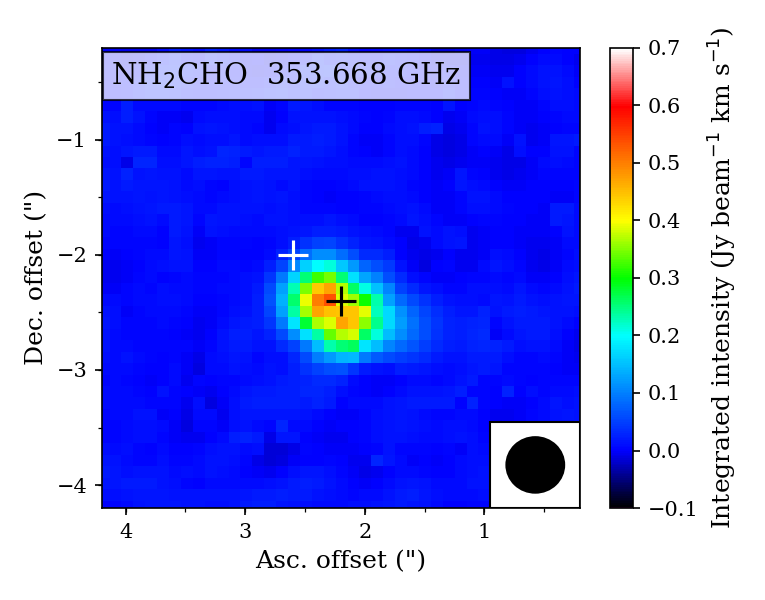} 
	}
\adjustbox{trim ={0.125\width} {0.13\height} {0.24\width} {0.04\height}, clip=true}{
	\includegraphics[scale=0.47]{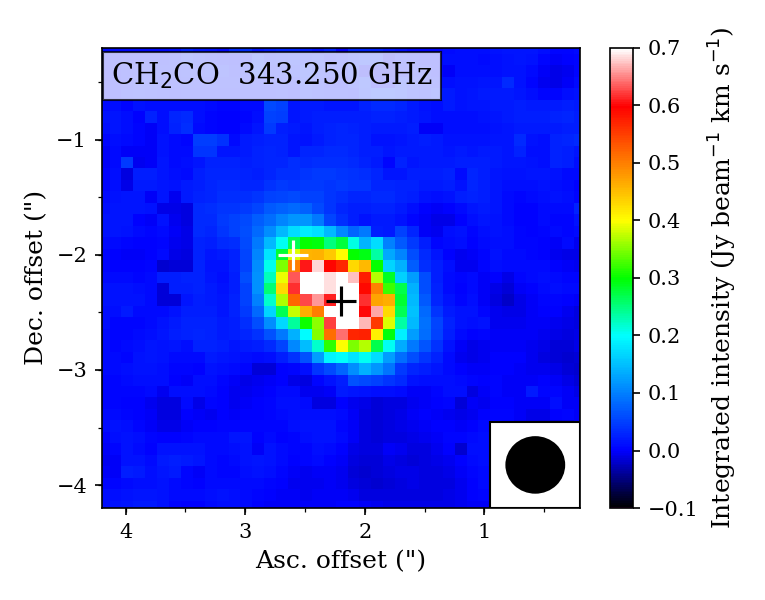} 
	}
\adjustbox{trim ={0.125\width} {0.13\height} {0.24\width} {0.04\height}, clip=true}{
	\includegraphics[scale=0.47]{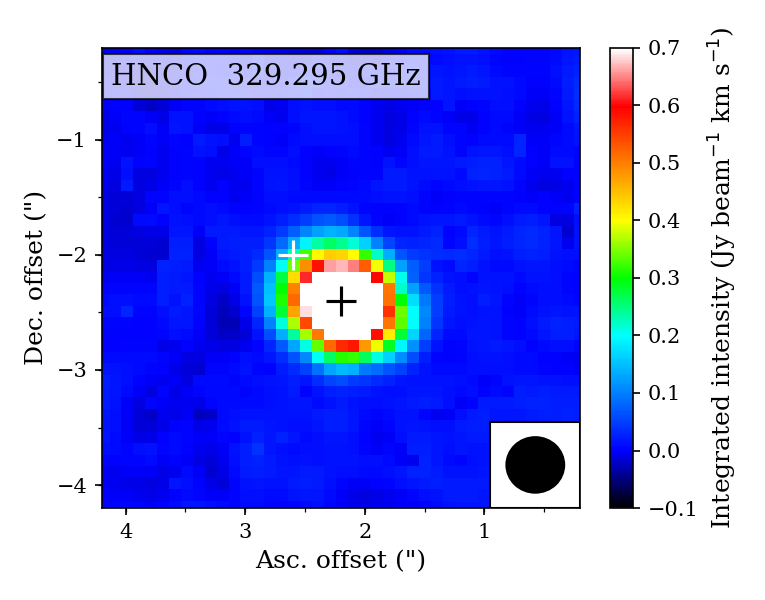} 
	}
\adjustbox{trim ={0.125\width} {0.13\height} {0.0\width} {0.04\height}, clip=true}{
	\includegraphics[scale=0.47]{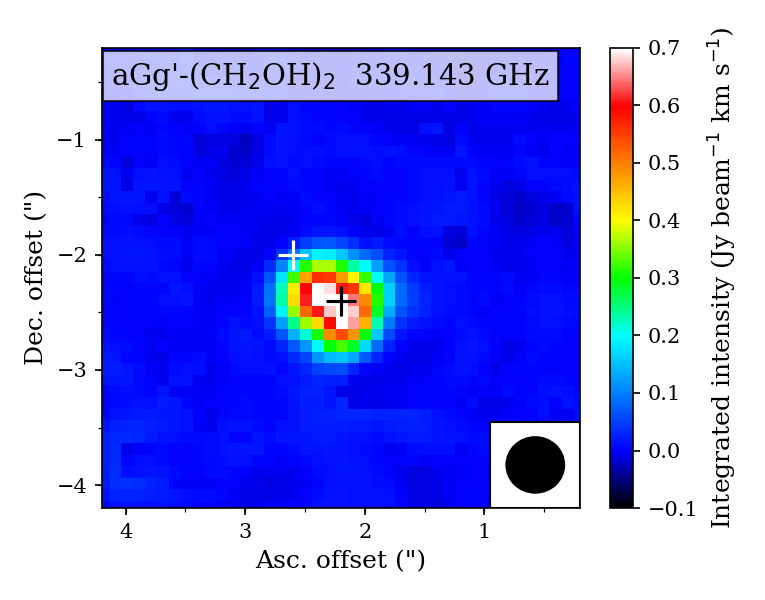} 
	}
\\
\adjustbox{trim ={0.0\width} {0.0\height} {0.24\width} {0.04\height}, clip=true}{
	\includegraphics[scale=0.47]{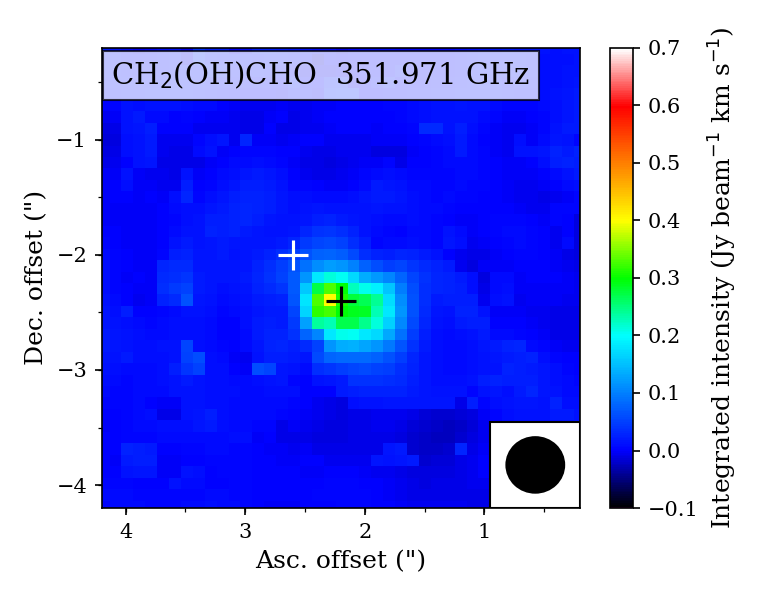} 
	}
\adjustbox{trim ={0.125\width} {0.0\height} {0.24\width} {0.04\height}, clip=true}{
	\includegraphics[scale=0.47]{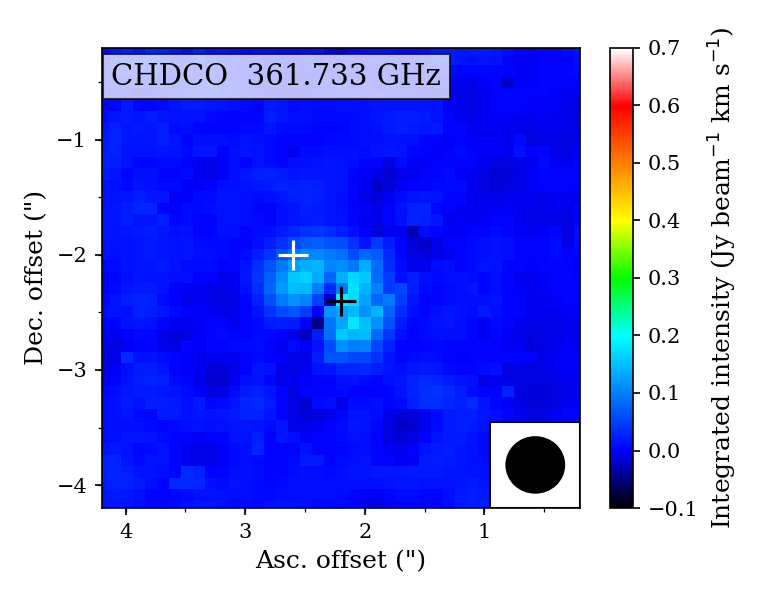} 
	}
\adjustbox{trim ={0.125\width} {0.0\height} {0.24\width} {0.04\height}, clip=true}{
	\includegraphics[scale=0.47]{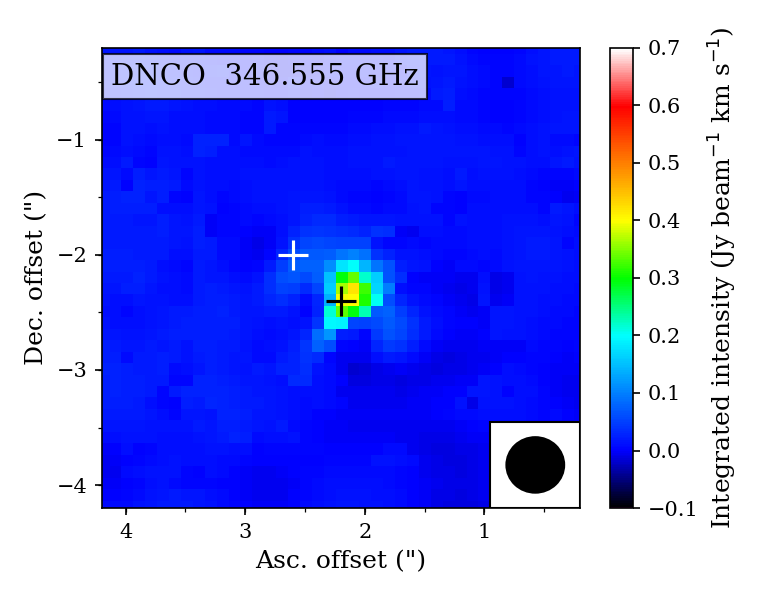} 
	}
\adjustbox{trim ={0.125\width} {0.0\height} {0.0\width} {0.04\height}, clip=true}{
	\includegraphics[scale=0.47]{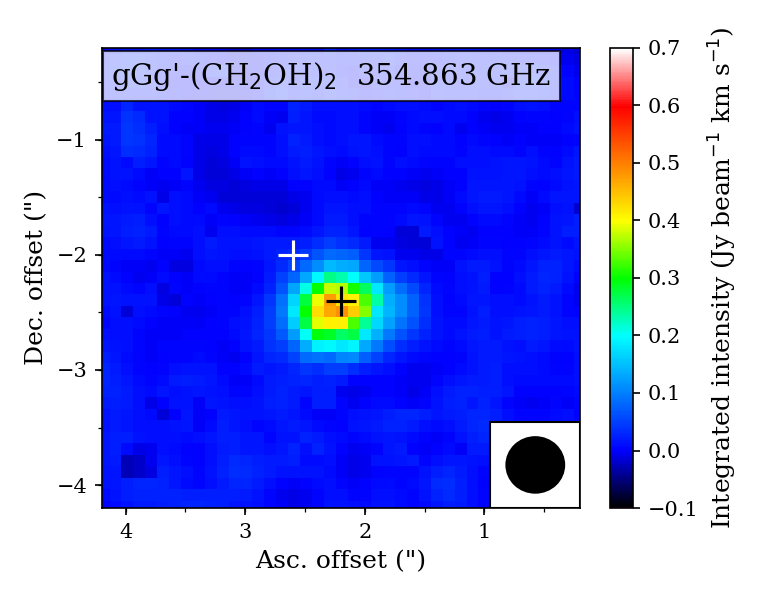} 
	}        
\caption{\label{fig-VINE}\small Representative velocity integrated emission maps of the detected COMs towards IRAS16293A. The black and white crosses indicate the continuum peak position and the offset position, respectively. The species and the line frequency are noted in the top-left corner and the 0{\farcs}5 beam is shown in the bottom right corner of each panels.}
\end{figure*}

The spatial distribution of the emission towards IRAS 16293 A and B can be very different depending on the species. The deconvolved source size of the IRAS 16293B continuum emission is estimated to be 0{\farcs}5 \citep{Jorgensen-2016}, which is similar to the beam size of the observations. Most of the detected species roughly follow the same distribution as the continuum and have the same source size. Concerning IRAS 16293A, the continuum emission, as well as the molecular line emission arising from the gas, is more extended in the NE-SW axis where a 6~km s$^{-1}$ velocity gradient was reported \citep{Pineda-2012, Favre-2014, Jorgensen-2016, Oya-2016}. Small variations in the spatial distribution of different species towards IRAS 16293A can be marginally resolved with the beam of PILS observations. 
However, the velocity gradient, which is associated with the very high line density that is present in the spectrum, makes it very difficult to integrate a single line across the source without contamination from nearby lines of other species.

\cite{Calcutt-2018b} developed a new method to map the emission, called the velocity integrated emission map (VINE map), which takes into account the velocity shift of the emitting gas when integrating over the molecular emission. At each pixel, the integration range is then shifted to the peak velocity of the line to be mapped. When there is no large velocity variation across the map, which is the case for IRAS 16293B for instance, the VINE map is equivalent to the standard integrated emission map (moment zero map). For IRAS 16293A, the velocity shift correction at each pixel of the VINE maps is based on the same bright transition $7_{3,5}-6_{4,4}$ of CH$_3$OH at 337.519~GHz as used by \cite{Calcutt-2018b}. This transition is particularly suited for tracing the hot corino while being well isolated from other lines. 
Figure \ref{fig-VINE} shows VINE maps of the main isotopologues, and few deuterated conformers were detected towards IRAS 16293A. 
In this study, the VINE maps were only used to check whether the emission distribution reaches the offset position, where we analysed the spectrum, as indicated by the white cross.

The emission of the species that have similar abundances towards IRAS 16293A and B is distributed along the velocity gradient direction and cover the offset position, whereas the emission distribution of the depleted species towards IRAS 16293A in comparison to IRAS 16293B is located around the continuum peak position of the respective sources and does not show the elliptical distribution of the continuum emission towards IRAS 16293A. Only the CH$_3$COOH emission distribution is as compact as the (CH$_2$OH)$_2$ emission distribution, while the abundance towards IRAS 16293A and B are similar. We checked the distribution emission towards IRAS 16293B as well as for this species and noticed that the emission does not cover the offset position used in the previous studies. Therefore, CH$_3$COOH seems depleted towards both IRAS 16293A and B. This is why CH$_3$COOH does not show a difference between IRAS 16293A and B since it belongs to the most compact region of the warm envelope.

Spatial differences between COMs were previously observed towards the Galactic Centre Sgr B2 region \citep{Hollis-2001} as well as towards high-mass protostars \citep[for example,][]{Calcutt-2014}. 
\citeauthor{Hollis-2001} observed the emission of CH$_3$OCHO and CH$_2$(OH)CHO, among others, at two different angular resolutions. They detected the two species with an $\sim$60" beam size using the NRAO 12 m telescope, whereas the Berkeley-Illinois-Maryland Association (BIMA) observations with a beam of $\sim$5" only revealed CH$_3$OCHO. They suggested that CH$_2$(OH)CHO was much more extended compared to CH$_3$OCHO and its emission was spatially filtered by the interferometric character of the observations. These findings are supported by the more recent study of \cite{Li-2017} towards an extended region of the Galactic Centre in which the authors reported cold gas emission of CH$_2$(OH)CHO $\sim$36 pc in width around Sgr B2(N). 
We note that, these two observations of the CH$_2$(OH)CHO spatial extent are specific to the Galactic Centre, where non-thermal desorption processes put the CH$_2$(OH)CHO in the gas phase at temperatures much lower than the thermal desorption temperature of this species.

\cite{Xue-2019} consistently observed  the three isomers CH$_3$COOH, CH$_3$OCHO, and CH$_2$(OH)CHO towards Sgr B2(N) using ALMA in Band 3 with an angular resolution of 1\farcs 6, which is sufficient to resolve the sources.
The spatial distribution of the species reveals that CH$_3$OCHO is more extended than both CH$_2$(OH)CHO and CH$_3$COOH, which is in agreement with the present study. \citeauthor{Xue-2019} suggest that the different formation pathways of the tree species through ice surfaces and gas-phase chemistry could explain the spatial differentiation. This interpretation was supported by the correlation with the distribution of the respective precursors especially for gas-phase formation, such as CH$_3$OCH$_3,$ which is the precursor of CH$_3$OCHO  \citep{Balucani-2015}, and C$_2$H$_5$OH the precursor of CH$_2$(OH)CHO and CH$_3$COOH \citep{Skouteris-2018}. 
The search for precursors through ice surfaces formation pathway is limited by the current infra-red absorption observations. Alternatively, the authors suggest that the effective desorption temperature of the species could explain their spatial distribution difference based on temperature-programmed desorption (TPD) experiments of C$_2$H$_4$O$_2$ isomers \citep{Burke-2015}. During the TPD, CH$_3$OCHO desorbed at $\sim$70~K whereas CH$_2$(OH)CHO and CH$_3$COOH were released in the gas phase at $\sim$110~K. In the warm-up model of star formation, this would result in a more extended emission of CH$_3$OCHO in comparison to the two other isomers.

\cite{Calcutt-2014} simultaneously detected CH$_3$OCHO and CH$_2$(OH)CHO, among other species, towards three high-mass sources, G31.41+0.31, G29.96--0.02, and G24.78+0.08A, which are located at 3.5, 7.1, and 7.7~kpc away, respectively. They find that CH$_3$OCHO was more extended than CH$_2$(OH)CHO towards G31 and that the two species had the same extent towards the two other sources. The spatial differentiation found in G31 could be similar to IRAS 16293; however, CH$_2$(OH)CHO and the other depleted species towards IRAS 16293A could also trace a deeper region in the envelope due to the excitation conditions. For example, CH$_3$OCHO lines are optically thicker than those of CH$_2$(OH)CHO and show a more extended spatial distribution. 

Also, such a spatial differentiation can be inferred indirectly. The first detection of glycolaldehyde by \cite{Jorgensen-2012} was performed with ALMA at $\sim$2" angular resolution: This is enough to separate the two components of the binary system but larger than the source size of both components. \citeauthor{Jorgensen-2012} find a relative abundance CH$_3$OCHO/CH$_2$(OH)CHO of 10-15 for both protostars on these scales and did not see the same evidence for depletion of CH$_2$(OH)CHO with respect to CH$_3$OCHO for IRAS 16293A in comparison to IRAS 16293B, as discussed above. However, it is consistent with the interpretation that CH$_2$(OH)CHO, as well as the other most depleted species towards the offset position from IRAS 16293A, are located in a more compact region of the hot corino. This is also supported by observations of other COMs, such as CH$_3$OCH$_3$, D$_2$CO \citep{Jorgensen-2011}, and C$_2$H$_5$OH \citep{Bisschop-2008}.

\subsection{Differentiation in rotational temperature}

The distribution of COMs towards IRAS 16293A suggests that a significant part of the emission is missing at the extracted spectrum position. However, the estimation of the rotational temperature is sensitive to the intensity variation of one line with respect to another line with a different upper energy level. Thus, the derivation of the rotational temperature is not affected by the apparent depletion due to the spatial extent and can be securely compared species-to-species. 

Towards IRAS 16293A, the rotational temperatures of the detected species and their isotopologues are between 90 and 180~K, where HNCO, H$_2$CO, CH$_2$(OH)CHO, (CH$_2$OH)$_2$, NH$_2$CHO, CH$_3$CHO, and C$_2$H$_5$OH are the hottest species. Although the distinction between those hot species and the other species is not pronounced, the hot species seem to be located closer to the protostars compared to the other species, showing a lower rotational temperature.

Using the PILS observations towards IRAS 16293B, \cite{Jorgensen-2018} detected the same species as those presented in this study and derived their rotational temperature. The authors find that the species located in the hot corino could be distinguished into two groups depending on their rotational temperature. Species such as CH$_3$OH, C$_2$H$_5$OH, CH$_3$OCHO, NH$_2$CHO, HNCO, CH$_2$(OH)CHO, and (CH$_2$OH)$_2$ were observed with an rotational temperature of 250--300~K, whereas CH$_2$CO, CH$_3$CHO, CH$_3$OCH$_3$, H$_2$CO, and c-H$_2$COCH$_2$ were found at 100--150~K. 
Using these results and the previous results of PILS observations towards IRAS 16293B from \cite{Coutens-2016, Jorgensen-2016} and \cite{Persson-2018}, \citeauthor{Jorgensen-2018} find a correlation between the rotational temperature and the desorption temperature of the oxygen-bearing COMs that are thought to mainly form on ice surfaces. This implies that the species that have a high rotational temperature (250--300~K) are more predominantly located in the hot and compact region of the hot corino, while the lower rotational temperature species belong to a more extended region inside the hot corino \citep[see Fig. 4 of][]{Jorgensen-2018}. By comparing this scenario to the difference in the spatial extent observed in IRAS 16293A, the compact species, such as CH$_2$(OH)CHO, (CH$_2$OH)$_2$, NH$_2$CHO, HNCO, and C$_2$H$_5$OH, are associated with high rotational temperature, whereas extended species, such as c-H$_2$COCH$_2$, CH$_3$OCH$_3$, and CH$_3$COCH$_3,$ have a low rotational temperature. The ordering of the species across the onion-like structure of the envelope is in agreement with the desorption temperatures observed in temperature-programmed desorption (TPD) experiments of pure and mixed ices \citep{Oberg-2009, Fedoseev-2015, Fedoseev-2014, Fedoseev-2017, Paardekooper-2016, Chuang-2016, Qasim-2019b, Qasim-2019a}.

In addition, the rotational temperature of CH$_3$COOH towards IRAS 16293B is consistent with its most compact location in the envelope of both sources. However, the correlation between the rotational temperature and the location does not work in IRAS 16293A for CH$_3$OH, CH$_3$OCHO, CH$_2$CO, H$_2$CO, CH$_3$CHO, and \emph{t}-HCOOH. 

Concerning CH$_3$OH and CH$_3$OCHO, these two species were found to be hot species in IRAS 16293B; however, they correspond to cold and extended species in IRAS 16293A. Their optically thick lines suggest the presence of two components towards IRAS 16293B that are indistinguishable due to the smaller source size with respect to the angular resolution: one is at a rotational temperature of 300 and another is at $\sim$125~K \citep{Jorgensen-2018}. In addition, the rotational temperature derived towards IRAS 16293A for the same two species is only consistent with the second component at 125~K, indicating that the bulk of the emission consists of the extended warm gas. From this, CH$_3$OH and CH$_3$OCHO are found to trace not only the central core but also the extended part of the hot corino. 

Regarding CH$_2$CO isotopologues, there are only a few lines in the range of the observations and most of those are blended with small species, such as $^{33}$SO at $\sim$343.08 GHz and CS at $\sim$342.88 GHz. The deuterated isotopologue CHDCO suggests that the low abundance towards IRAS 16293A compared to IRAS 16293B is due to the spatial extent differences between the two sources. 
The emission distribution of H$_2$CO was found to be much more complex and extended across the binary protostars and it traces both the hot corino region and the outflow interface with the bridge structure between IRAS 16293A and B and the E-W outflow emerging from IRAS 16293A \citep{vanderWiel-2019}.

The few unblended, but optically thick, lines of CH$_3$CHO show a large distribution towards both sources, suggesting that the bulk of the emission corresponds to the extended part of the hot corino, despite the abundance difference between sources A and B. The three-phase chemical model of \cite{Garrod-2013} shows that the abundance of CH$_3$CHO in the gas phase increases during the evolution of the hot corino, which occurs even before the species desorbs from the ice grain surfaces. This suggests that gas-phase formation paths significantly contribute to the formation of CH$_3$CHO at relatively low temperatures before the bulk of the species desorb from the ice. Concerning the abundance observed towards IRAS 16293 binary, if the gas-phase formation paths  effectively produce CH$_3$CHO in the outer part of the hot corino, it should increase the abundance of CH$_3$CHO towards the most luminous source with respect to the other component. However, the luminosity of IRAS 16293A was estimated to be $\sim18$~L$_\odot$, which is six times higher than IRAS 16293B \citep{Jacobsen-2018}. 

\begin{figure}[t]
\centering
  \includegraphics[scale=0.83]{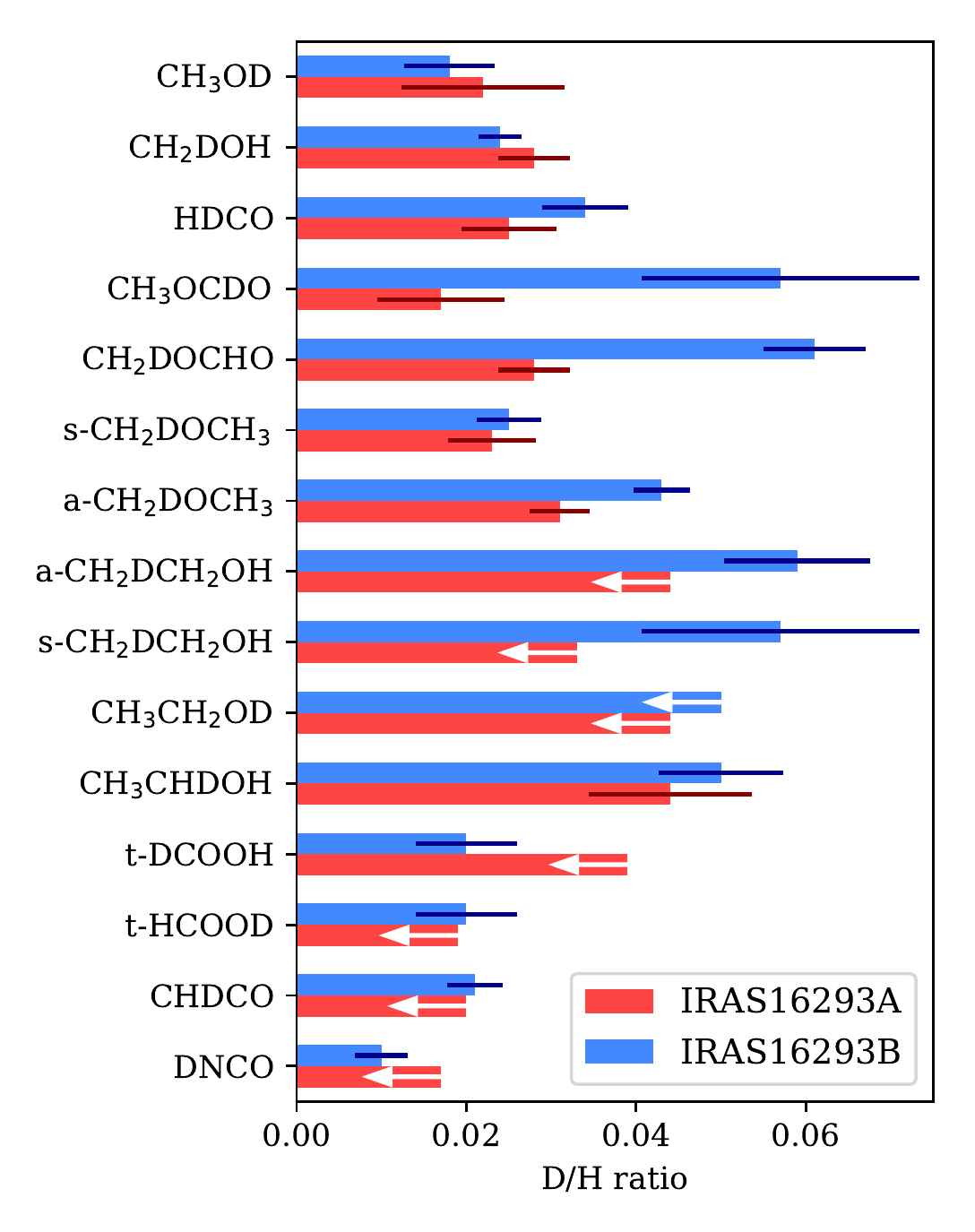}
    \caption{D/H ratio of the deuterated species detected towards IRAS 16293A in red and IRAS 16293B in blue. The D/H ratio is calculated from the abundance ratio, with the statistical correction due to the chemical group \citep[see Appendix B of ][]{Manigand-2018}. D/H ratios towards IRAS 16293B are taken from \cite{Coutens-2016}, \cite{Lykke-2017}, \cite{Persson-2018} and \cite{Jorgensen-2018}.}
    \label{fig-DHratios}
\end{figure}

In summary, the analysis of the spatial extent, the abundances, and the rotational temperatures suggests that the hot corinos show a stratification of the COMs along the distance to the forming star. This spatial differentiation towards IRAS 16293A could be linked to the rotational temperature of the species observed towards IRAS 16293B in the way that species at 300~K are more compact than the species at 125~K. Although this spatial differentiation is not resolved for IRAS 16293B, the rotational temperature suggests its presence. Towards IRAS 16293A, the difference in rotational temperature between compact and extended species is less pronounced, which could be due to the geometry of the source (inclination, size) or the presence of the nearly edge-on disc \citep{Pineda-2012, Favre-2014, Girart-2014}.

\subsection{D/H ratio}

The sensitivity of the ALMA observations makes the detection of many deuterated isotopologues of the oxygen-bearing species possible. Figure~\ref{fig-DHratios} compares the D/H ratios of these species derived from the estimated column densities found towards the offset position of IRAS 16293A and reported in the previous studies of IRAS 16293B \citep{Coutens-2016, Lykke-2017, Persson-2018, Jorgensen-2018} on a linear scale. In general, the D/H ratios of IRAS 16293A species are found to be as high as those of the species of IRAS 16293B. However, due to the higher uncertainty in the column densities and the higher number of upper limits, it is difficult to observe a trend in the deuteration itself. Nevertheless, there is apparently no direct correlation between the D/H ratios and the stratification suggested by the rotational temperatures. This could suggest that contrary to the rotational temperatures and the spatial extents, which give a clue as to the structure of the hot corino and thus the warm chemistry in situ, the D/H ratio gives information about the history of the envelope during the pre-stellar phase when the gas was cold enough to let the D/H ratio of the frozen species increase. This is supported by the D/H ratio of the doubly-deuterated isotopologues D$_2$CO of $2.0 \pm 0.4\times10^{-1}$ and CHD$_2$OCHO of $8.2\pm0.6\times10^{-2}$ \citep{Manigand-2018} found towards IRAS 16293A. 

Using the result reported in Table \ref{tab:newdetect} and the column density of CH$_3$CHO measured towards IRAS16293B \citep{Lykke-2017, Jorgensen-2018}, which is $1.2\times10^{17}$ cm$^{-2}$, the D/H ratio of CH$_2$DCHO and CH$_3$CDO are 1.4 and 4.7\%, respectively. The upper limits of the D/H ratio of CH$_2$DCHO and CH$_3$CDO towards IRAS 16293A are $<6$ and $<17$\%, respectively. The D/H ratio of the groups CH$_3$-- and HCO-- are significantly different for CH$_3$CHO even after applying the statistical correction for both sources. Assuming that the main formation route is due to the addition of those two radicals on the ice surface during the warm-up phase, then the difference in deuteration may indicate that the deuteration enhancement is different for CH$_3$ and HCO, which are already in the prestellar phase when the temperature is low enough to favour the deuteration enhancement of H$_3^+$ in the gas phase. However, the other species thought to be formed on ice surfaces do not show such a difference in the D/H ratio between CH$_3$-- or CH$_2$-- and HCO-- groups, such as CH$_3$OCHO and CH$_2$(OH)CHO \citep{Jorgensen-2016}. This suggests a selective process, independent of the luminosity difference between IRAS 16293A and B, which increases the deuteration of HCO-- or decreases those of CH$_3$-- without significantly impacting the other species. 

As a side note, the high intensity of the candidate lines found for CHD$_2$OH in the range of the observations and the enhancement of the D/H ratio of the two doubly-deuterated isotopologues D$_2$CO and CHD$_2$OCHO with respect to their respective singly-deuterated isotopologues, HDCO and CH$_2$DOCHO, suggest a roughly equally high column density for CHD$_2$OH. This stresses the importance of future studies on the spectroscopy of CHD$_2$OH and even more its deuterated conformers CD$_3$OH and CD$_3$OD.

\section{Conclusions}
In this study, we analysed the molecular content of the protostar IRAS 16293A at the hot corino scale and compared the abundances to those measured towards its protostar companion IRAS 16293B. Numerous O-bearing species have been detected, along with their rarer isotopologues. The main findings of this work are summarised below.
\begin{enumerate}
\item The abundances with respect to CH$_3$OH of half of the species are significantly lower towards the 0\farcs 6 offset position from IRAS 16293A in comparison to the 0\farcs 5 offset position from IRAS 16293B in spite of its higher luminosity.
\item The cross-correlation of the main isotopologue abundances highlights a selective differentiation depending on the species observed. Different categories are identified whether the species abundances are similar or significantly different between IRAS 16293A and B. 
\item The first category, including CH$_3$OH, CH$_3$OCHO, c-C$_2$H$_4$O, CH$_3$COCH$_3$, CH$_3$COOH, CH$_3$OCH$_3$, \emph{t}-C$_2$H$_5$OCH$_3$, and CH$_3$OCH$_2$OH, corresponds to the species that have a similar abundance towards both IRAS 16293A and B. These species show an extended spatial distribution across IRAS 16293A and have a relatively low rotational temperature of $\sim$125~K towards IRAS 16293B, except for CH$_3$COOH, CH$_3$OH and CH$_3$OCHO. 
CH$_3$COOH spatial distribution is much more compact towards both IRAS 16293A and B than the other extended species and the rotational temperature found towards IRAS 16293B, which is unambiguously $\sim$300~K. 
In addition, CH$_3$OH and CH$_3$OCHO show attributes of both compact and extended regions as suggested by their extended distribution and their high desorption temperature. Their emission is suspected to trace both regions in the hot corino. 
\item The second category concerns the species showing a significantly lower abundance towards IRAS 16293A with respect to IRAS 16293B, which are HNCO, C$_2$H$_5$CHO, C$_2$H$_5$OH, \emph{t}-HCOOH, NH$_2$CHO, CH$_2$CO, H$_2$CO, CH$_3$CHO, CH$_2$(OH)CHO, and (CH$_2$OH)$_2$. These species have a more compact spatial distribution towards IRAS 16293A compared to the first category of species. In addition, they tend to have a high rotational temperature of$\sim$300~K, especially for those that are the most compact, such as CH$_2$(OH)CHO and (CH$_2$OH)$_2$. However, the spatial distribution difference is not resolved towards IRAS 16293B, the rotational temperature difference between compact and extended species is more pronounced though. H$_2$CO, CH$_2$CO, and CH$_3$CHO do not fit into any category as they have a roughly compact spatial emission but a low rotational temperature. 
\item In the search for O-bearing species, we report the new detection of \emph{t}-C$_2$H$_5$OCH$_3$ and CH$_3$OCH$_2$OH towards IRAS 16293B. Their abundances with respect to CH$_3$OH are consistent with the previous detection of these species towards the high mass protostellar regions Sgr B2(N) and Orion KL. Upper limits of their column density are derived towards IRAS 16293A.
\item The D/H ratio is not correlated to the structure of the hot corino itself; however, it is the result of what happens during the formation of COMs in the pre-stellar phase. The multiply-deuterated COMs seem to have a systematically higher D/H ratio compared to singly-deuterated species. More observations of different sources, supported by future spectroscopic studies of other multiply-deuterated COMs, are required to confirm this trend. In addition, we report the identification of 23 CHD$_2$OH transitions at 0.8 mm wavelength, which have not been observed so far. The intensity of CHD$_2$OH lines suggests a high abundance and D/H ratio, as it is for CHD$_2$OCHO. The identification of these CHD$_2$OH transitions highlights the need for spectroscopic data for CHD$_2$OH and CD$_3$OH to derive their rotational temperatures and their column densities and to get their D/H ratios. 
\end{enumerate}

As discussed in this paper, abundance variations between IRAS 16293A and B suggest that the spatial distributions are different for the species associated with low or high rotational temperatures.
Whether the origins of these differences are physical or chemical, future higher angular observations will be critical in resolving the innermost structure of the two protostars.

\begin{acknowledgements}
The authors wish to thank the anonymous referee for the constructive comments that significantly improved the paper.
The authors are grateful to Brett McGuire and Roman Motiyenko for providing the line strengths of CH$_3$OCH$_2$OH spectroscopic data. The authors acknowledge Gleb Fedoseev for the valuable discussion about the desorption temperatures from the TPD experiments of the O-bearing COMs and Troels C. Petersen for discussions of the statistical methods.
This paper makes use of the following ALMA data: ADS/JAO.ALMA{\#}2012.1.00712.S and ADS/JAO.ALMA{\#}2013.1.00278.S. ALMA is a partnership of ESO (representing its member states), NSF (USA) and NINS (Japan), together with NRC (Canada), NSC and ASIAA (Taiwan), and KASI (Republic of Korea), in cooperation with the Republic of Chile. The Joint ALMA Observatory is operated by ESO, AUI/NRAO and NAOJ.
The group of J.K.J. acknowledges support from the H2020 European Research Council (ERC) (grant agreement No 646908) through ERC Consolidator Grant ???S4F???. Research at Centre for Star and Planet Formation is funded by the Danish National Research Foundation. 
A.C. postdoctoral grant is funded by the ERC Starting Grant 3DICE (grant agreement 336474).
M.N.D. is supported by the Swiss National Science Foundation (SNSF) Ambizione grant 180079, the Center for Space and Habitability (CSH) Fellowship and the IAU Gruber Foundation Fellowship.
This research has made use of NASA's Astrophysics Data System and VizierR catalogue access tool, CDS, Strasbourg, France \citep{Vizier}, as well as community-developed core Python packages for astronomy and scientific computing including 
Astropy \citep{Astropy}, 
Scipy \citep{Scipy}, 
Numpy \citep{Numpy} and 
Matplotlib \citep{Matplotlib}. 
\end{acknowledgements}

\bibliographystyle{aa} 
\bibliography{biblio} 

\appendix

\section{\label{sec-stat}Statistical distance}

The statistical distance of the abundance of a species $X$ relative to a species $Y$ between IRAS 16293A and IRAS 16293B is calculated as follows:
\begin{equation}
S_{X/Y} = \frac{\left(\frac{N_X}{N_Y}\right)_\text{A} - \left(\frac{N_X}{N_Y}\right)_\text{B}}{ \sqrt{\sigma_{\text{A}}^2 + \sigma_{\text{B}}^2}}
\end{equation}
in which $\sigma_\text{A}$ and $\sigma_\text{B}$ are the uncertainties of the column density ratios $\left(\frac{N_X}{N_Y}\right)_\text{A}$ and $\left(\frac{N_X}{N_Y}\right)_\text{B}$, respectively. The value of $S_{X/Y}$ corresponds to the significance of the distance in terms of the number of $\sigma$. 

\section{\label{sec-MCerror}Monte-Carlo error estimation}

Because of the non-linearity of the model used in this study and the non-standard $\chi^2$ minimisation method, the uncertainties on the excitation temperature and the column density are not derivable from the covariance matrix, usually calculated with the $\chi^2$ minimisation methods. In this case, the Monte Carlo (MC) simulation becomes a convenient tool to estimate the uncertainties from the posterior probability distributions of each parameter of the model. In this section, we describe the bootstrap algorithm, which is a specific class of Monte Carlo simulations, that was tested to estimate the uncertainties of the parameters of the model.

Consider the data fitted as a vector $\vec{S}$ of $N$ elements, which are the data points:
\begin{equation}
\vec{S} = \left(s_0,~s_1,~s_2, \cdots\right)
.\end{equation}
A Gaussian noise $\vec{\epsilon}$ is defined with the same logic:
\begin{equation}
\vec{\epsilon} =\left(\epsilon_0,~\epsilon_1,~\epsilon_2,\cdots\right)
.\end{equation}
The $\chi^2$ minimisation can be represented by a function $F_{\chi^2}$ that transforms a data vector $\vec{S}$ into a parameter vector $\vec{P}$, which is the best fit parameters of the model for the given data. The inverse function $F^{-1}$ is thus the fitting model, that is, the synthetic spectrum in the present study:
\begin{eqnarray}
\vec{P}=&\left(p_0,~p_1,~p_2, \cdots\right) \\
F_{\chi^2}:&\vec{S} \longrightarrow F_{\chi^2}(\vec{S}) = \vec{P} \\
F^{-1}:&\vec{P} \longrightarrow F^{-1}(\vec{P}) = \vec{\tilde{S}}
\end{eqnarray}
in which $\vec{\tilde{S}}$ is the closest synthetic spectrum to the data $\vec{S}$.
The MC method used in this study is a simple application of the Markov-Chain MC simulations where: the prior distribution is a group of normally distributed noise $\left\{\vec{\epsilon}_i\right\}_{N_{MC}}$ of a width equivalent to the uncertainty $\sigma$ of the data, with $N_{MC}$ the total number of walkers; the sampler  is the $\chi^2$ minimisation function $F_{\chi^2}$; the probability for a walker to jump to a worse position in the parameter space is null; and only the final position of each walker is kept.

In this particular case, the posterior probability distributions are fully dependent on the prior distributions. From the MC simulation, we get a group of parameter vectors:
\begin{eqnarray}
\mathcal{P} &=& \left\{ F_{\chi2}\left(\vec{S} + \vec{\epsilon_i} \right)\right\}_{N_{MC}} \\
&=& \left\{ (p_0,~p_1,~p_2,\cdots)_0,~(p_0,~p_1,~p_2,\cdots)_1,\cdots\right\}_{N_{MC}} \\
\mathcal{P}^T&=& \left(\left\{p_0\right\}_{N_{MC}},~\left\{p_1\right\}_{N_{MC}},~\left\{p_2\right\}_{N_{MC}},\cdots\right) \\
\mathcal{P}^T&=& \left(\mathcal{P}_0,~\mathcal{P}_1,~\mathcal{P}_2,\cdots\right)
\end{eqnarray}
where $\mathcal{P}_i$ is the group of parameters $p_i$.
The distribution of $\mathcal{P}_i$, called the posterior probability distribution of $p_i$, gives the needed information for the fitted parameters. It is maximal for the best fit value and the width at half-maximum of each side of the distribution corresponds to the negative and positive uncertainties on $p_i$. 

\section{\label{sec-spectro}Laboratory spectroscopic data}

Most of the analysed species do not assume any excited vibrational or torsional level contributions to the partition function at low rotational temperatures. However, they are all detected at rotational temperatures above 100 K, where the vibrational and torsional contribution can be significant. Thus, it is necessary to detail the spectroscopic dataset used for each species. In this study, a vibrational correction factor $>1.1$ is considered significant and is not applied to the column density derived from the synthetic spectrum fit if it lies below this value.

Spectroscopic data for CH$_3^{18}$OH are provided in the Cologne Database of Molecular Spectroscopy\footnote{\href{https://www.astro.uni-koeln.de/cdms}{https://www.astro.uni-koeln.de/cdms}} (CDMS, \cite{cdms_2} and \cite{cdms_1,cdms_0}) from the spectroscopic analyses of \cite{spec-CH3-18-OH_0} and \cite{spec-CH3-18-OH_1}. The entry of $^{13}$CH$_3$OH, from the CDMS catalogue, is based on \cite{spec-13-CH3OH_0} and \cite{spec-13-CH3OH_1}. 
The dipole moment is assumed to be the same as of the main isotopomer, and the partition function takes only the permanent dipole moment into account. The spectroscopy of CH$_2$DOH is taken from the Jet Propulsion Laboratory catalogue\footnote{\href{https://spec.jpl.nasa.gov/}{https://spec.jpl.nasa.gov/}} \citep[JPL,][]{jpl_0}, where the entry is based on \cite{spec-CH2DOH_0}.
The frequency list for CHD$_2$OH is based on \cite{Ndao-2016}, with additional transitions in the range of the PILS observations from \cite{Mukhopadhyay-2016}. The data do not include line strengths or the partition function; however, the frequencies are assigned with uncertainty much lower than the resolution of the observations used in this study.

The spectroscopic data for CH$_3$OCHO isotopologues, except for CH$_3$O$^{13}$CHO, are described in \cite{Manigand-2018}, and references therein, from which the derived column densities for source A are taken. The CDMS entry for CH$_3$O$^{13}$CHO is based on \cite{spec-CH3O-13-CHO_0}, with measured data in the range of the PILS survey from \cite{spec-CH3O-13-CHO_1}, \cite{spec-CH3O-13-CHO_2}, and \cite{spec-CH3OCHO_3}. The vibrational contribution of CH$_3$O$^{13}$CHO is already included in the CDMS entry \citep{Favre-2014}.

The data for CH$_3$OCH$_3$ are provided by the CDMS catalogue. The entry is based on \cite{spec-CH3OCH3_0} with additional data in the range of our survey from \cite{spec-CH3OCH3_1}. The mono-deuterated dimethyl ether (CH$_2$DOCH$_3$) exists in two forms, symmetric and asymmetric. The data of both conformers are taken from \cite{spec-CH2DOCH3_0}. The data of CH$_3$O$^{13}$CH$_3$ are taken from \cite{spec-CH3O-13-CH3_0} and from the Vizier Online Data Catalogue, which is from the Centre de Donn{\'e}es Astronomiques de Strasbourg\footnote{\href{http://cdsarc.u-strasbg.fr/}{http://cdsarc.u-strasbg.fr/}} (CDS). The contribution from excited vibrational levels is estimated to be 1.15 at T$_{ex}$~=~100~K for the $^{13}$C and deuterated isotopologues. 

Spectroscopic data for CH$_3$CHO can be found in the JPL catalogue. The entry is based on \cite{spec-CH3CHO_0}, with experimental transition frequencies in the range of our survey from \cite{spec-CH3CHO_1}. The CDMS entry for CH$_3$CDO is based on \cite{Coudert-2019}. The fit and the experimental frequencies used are described in \cite{Elkeurti-2010}, with additional data below 50~GHz taken from \cite{Martinache-1989}. The partition function includes the first three excited torsional mode contributions and is reliable at 140~K. The spectroscopy of the other deuterated isotopologue CH$_2$DCHO is also described in \cite{Coudert-2019} with additional data taken from \cite{Turner-1976} and \cite{Turner-1981}. The contribution of the lowest torsional mode and the small amplitude vibration are included in the partition function. The partition function converged up to about 100~K. Therefore, vibrational corrections are neglected at 140~K.

C$_2$H$_5$OH isotopologues exist under two isomeric forms, anti and gauche, depending on the OH group torsion. In addition, the gauche conformer has two degenerated states. The extended analysis of the vibrational ground state is provided by \cite{Pearson-2008}. However, \cite{Muller-2016} find that the predicted intensities do not fit the Sgr B2(N2) emission at 3 mm, and they provide a corrected entry in the CDMS catalogue. This entry combines the anti and the gauche conformer transitions and includes data from \cite{Pearson-1995, Pearson-1996} in the range of the PILS survey. $^{13}$C and deuterated C$_2$H$_5$OH isotopologues spectroscopic data are taken from  corresponding CDMS entries, based on \cite{Bouchez-2012} and \cite{Walters-2015}, respectively. However, these data consider only the anti conformer. The column density has to be multiplied by a factor of $\sim$2.32 at T$_{ex}$ = 135 K to take the gauche conformer into account \citep{Muller-2016, Jorgensen-2018}. The contribution from excited vibrational levels is dominated by the two torsional modes of the methyl group \citep{Durig-1990}\footnote{It is the combination of four torsional transitions at 244.4 and 231.1 cm$^{-1}$ of the trans conformer and 490.0 and 453.0 cm$^{-1}$ of the gauche conformer.}. The vibrational correction factor is 1.24 for the main isotopologue \citep{Durig-2011} and is applicable for the $^{13}$C and deuterated isotopologues.

Spectroscopic data of the H$_2$CO isotopologues are taken from the corresponding CDMS entries and are the result of several contributions over the past four decades. H$_2$CO, H$_2$C$^{18}$O and H$_2$C$^{17}$O entries are based on \cite{Muller-2017}, the H$_2^{13}$CO entry is based on \cite{Muller-2000c}, and HDCO and D$_2$CO entries are based on \cite{Zakharenko-2015}. Additional data were taken from \cite{Brunken-2003}, \cite{Bocquet-1996} and \cite{Cornet-1980} for H$_2$CO as well as a few transition frequencies of H$_2$C$^{18}$O and H$_2^{13}$CO. \cite{Muller-2000a} provided the ground state combination differences employed in the spectroscopic analysis of H$_2$CO. For the two deuterated isotopologues HDCO and D$_2$CO, the data were taken from \cite{Lohilahti-2004}, \cite{Bocquet-1999} and \cite{Dangoisse-1978}. Additional H$_2$C$^{18}$O transition frequencies were taken from  \cite{Muller-2000b}. The dipole moments were measured by \cite{Fabricant-1977} and \cite{Johns-1977}. The dipole moments of H$_2$C$^{17}$O and H$_2$C$^{18}$O are assumed to be the same as the H$_2$CO one. The vibration state energies are way too high to significantly affect the partition function at 155~K.

The publicly available JPL entry for CH$_3$COCH$_3$, also called propanone, is based on \cite{Groner-2002}, with the rotational partitioning and the line strength calculation. This study used additional data from \cite{Oldag-1992}, \cite{Vacherand-1986} and \cite{Peter-1965} for transition frequencies up to 300 GHz. In the present study, we used an updated version of the spectroscopic data, which will be available on CDMS. This entry is based on the study of \cite{Ordu-2019}, which includes the previous linelists as well as those from the first two torsional excited states taken from \cite{Morina-2019} and \cite{Groner-2006}. The contribution from excited vibrational level $\nu$=1 is taken into account in the harmonic approximation and thus in the partition function. 

c-H$_2$COCH$_2$ data are provided by the CDMS catalogue. The entry was based on experimental data taken from \cite{Hirose-1974}, \cite{Creswell-1974} and \cite{Pan-1998}. Much more additional data from \cite{Medcraft-2012} enriched the set, extending the frequency range from 358 GHz to 4 THz. The dipole moment is taken from \cite{Cunningham-1951}. The vibrational correction factor is insignificant at 95~K.

The rotational spectrum of CH$_3$COOH was first extensively analysed by \cite{Ilyushin-2008}, which was based on a fair amount of measured spectroscopic data from 8 to 358 GHz taken from \cite{Tabor-1957}, \cite{Krisher-1971}, \cite{VanEijck-1981}, \cite{VanEijck-1983}, \cite{Demaison-1982}, \cite{Wlodarczak-1988}, \cite{Ilyushin-2001} and \cite{Ilyushin-2003}. The present study uses the recent datafile provided by \cite{Ilyushin-2013}. This last study considerably extended the number of measured transition frequencies up to 845 GHz and observed the first two torsional excited states. The partition function takes into account these states and more excited torsional up to $v_\text{t}=8$ in the vibrational ground state. The low vibration states are too energetically high to make the vibrational correction factor significant at 110~K.

The JPL catalogue provides the line list for CH$_2$(OH)CHO. The entry is based on the rotational lines collection of \cite{Carroll-2010}. Previous laboratory measurements were carried out by \cite{Marstokk-1970}, \cite{Marstokk-1973} and \cite{Butler-2001} and served as the base for the latest study. The partition function was previously calculated by \cite{WidicusWeaver-2005} over a measured frequency range from 72 to 122.5 GHz and included the ground state and three partially populated vibrational states at room temperature. \cite{Carroll-2010} find a systematic bias in the lines with K$_\text{a}\sim$~28$-$31 and they corrected iteratively the assignment in the JPL line list. 

NH$_2$CHO data used in this study are provided by the CDMS catalogue. The dipole moment was measured by \cite{Kurland-1957}. The observed experimental frequencies are taken from \cite{Kukolich-1971}, \cite{Nielsen-1973}, \cite{Hirota-1974}, \cite{Gardner-1980}, \cite{Moskienko-1991}, \cite{Vorobeva-1994}, \cite{Blanco-2006} and \cite{Kryvda-2009}. The latest study \citep{Motiyenko-2012} considerably extended the frequency range of measured transitions up to 930 GHz. The partition function does not include any contributions from the excited vibrational state. However, the lowest vibration state energy of 288~cm$^{-1}$ leads to the highest vibration correction factor, which is lower than 1.1 at 145~K. 

Spectroscopic data of the HCOOH isotopologues are provided in the CDMS catalogue for \textit{t}-HCOOH and by JPL for \textit{t}-H$^{13}$COOH, \textit{t}-DCOOH, and \textit{t}-HCOOD. The CDMS entry for the main isotopologue is based on \cite{Winnewisser-2002} with additional data from previous studies \citep{Auwera-1992, Willemot-1980}. The JPL entries are based on the same study of \cite{Lattanzi-2008} gathering the analysis of these three isotopologues. Experimental measurements were taken from \cite{Bellet-1971} and \cite{Lerner-1957} for the three entries, and \cite{Wellington-1980}, \cite{Winnewisser-2002} only for the \textit{t}-H$^{13}$COOH entry.  The dipole moment is assumed to be the same as the main isotopologue, which was measured by \cite{Kuze-1982}. The excited vibrational state, especially $\nu_7$ and $\nu_9,$ was also measured for the $^{13}$C and deuterated isotopologues in the studies of \cite{Baskakov-2006}, \cite{Baskakov-1999} and \cite{Baskakov-1996}. Vibration correction factors are neglected at 90~K for both the main and deuterated species.

The CDMS entries for CH$_2$CO isotopologues are based on \cite{Guarnieri-2003} and on \cite{Guarnieri-2005} for the deuterated isotopologues (CHDCO). Laboratory measurements of the different isotopologues are taken from \cite{Johnson-1952}, \cite{Brown-1990}, \cite{Johns-1992} and \cite{Sutter-2000}. The dipole moments were calculated by \cite{Fabricant-1977}. Vibrations contributions are negligible at 135~K for both isotopologues.

(CH$_2$OH)$_2$ has three coupled rotors along the molecular chain. It can exist in a total of ten stable configurations \citep{Christen-2001} based on the orientation of the two OH groups at both ends of the carbon chain. Most of the transitions, especially the low-energy rotational transitions, are emitted from the two lowest fundamental state conformers \textit{aGg'}-(CH$_2$OH)$_2$ and \textit{gGg'}-(CH$_2$OH)$_2$, the second one lying $~$290~K above \textit{aGg'} \citep{Muller-2004}. The spectroscopic data used here are taken from the CDMS entries. Experimental measurements and extensive analysis were carried out by \cite{Christen-1995} and \cite{Christen-2003} for \textit{aGg'} and by \cite{Christen-2001} and \cite{Muller-2004} for \text{gGg'}, up to 370 and 579~GHz, respectively. Concerning the \text{gGg'} conformer, the line strength and partition function were treated as if the conformer was the lowest conformer energy. The ground state energy of $~$290~K contributes to the line strengths by a factor of 2.35 at 145~K \citep{Muller-2004}. 
The vibrational correction factor at 145~K of 1.75, which includes the interactions between \textit{gGg'} and \textit{aGg'}, were calculated in the anharmonic approximation from \cite{Boussessi-2016} for both conformers.

\textit{t}-C$_2$H$_5$OCH$_3$ data were taken from a preliminary CDMS entry. They are based on \cite{Fuchs-2003}, with significant additional data from \cite{Kobayashi-2016} as well as the experimental measurements from \cite{Hayashi-1982}, \cite{Tsunekawa-2003}, and  \cite{Tsunekawa-2011}. The dipole moment was measured by \cite{Hayashi-1975}. Torsional and vibrational transitions were studied by \cite{Senent-2009}, providing a future basis for determining the contribution of higher vibrational state energies to the partition function. The partition function and the line strength were revised by H.~S.~P. M{\"u}ller. With the lowest vibrations at 288~cm$^{-1}$, the vibrational correction factor is negligible.

Regarding HNCO isotopologues, the spectroscopic data are taken from the CDMS entry for HNCO, and from the JPL entry for DNCO. The CDMS entry is based on the work of \cite{Lapinov-2007}, using additional data from \cite{Kukolich-1971}, \cite{Hocking-1975} and \cite{Niedenhoff-1995}. The JPL entry was last updated in 1987 and only includes the frequencies from \cite{Hocking-1975}. However, the frequency uncertainties are lower than 0.4 MHz in the range of the PILS observations, except for the transitions at 346.027, 346.176, 346.380, and 346.541 GHz. These transitions correspond to very high upper state energies (i.e. $>$ 1000 K) and they are not populated in the temperature and density ranges of IRAS 16293; therefore, the spectroscopy is reliable. We note that the HNCO partition function does not take into account the spin multiplicity of the $^{14}$N nucleus. Vibration correction factors are negligible at 180~K due to the high lowest vibration contributions of 577 and $\sim$460~cm$^{-1}$ for HNCO and DNCO, respectively.

CH$_3$OCH$_2$OH spectroscopy has been investigated very recently because of its lower dipole moment, which makes this species more difficult to detect in space. The data used are based on the work of \cite{Motiyenko-2018}. They acquired the first measurements of rotation transitions between 150 and 460 GHz. The line strength and the partition function have not been published yet; however, they have been kindly provided by B. McGuire under the approval of R.~A. Motiyenko. Based on the calculated vibrations (harmonic) from \cite{Hays-2013}, the vibrational correction factor is 1.55 at 130~K.

The spectroscopic data of C$_2$H$_5$CHO were found on the CDMS database. The CDMS entry is largely based on the recent study of \cite{Zingsheim-2017} with additional data taken from \cite{Hardy-1982} and \cite{Demaison-1987}, though they were remeasured in the most recent study. The dipole moment was measured by \cite{Butcher-1964}. The partition function takes into account the ground vibrational state alone. The vibration correction factor is 1.42 at 120~K \citep{Durig-1980}.

\onecolumn

\section{Identification of CHD$_2$OH transitions}

\begin{figure}[h!]
\centering
\includegraphics[scale=0.5]{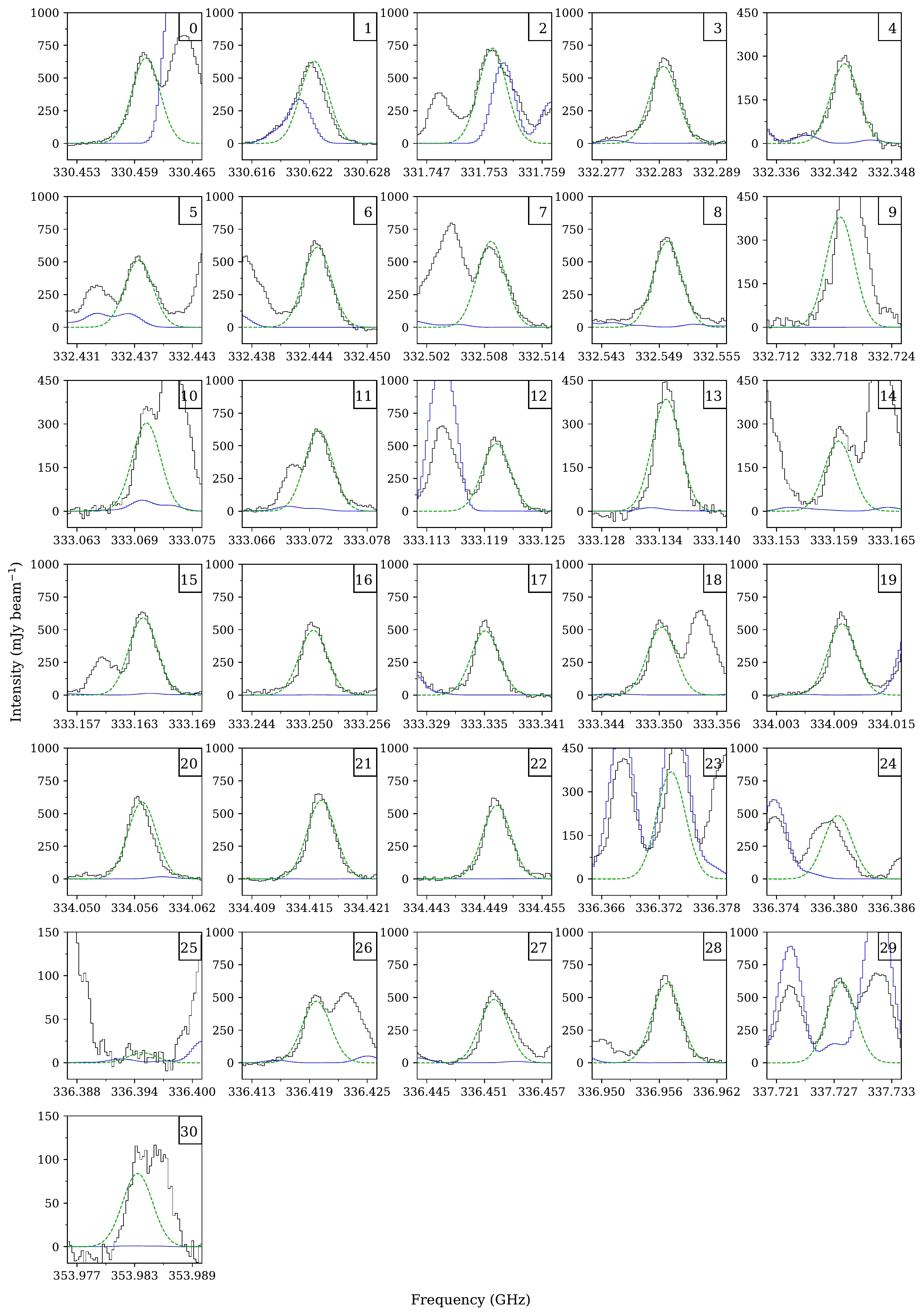}
\caption{\label{App-CHD2OH-idA-fig}Selection of CHD$_2$OH identified lines in the data (in black) with the Gaussian fit in green dashed line, with fixed central velocity and FWHM. The same reference spectrum used in the analysis is represented in blue. The squared number in the top-right corner corresponds to the identification number of the transition, i.e. the panel number in Table \ref{App-CHD2OH-id-tab}.}
\end{figure}

\begin{figure}[h!]
\centering
\includegraphics[scale=0.55]{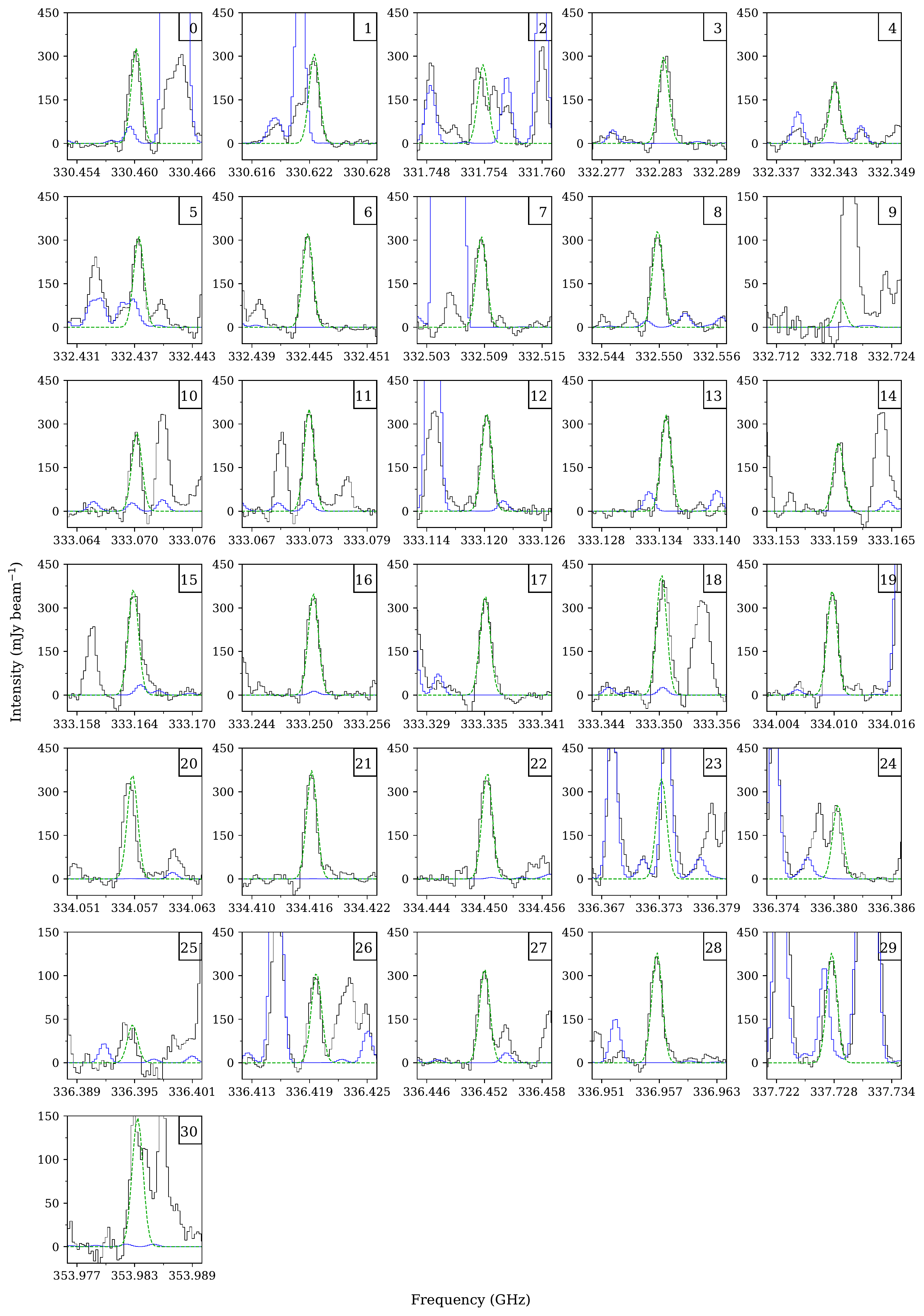}
\caption{\label{App-CHD2OH-idB-fig}Selection of CHD$_2$OH identified lines in the data (in black) with the Gaussian fit in green dashed line, with fixed central velocity and FWHM. The same reference spectrum used in the analysis is represented in blue. The squared number in the top-right corner corresponds to the identification number of the transition, i.e. the panel number in Table \ref{App-CHD2OH-id-tab}.}
\end{figure}

The identified lines of CHD$_2$OH are given in this section, in Table \ref{App-CHD2OH-id-tab}. The unblended lines alone were fitted using a Gaussian line profile and plotted in Figures \ref{App-CHD2OH-idA-fig} and \ref{App-CHD2OH-idB-fig}.

\begin{table}[h!]
\centering
\caption{\label{App-CHD2OH-id-tab}List of the CHD$_2$OH identified lines. The quantum numbers are taken from \cite{Mukhopadhyay-2016}.}
\begin{tabular}{ccccc}
\hline\hline
Transition & Frequency & panel number\tablefootmark{b} & $W_\text{IRAS 16293A}$ & $W_\text{IRAS 16293B}$ \\
(J'', K'') $\rightarrow$ (J', K')\tablefootmark{a} & (GHz) & & (Jy beam$^{-1}$ km s$^{-1}$) & (Jy beam$^{-1}$ km s$^{-1}$) \\
\hline
(7, 1 $+$ e$_1$) $\rightarrow$ (8, 1 $+$ e$_1$) &    330.4602 &  0 & -- & $ 0.362\pm 0.013$ \\
(7, 0o$_1+$) $\rightarrow$ (8, 0o$_1+$) &    330.6225 &  1 & -- & -- \\
(7, 0e$_0+$) $\rightarrow$ (8, 0e$_0+$) &    331.7538 & 2 & -- & -- \\
(7, 0e$_1+$) $\rightarrow$ (8, 0e$_1+$) &    332.2835 &  3 & $ 2.129\pm 0.078$ & $ 0.314\pm 0.014$ \\
(7, 7 $\pm$ e$_1$) $\rightarrow$ (8, 7 $\pm$ e$_1$) &    332.3431 &  4 & $ 0.938\pm 0.079$ & $ 0.239\pm 0.014$ \\
(7, 6 $\pm$ e$_1$) $\rightarrow$ (8, 6 $\pm$ e$_1$) &    332.4374 &  5 & -- & $ 0.349\pm 0.013$ \\
(7, 2 $-$ e$_1$) $\rightarrow$ (8, 2 $-$ e$_1$) &    332.4448 &  6 & $ 2.139\pm 0.094$ & $ 0.353\pm 0.016$ \\
(7, 5 $\pm$ e$_1$) $\rightarrow$ (8, 5 $\pm$ e$_1$) &    332.5087 &  7 & -- & $ 0.353\pm 0.015$ \\
(7, 4 $\pm$ e$_1$) $\rightarrow$ (8, 4 $\pm$ e$_1$) &    332.5498 &  8 & $ 2.382\pm 0.081$ & $ 0.370\pm 0.014$ \\
(7, 2 $+$ e$_1$) $\rightarrow$ (8, 2 $+$ e$_1$) &    332.7186 &  9 & -- & -- \\
(7, 7e$_0\pm$) $\rightarrow$ (8, 7e$_0\pm$) &    333.0702 &  10 & -- & $ 0.286\pm 0.014$ \\
(7, 2 $-$ o$_1$) $\rightarrow$ (8, 2 $-$ o$_1$) &    333.0730 & 11 & -- & $ 0.391\pm 0.016$ \\
(7, 5 $\pm$ o$_1$) $\rightarrow$ (8, 5 $\pm$ o$_1$) &    333.1202 &  12 & $ 2.290\pm 0.107$ & $ 0.376\pm 0.015$ \\
(7, 6 $\pm$ o$_1$) $\rightarrow$ (8, 6 $\pm$ o$_1$) &    333.1347 &  13 & $ 1.195\pm 0.081$ & $ 0.350\pm 0.016$\\
(7, 6e$_0\pm$) $\rightarrow$ (8, 6e$_0\pm$) &    333.1595 & 14 & -- & $ 0.254\pm 0.015$ \\
(7, 7 $\pm$ o$_1$) $\rightarrow$ (8, 7 $\pm$ o$_1$) &    333.1639 & 15 & $ 2.403\pm 0.091$ & $ 0.442\pm 0.016$ \\
(7, 5e$_0\pm$) $\rightarrow$ (8, 5e$_0\pm$) &    333.2504 & 16 & $ 1.645\pm 0.082$ & $ 0.396\pm 0.015$ \\
(7, 3 $+$ e$_1$) $\rightarrow$ (8, 3 $+$ e$_1$) &    333.3351 & \multirow{ 2}{*}{$\left.{\color{white}\begin{matrix}i\\i\end{matrix}}\right\}17$} & \multirow{ 2}{*}{$ 1.579\pm 0.054$} & \multirow{ 2}{*}{$ 0.371\pm 0.013$} \\
(7, 3 $+$ o$_1$) $\rightarrow$ (8, 3 $+$ o$_1$) &    333.3361 & & & \\
{\Large\textcolor{white}{I}}(7, 3 $-$ o$_1$) $\rightarrow$ (8, 3 $-$ o$_1$){\Large\textcolor{white}{I}} &    333.3502 & \multirow{ 3}{*}{$\left.{\color{white}\begin{matrix}i\\i\\i\end{matrix}}\right\}18$} & \multirow{ 3}{*}{--} & \multirow{ 3}{*}{$ 0.467\pm 0.013$} \\
(7, 4e$_0\pm$) $\rightarrow$ (8, 4e$_0\pm$) &    333.3502 & & & \\
(7, 3 $-$ e$_1$) $\rightarrow$ (8, 3 $-$ e$_1$) &    333.3509 & & & \\
{\Large\textcolor{white}{I}}(7, 2 $-$ e$_0$) $\rightarrow$ (8, 2 $-$ e$_0$){\Large\textcolor{white}{I}} &    334.0098 & 19 & $ 1.904\pm 0.080$ & $ 0.389\pm 0.014$ \\
(7, 2 $+$ e$_0$) $\rightarrow$ (8, 2 $+$ e$_0$) &    334.0568 & 20 & $ 1.849\pm 0.074$ & $ 0.362\pm 0.012$ \\
(7, 1 $-$ e$_1$) $\rightarrow$ (8, 1 $-$ e$_1$) &    334.4162 & 21 & $ 1.929\pm 0.069$ & $ 0.407\pm 0.013$ \\
(7, 2 $+$ o$_1$) $\rightarrow$ (8, 2 $+$ o$_1$) &    334.4503 & 22 & $ 1.856\pm 0.079$ & $ 0.429\pm 0.015$ \\
(4, 3o$_1$) $\rightarrow$ (4, 4e$_0$) &    336.3732 & 23 & -- & -- \\
(5, 3o$_1$) $\rightarrow$ (5, 4e$_0$) &    336.3804 & 24 & -- & -- \\
(6, 3o$_1$) $\rightarrow$ (6, 4e$_0$) &    336.3948 & 25 & $ 0.107\pm 0.207$ & $ 0.049\pm 0.013$ \\
(7, 3o$_1$) $\rightarrow$ (7, 4e$_0$) &    336.4197 & 26 & -- & $ 0.341\pm 0.013$ \\
(8, 3o$_1$) $\rightarrow$ (8, 4e$_0$) &    336.4520 & 27 & $ 1.806\pm 0.078$ & $ 0.367\pm 0.014$ \\
(7, 1 -- e$_0$) $\rightarrow$ (8, 1 -- e$_0$) &    336.9568 & 28 & $ 2.042\pm 0.072$ & $ 0.414\pm 0.013$ \\
(7, 1 $-$ o$_1$) $\rightarrow$ (8, 1 $-$ o$_1$) &    337.7278 & 29 & -- & -- \\
(6, 5e$_1$) $\rightarrow$ (7, 6e$_1$) &    353.9833 & 30 & -- & -- \\
\hline
\end{tabular}
\tablefoot{
\tablefoottext{a}{e$_0$, o$_1$ and e$_1$ corresponds to the three first torsional states. The notation of the quantum numbers is directly taken from \cite{Mukhopadhyay-2016}.}
\tablefoottext{b}{This corresponds to the number indicated in the top-right corner of each panel in Figures \ref{App-CHD2OH-idA-fig} and \ref{App-CHD2OH-idB-fig}.}
\tablefoottext{c}{$W$ is the integrated intensity of the unblended lines, summed over the range [$\nu_0 - \Delta\nu_\text{FWHM}$, $\nu_0 + \Delta\nu_\text{FWHM}$], given $\Delta\nu_\text{FWHM} = \nu_0\frac{FWHM}{\text{c}}$, with FWHM equal to 2.2 and 0.8~km~s$^{-1}$ for IRAS 16293A and B, respectively.}
}
\end{table}

\section{Synthetic spectra}

\begin{figure}[h!]
{CH$_3$OH}\\
\includegraphics[scale=0.52]{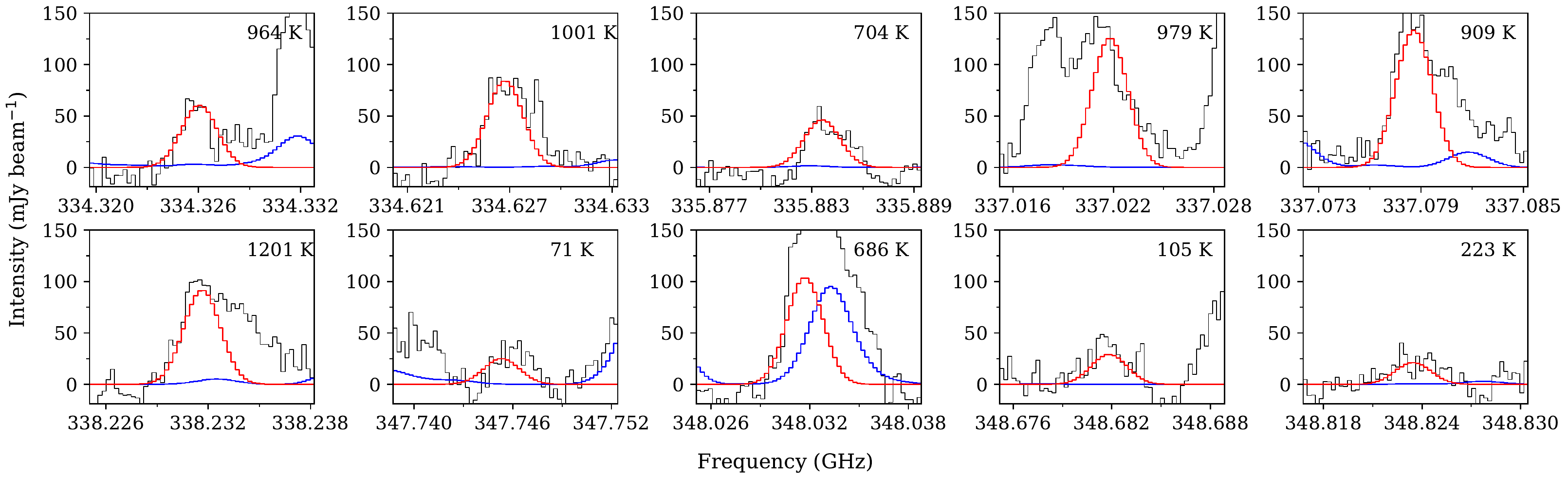} 

{$^{13}$CH$_3$OH}\\
\includegraphics[scale=0.52]{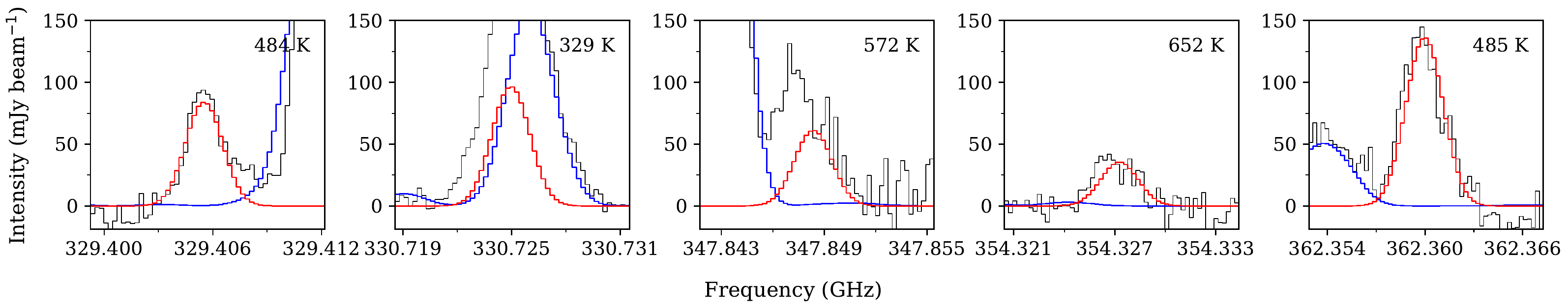} 

{CH$_3^{18}$OH}\\
\includegraphics[scale=0.52]{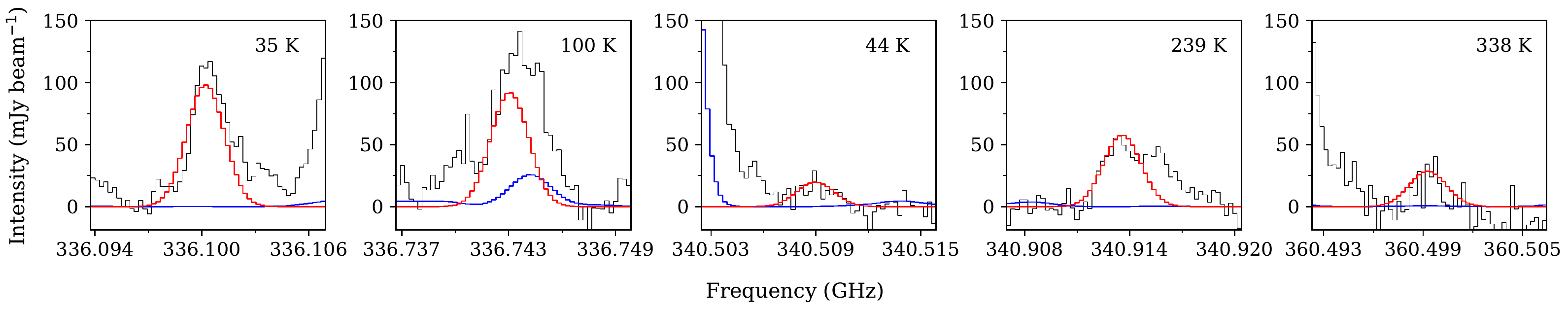} 

{CH$_2$DOH}\\
\includegraphics[scale=0.52]{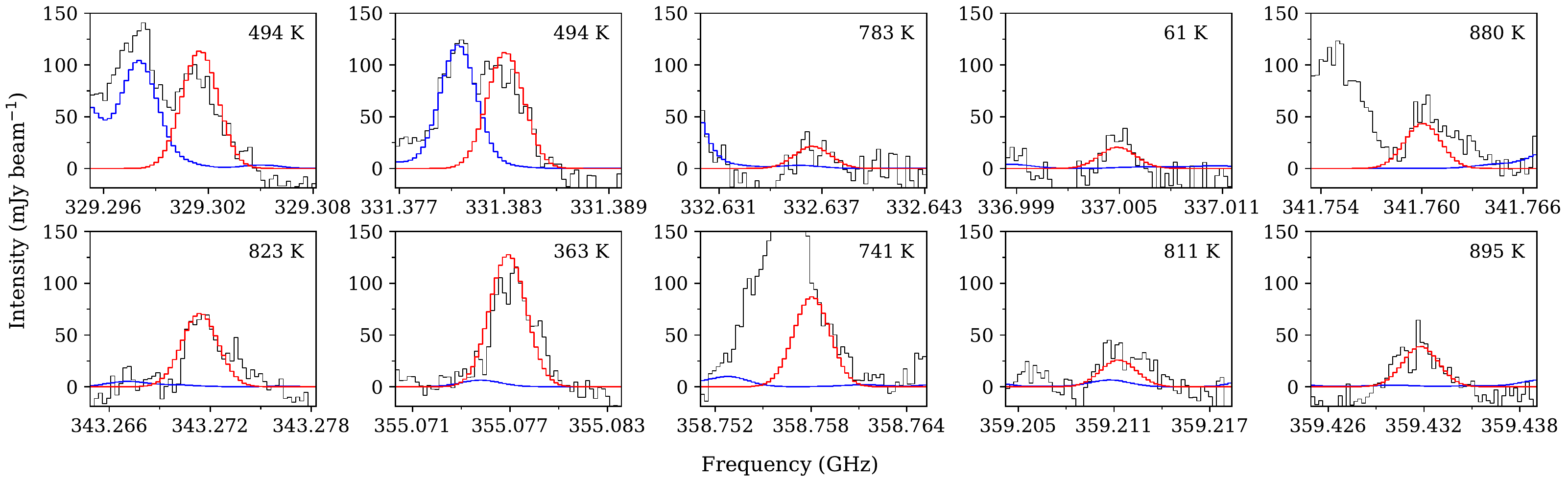} 

{CH$_3$OD}\\
\includegraphics[scale=0.52]{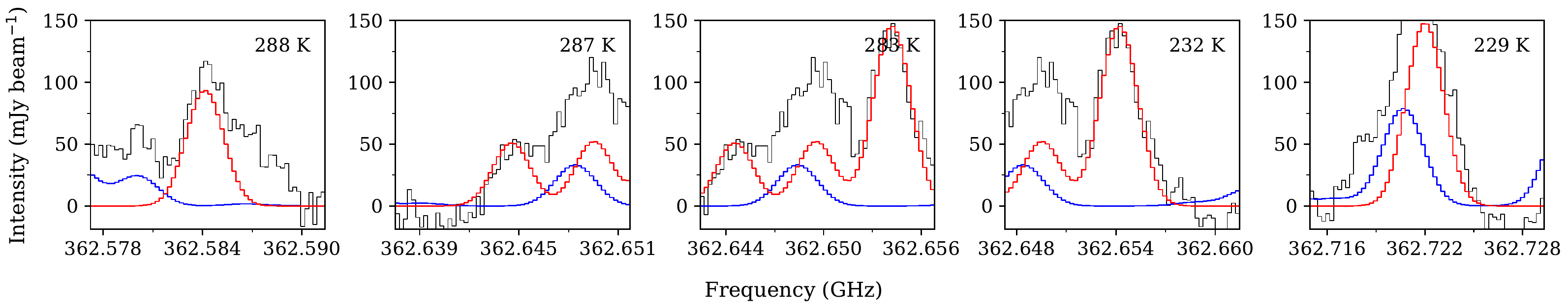} 
\caption{Representative selection of transitions of CH$_3$OH isotopologues towards IRAS 16293 A. The synthetic spectra is over-plotted in red, the reference spectrum in blue, and the data in black. The upper energy level of the transition is indicated in the top-left corner.}
\end{figure}

\newpage
\begin{figure}[h!]
{CH$_3$OCH$_3$}\\
\includegraphics[scale=0.54]{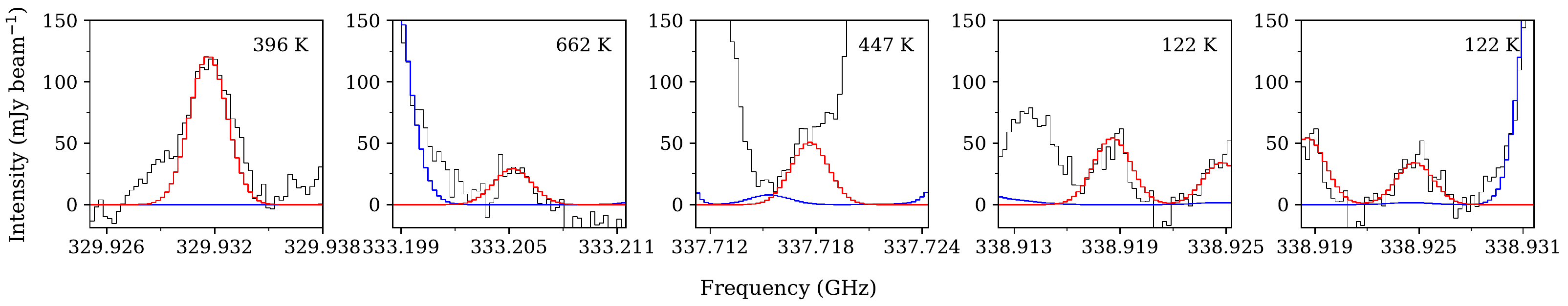} 

{CH$_3$O$^{13}$CH$_3$}\\
\includegraphics[scale=0.54]{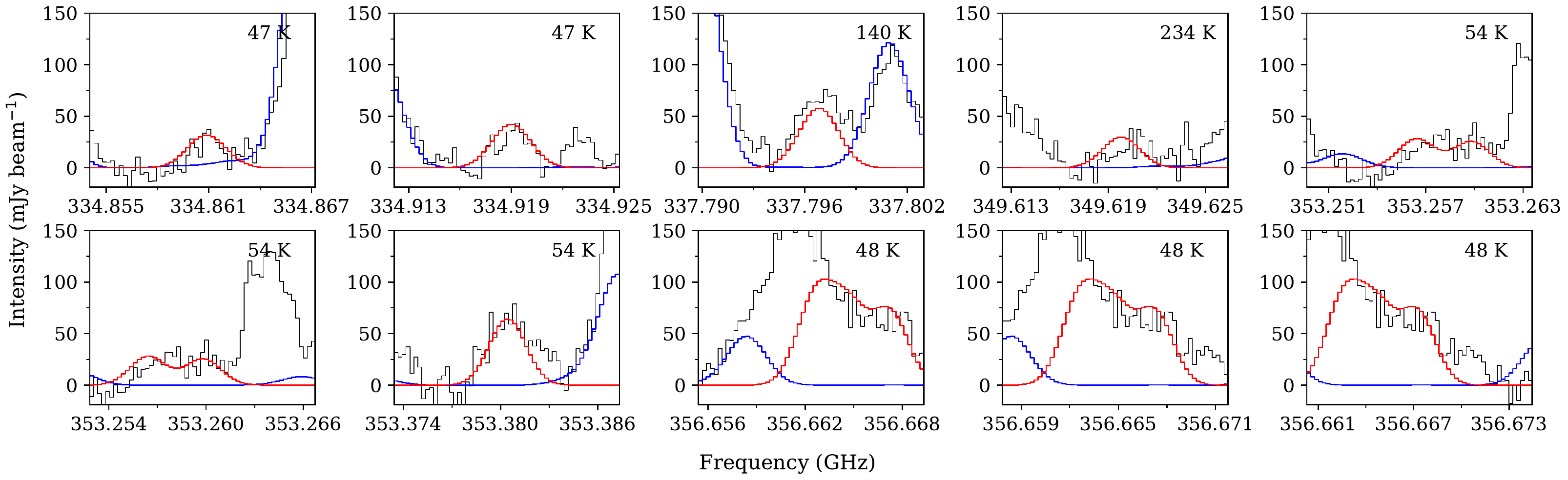} 

{a-CH$_2$DOCH$_3$}\\
\includegraphics[scale=0.54]{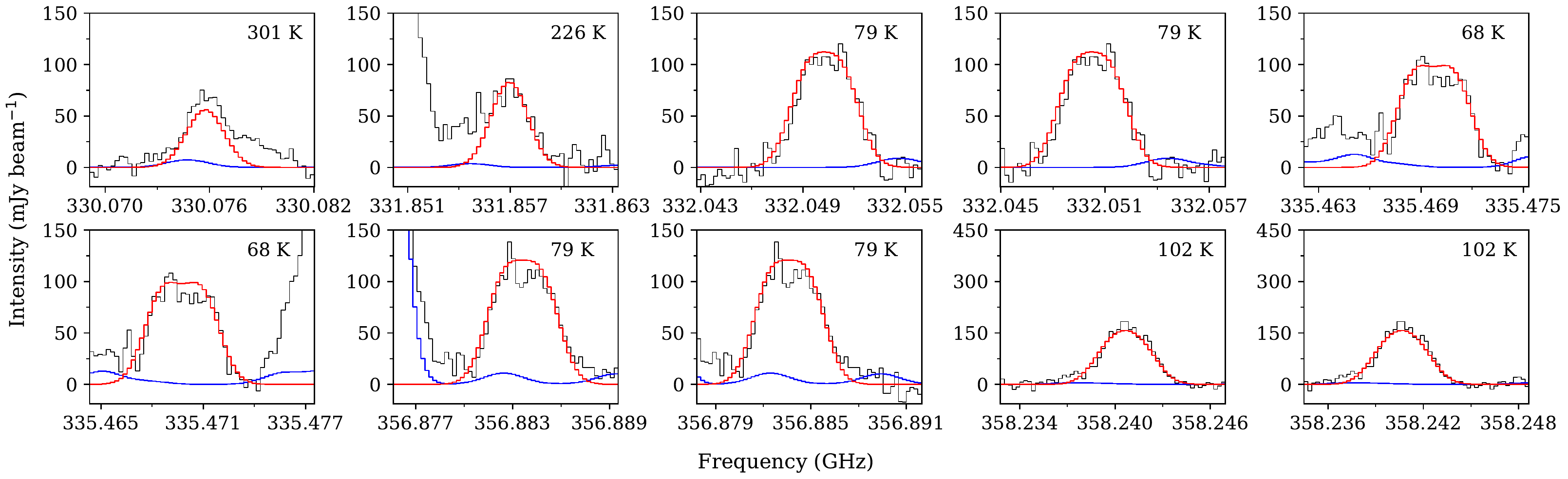} 

{s-CH$_2$DOCH$_3$}\\
\includegraphics[scale=0.54]{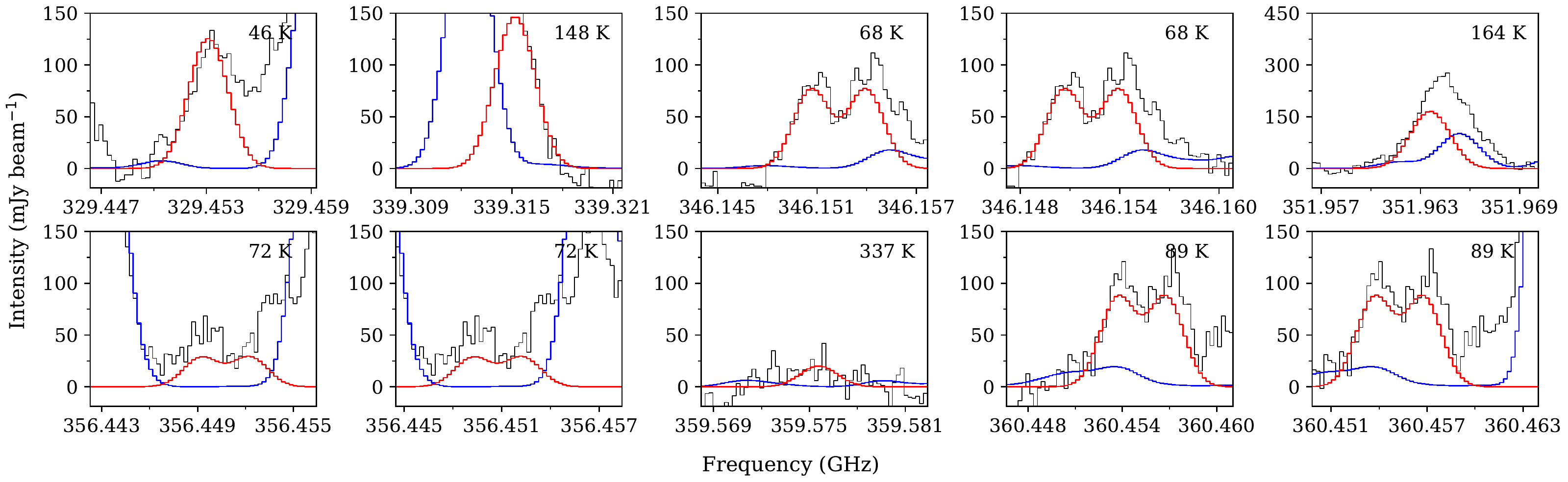} 
\caption{Representatives selection of transitions of CH$_3$OCH$_3$ isotopologues towards IRAS 16293 A. The synthetic spectra is over-plotted in red, the reference spectrum in blue, and the data in black. The upper energy level of the transition is indicated in the top-left corner.}
\end{figure}

\newpage
\begin{figure}[h!]
{H$_2$CO}\\
\includegraphics[scale=0.54]{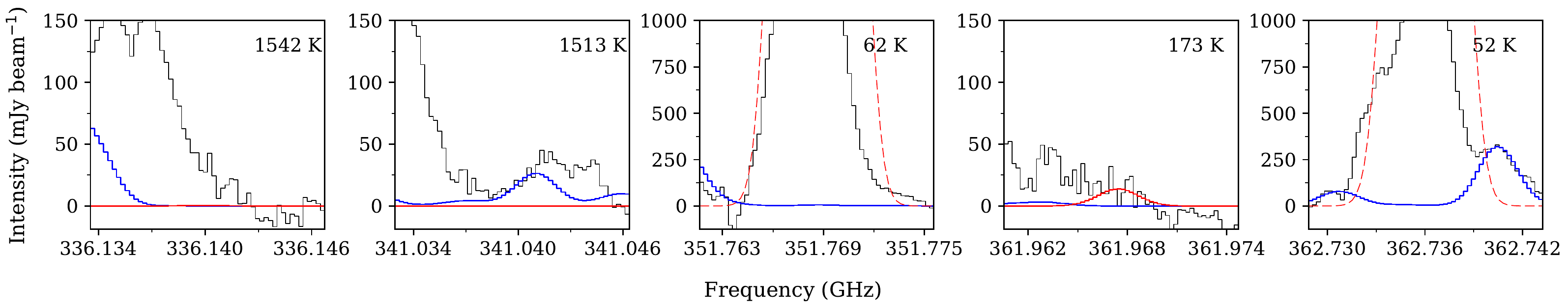}

{H$_2^{13}$CO}\\
\includegraphics[scale=0.54]{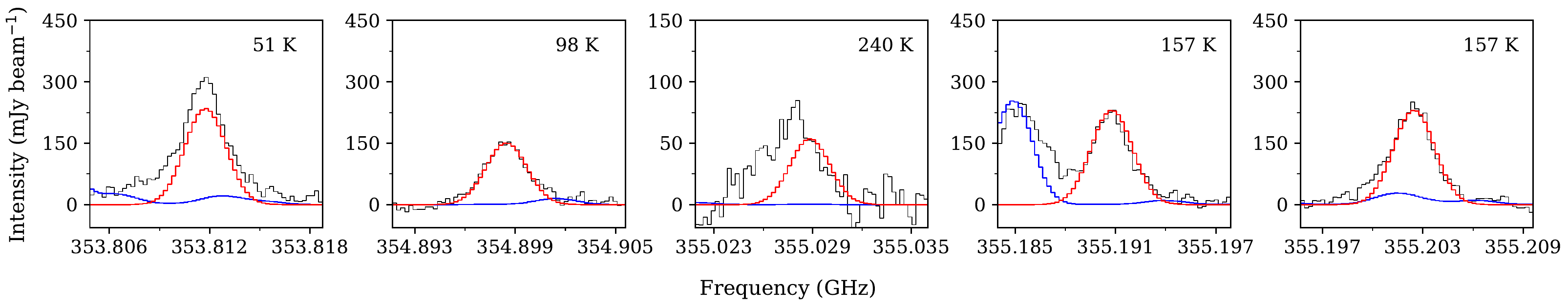}

{H$_2$C$^{18}$O}\\
\includegraphics[scale=0.54]{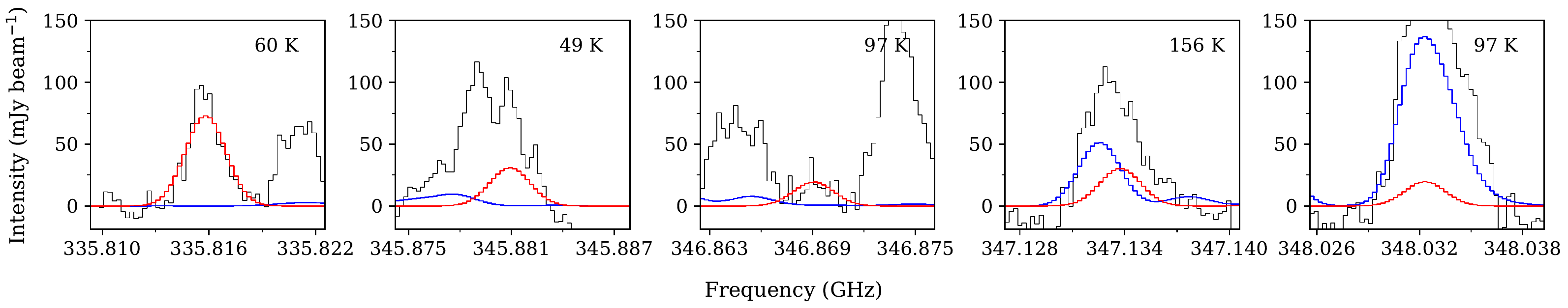}

{HDCO}\\
\includegraphics[scale=0.54]{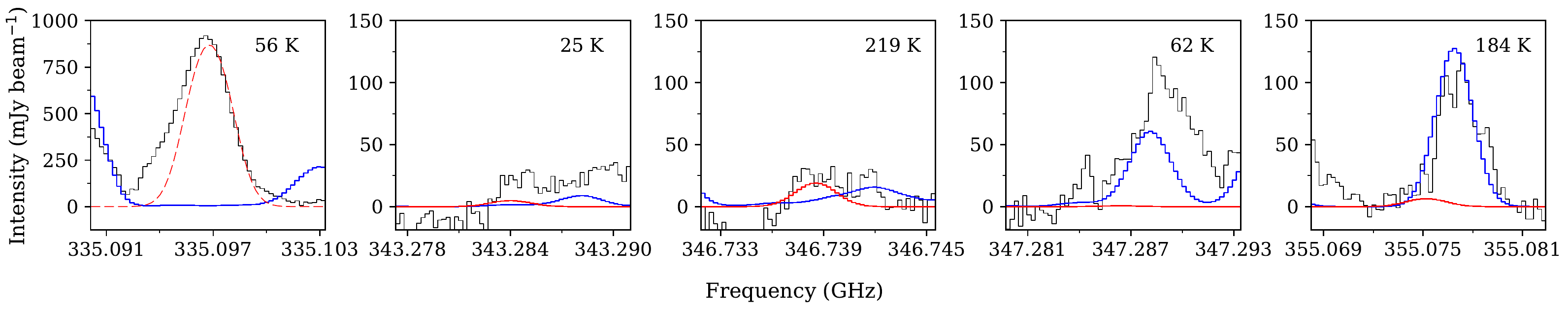}

{D$_2$CO}\\
\includegraphics[scale=0.54]{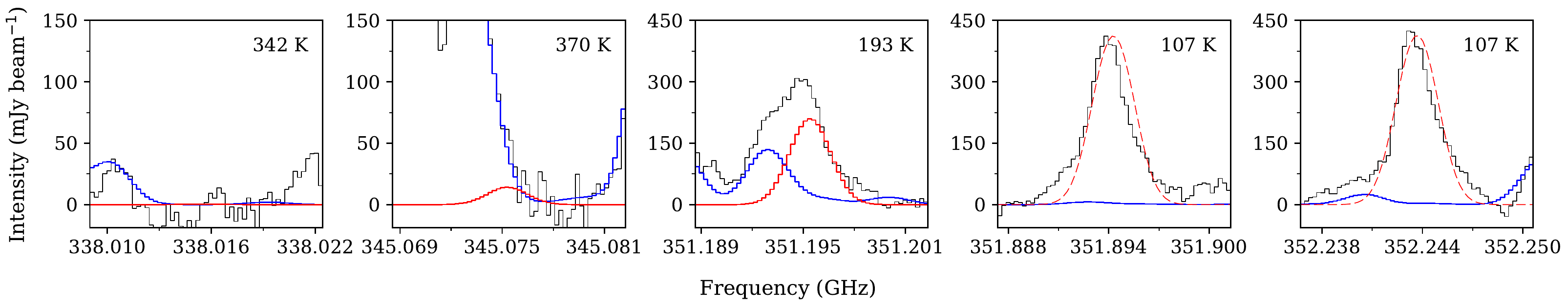}
\caption{Representative selection of transitions of H$_2$CO isotopologues towards IRAS 16293A. The synthetic spectra is over-plotted in red, the reference spectrum in blue, and the data in black. The dashed curves represent optically thick lines. The upper energy level of the transition is indicated in the top-left corner.}
\end{figure}
\newpage

\begin{figure}[h!]
{C$_2$H$_5$OH}\\
\includegraphics[scale=0.54]{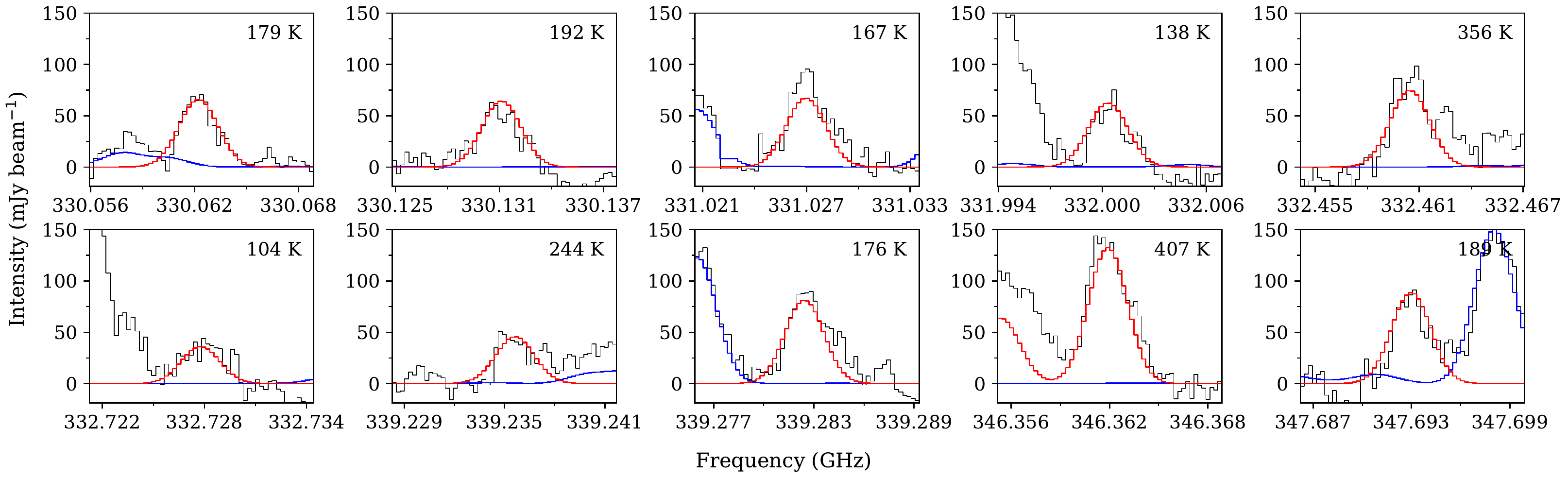}

{\textit{a}-CH$_3$CHDOH}\\
\includegraphics[scale=0.54]{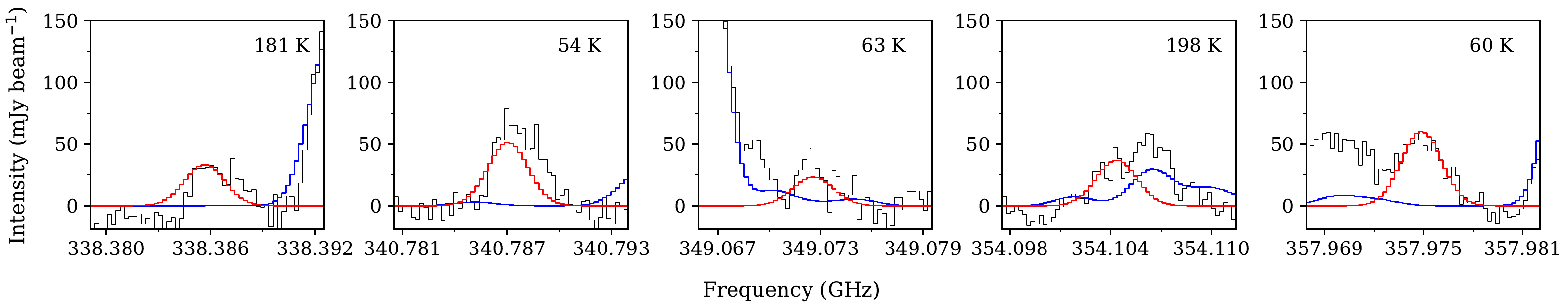}

{a-CH$_3$CH$_2$OD}\\
\includegraphics[scale=0.54]{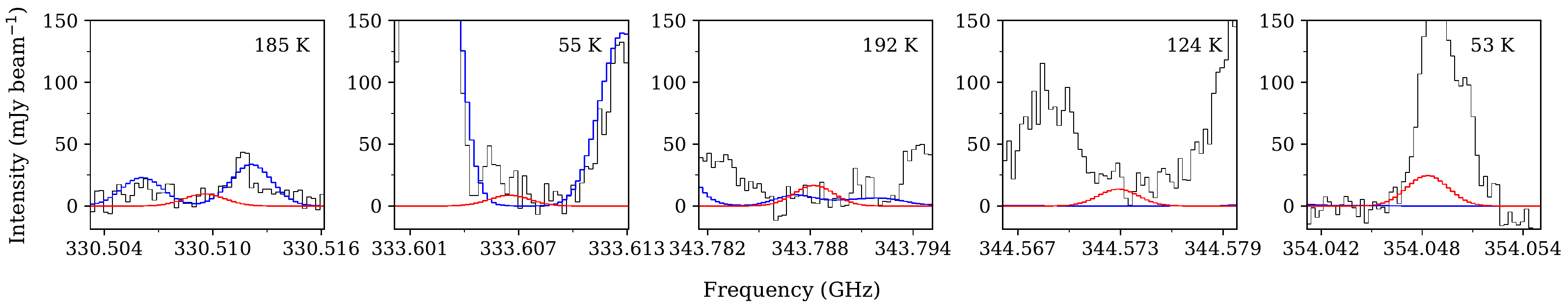}

{a-a-CH$_2$DCH$_2$OH}\\
\includegraphics[scale=0.54]{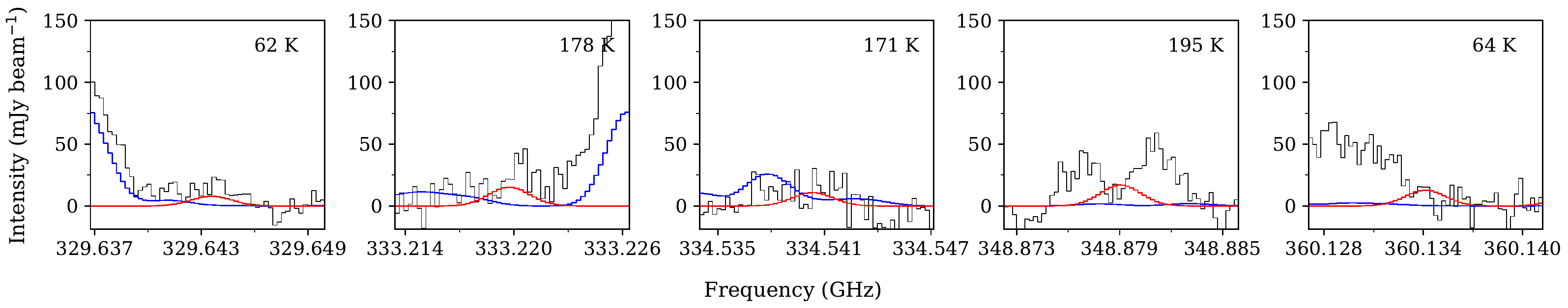}

{a-s-CH$_2$DCH$_2$OH}\\
\includegraphics[scale=0.54]{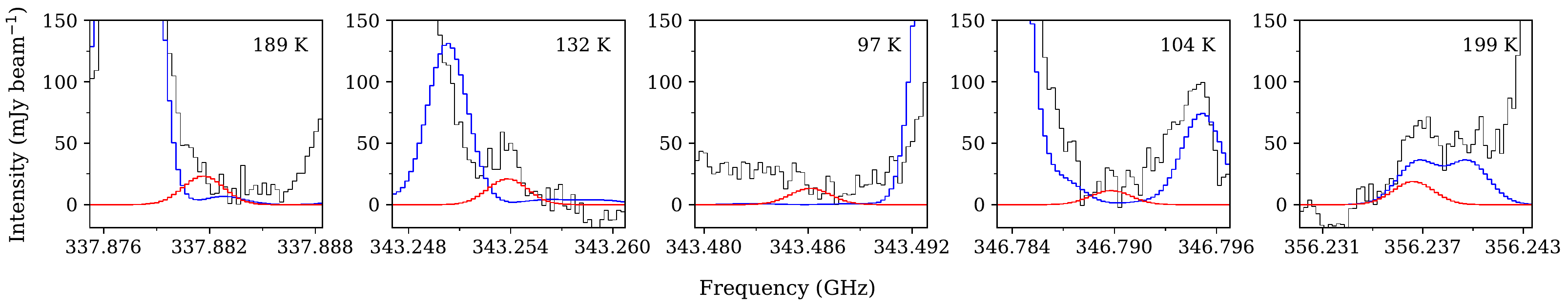}
\caption{Representative selection of transitions of C$_2$H$_5$OH isotopologues towards IRAS 16293 A. The synthetic spectra is over-plotted in red, the reference spectrum in blue, and the data in black. The upper energy level of the transition is indicated in the top-left corner.}
\end{figure}

\newpage
\noindent
\begin{figure}[h!]
{t-HCOOH}\\
\includegraphics[scale=0.54]{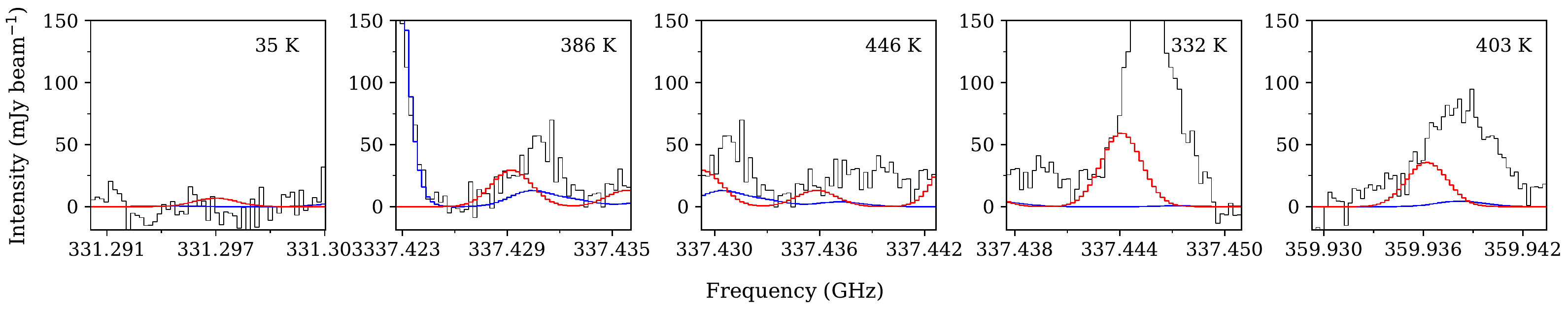}

{t-H$^{13}$COOH}\\
\includegraphics[scale=0.54]{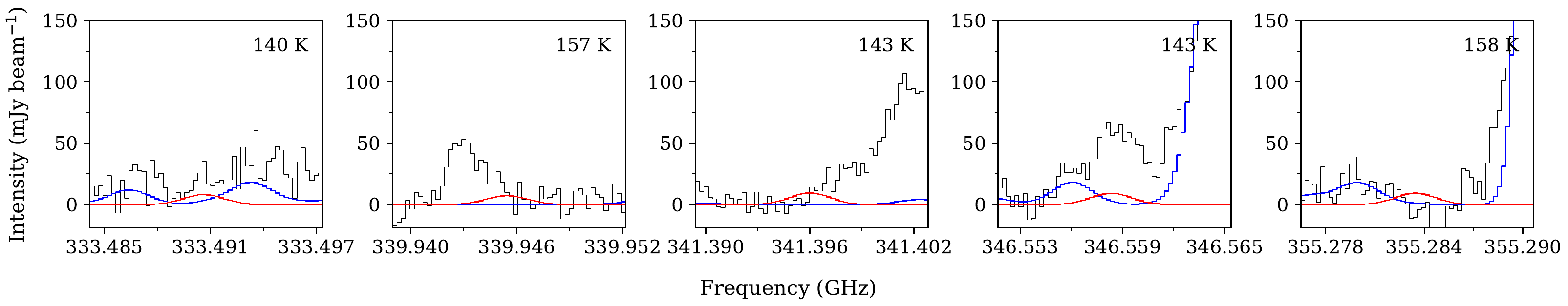}

{t-DCOOH}\\
\includegraphics[scale=0.54]{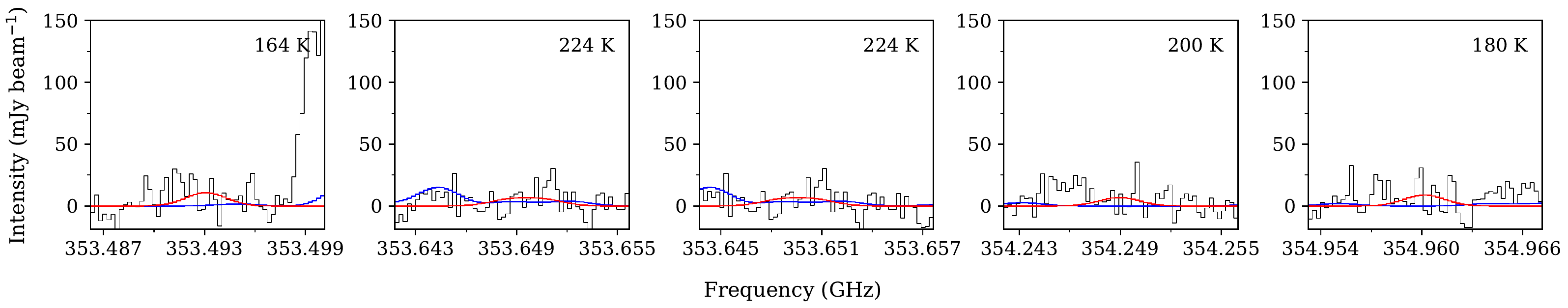}

{t-HCOOD}\\
\includegraphics[scale=0.54]{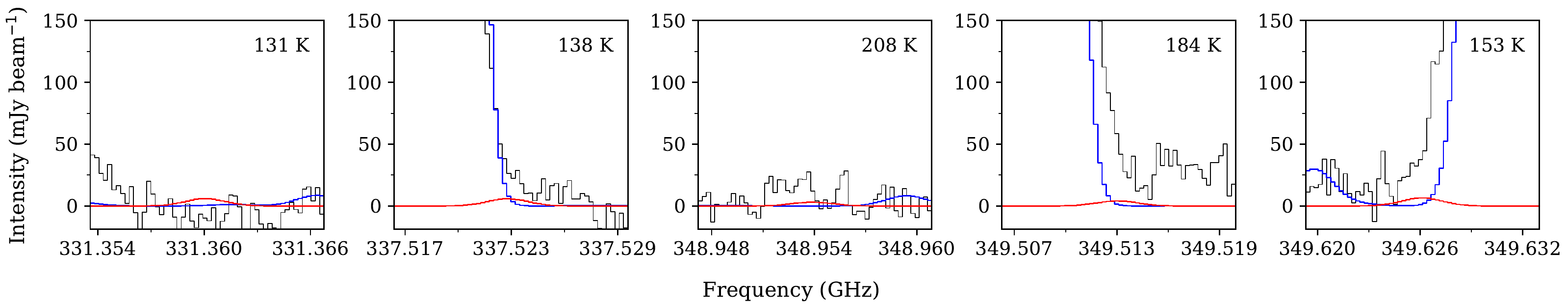}

{CH$_3$COOH}\\
\includegraphics[scale=0.54]{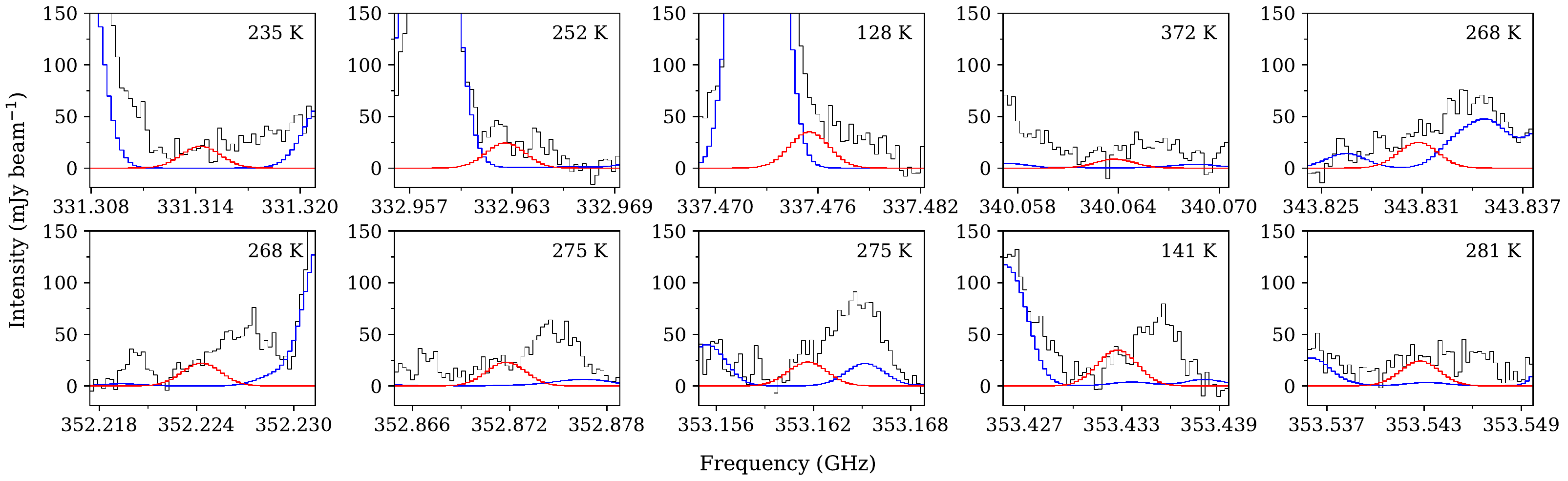}
\caption{Representative selection of transitions of t-HCOOH isotopologues and CH$_3$COOH towards IRAS 16293 A. The synthetic spectra is over-plotted in red, the reference spectrum in blue, and the data in black. The upper energy level of the transition is indicated in the top-left corner.}
\end{figure}

\newpage
\begin{figure}[h!]
{CH$_2$CO}\\
\includegraphics[scale=0.54]{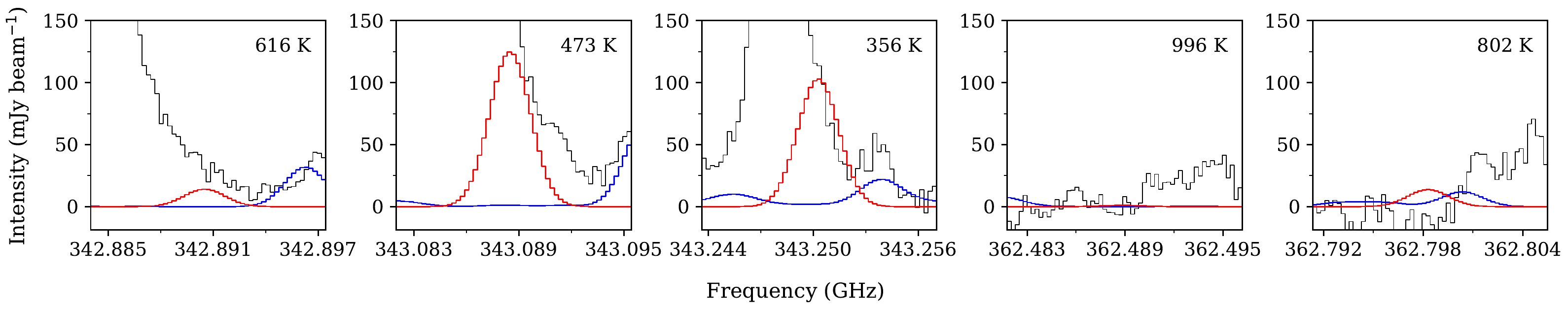}

{CHDCO}\\
\includegraphics[scale=0.54]{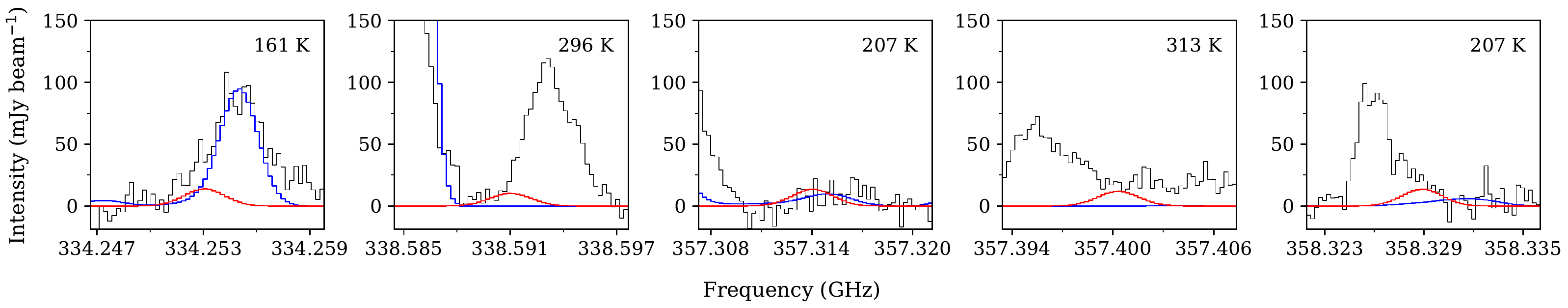}

{HNCO}\\
\includegraphics[scale=0.54]{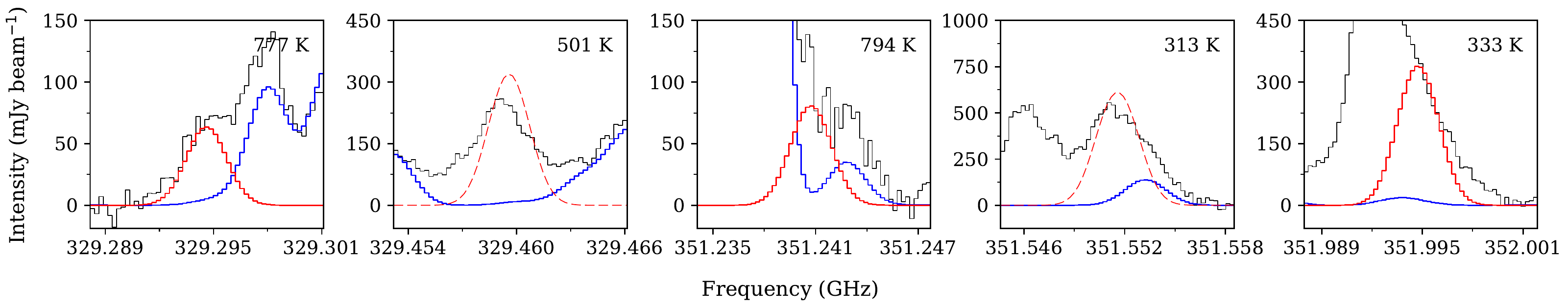}

{DNCO}\\
\includegraphics[scale=0.54]{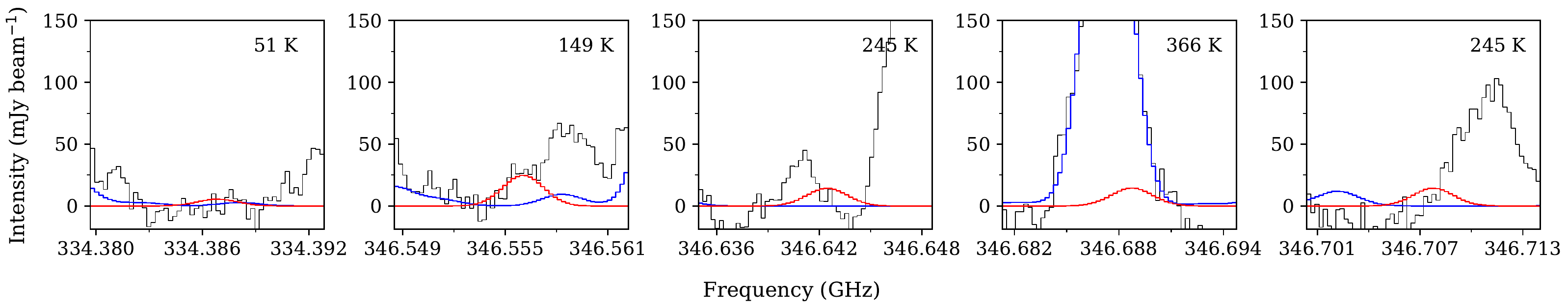}

{CH$_3$CHO}\\
\includegraphics[scale=0.54]{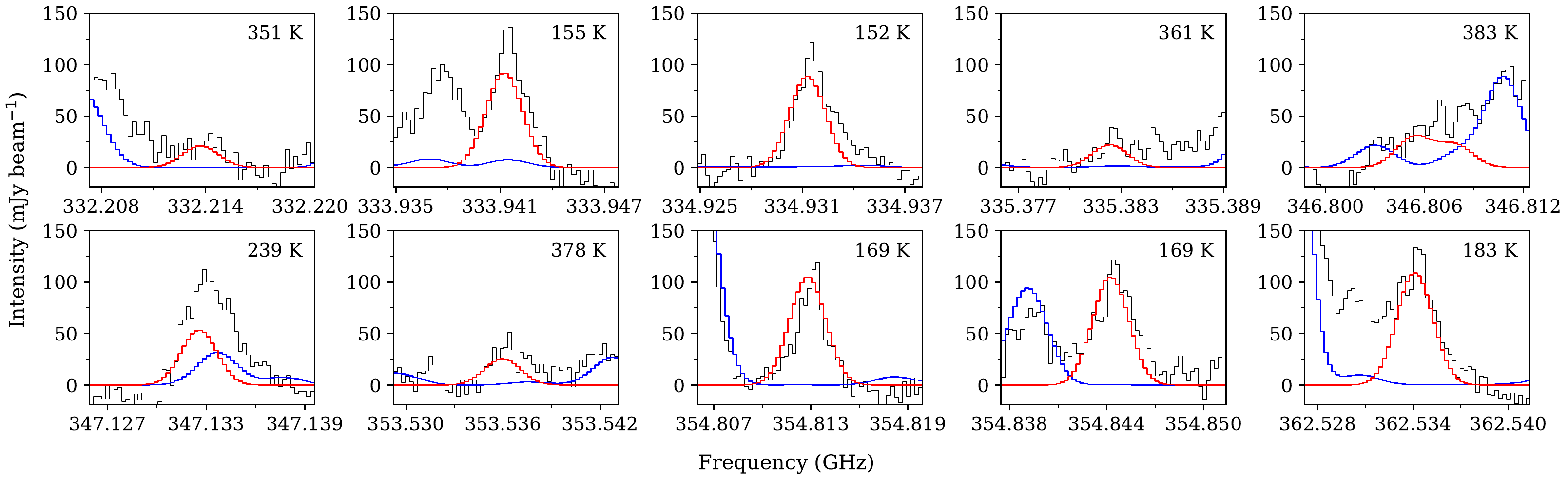}
\caption{Representative selection of transitions of CH$_2$CO and HNCO isotopologues and CH$_3$CHO towards IRAS 16293 A. The synthetic spectra is over-plotted in red, the reference spectrum in blue, and the data in black. The upper energy level of the transition is indicated in the top-left corner.}
\end{figure}

\newpage
\begin{figure}[h!]
{CH$_3$COCH$_3$}\\
\adjustbox{trim ={0.02\width} {0.0\height} {0.0\width} {0.0\height}, clip=true}{
    \includegraphics[scale=0.47]{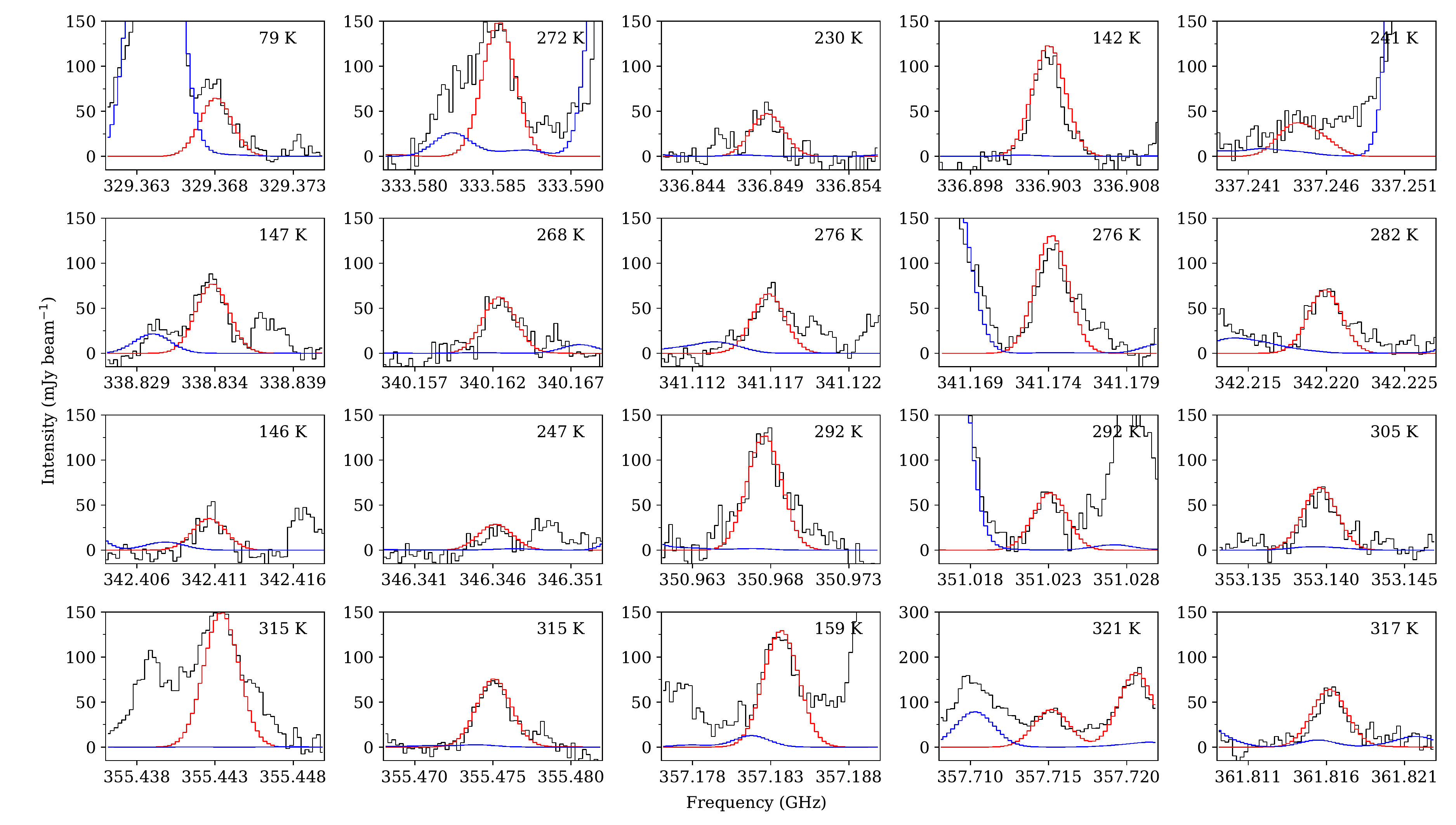}}\\
{c-H$_2$COCH$_2$}\\
\includegraphics[scale=0.54]{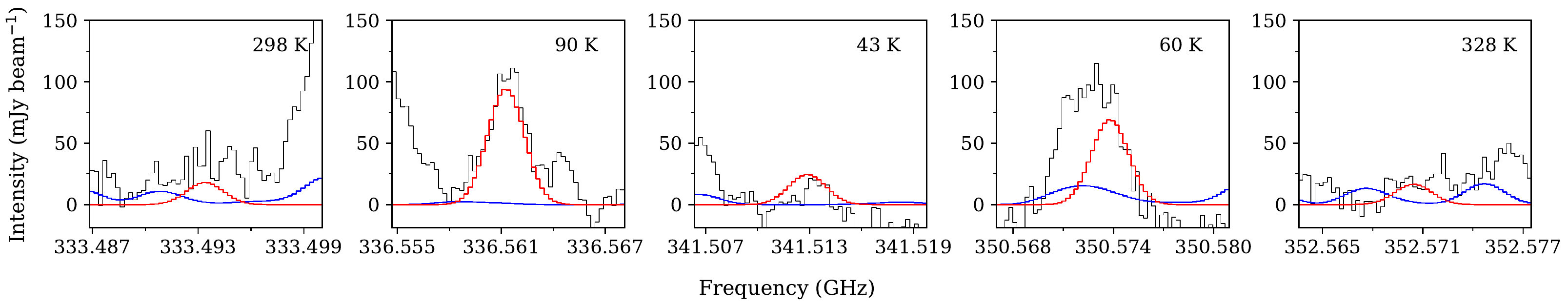}

{CH$_3$O$^{13}$CHO}\\
\includegraphics[scale=0.54]{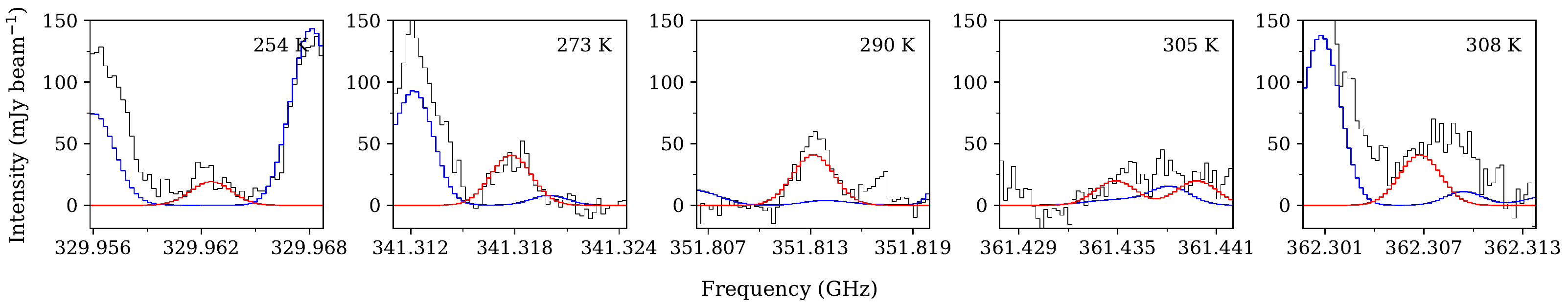}
{CH$_2$(OH)CHO}\\
\includegraphics[scale=0.54]{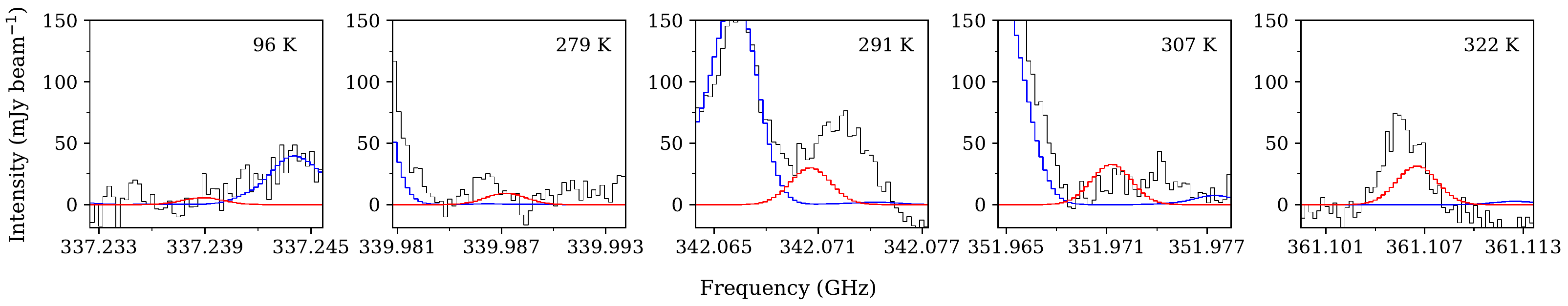}
\caption{Representative selection of transitions of CH$_3$COCH$_3$, c-H$_2$COCH$_2$, CH$_3$O$^{13}$CHO, and CH$_2$(OH)CHO towards IRAS 16293 A. The synthetic spectra is over-plotted in red, the reference spectrum in blue, and the data in black. The upper energy level of the transition is indicated in the top-left corner.}
\end{figure}

\newpage
\begin{figure}[h!]
{aGg'-(CH$_2$OH)$_2$}\\
\includegraphics[scale=0.54]{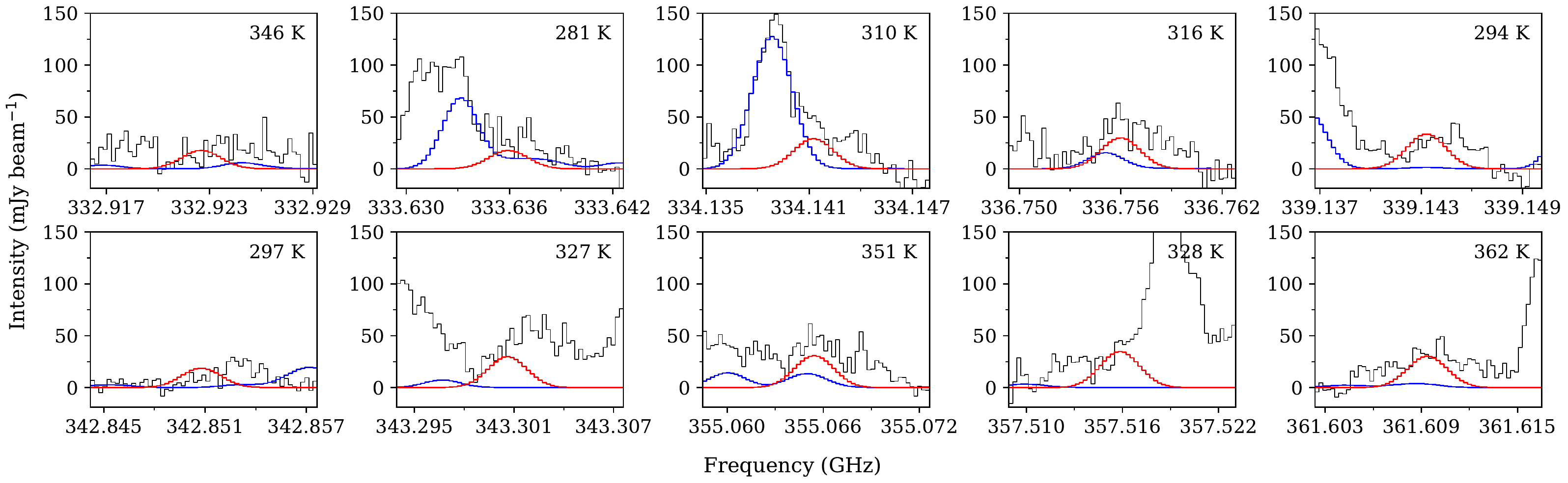}

{gGg'-(CH$_2$OH)$_2$}\\
\includegraphics[scale=0.54]{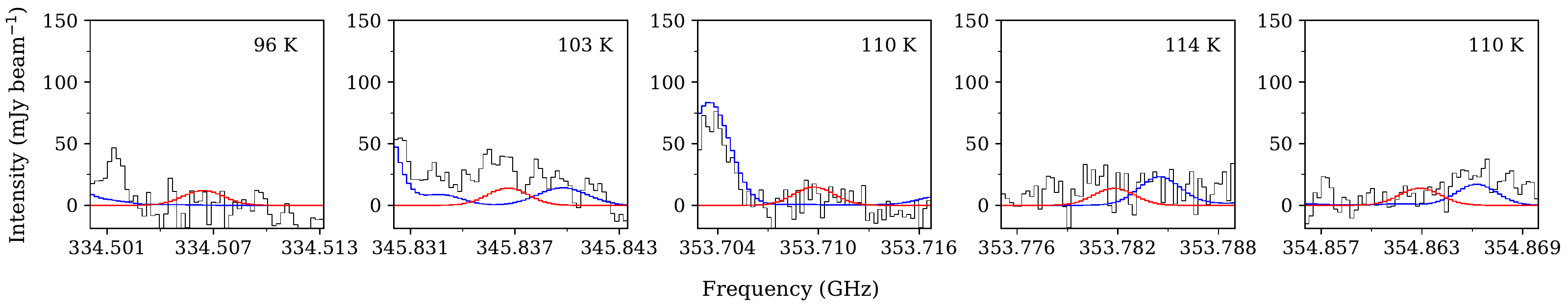}

{t-C$_2$H$_5$OCH$_3$ {(towards IRAS 16293A)}}\\
\includegraphics[scale=0.54]{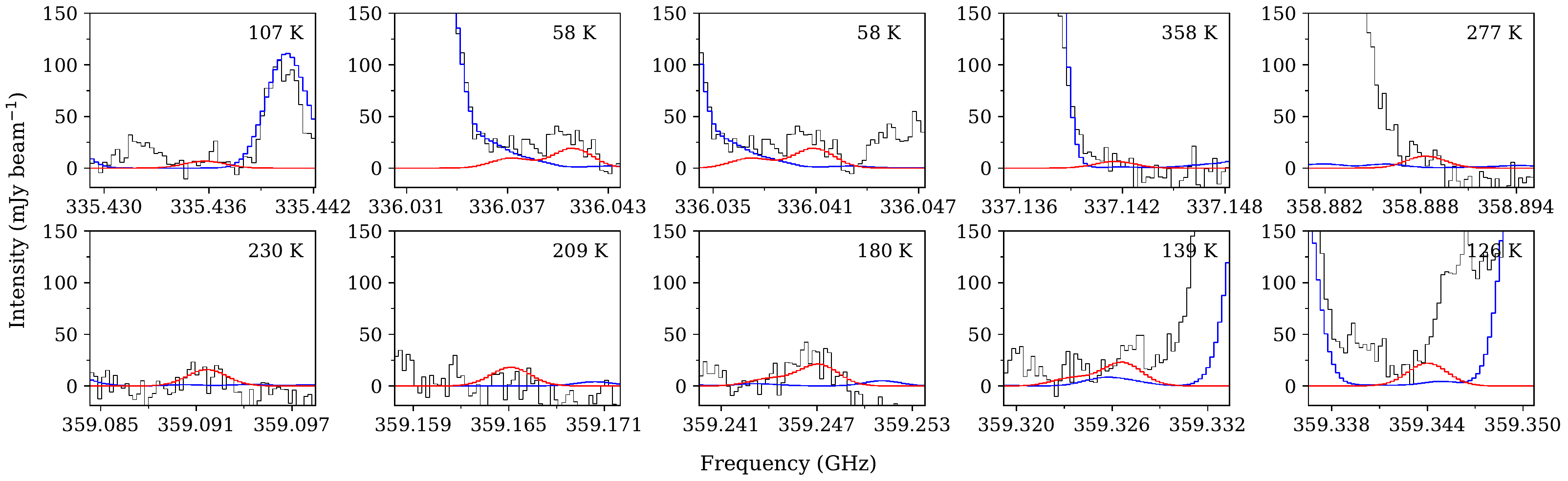}

{t-C$_2$H$_5$OCH$_3$ {(towards IRAS 16293B)}}\\
\includegraphics[scale=0.54]{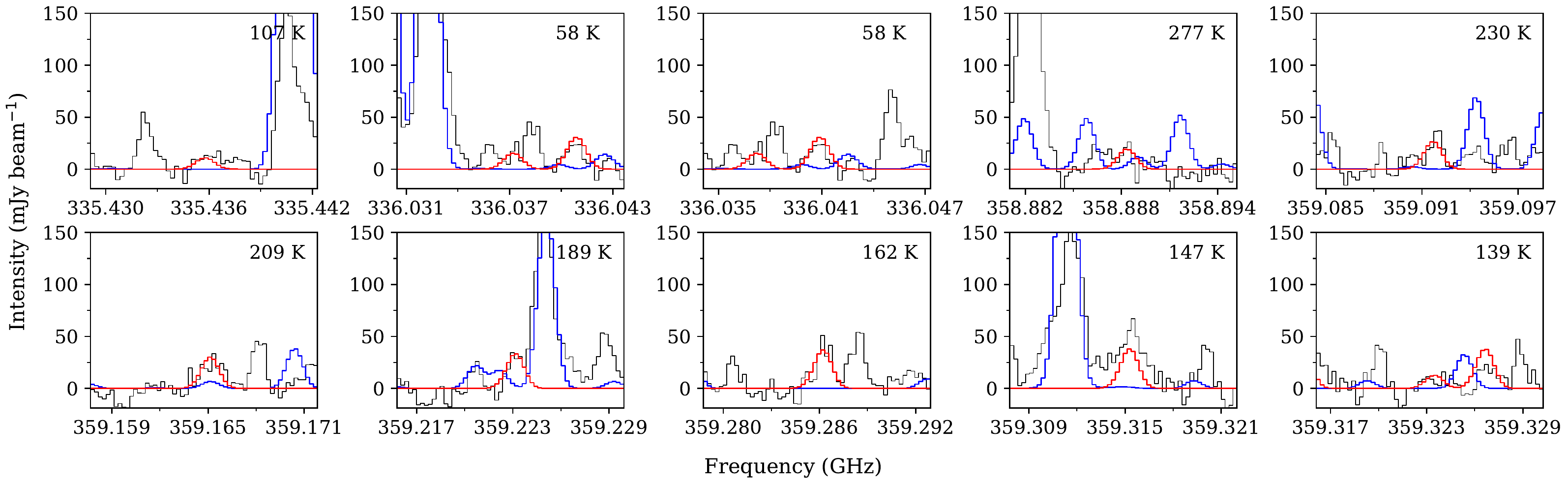}
\caption{Representative selection of transitions of the two energetically lowest conformers of (CH$_2$OH)$_2$ towards IRAS 16293A and t-C$_2$H$_5$OCH$_3$ towards IRAS 16293 A and B. The synthetic spectra is over-plotted in red, the reference spectrum in blue, and the data in black. The upper energy level of the transition is indicated in the top-left corner.}
\end{figure}

\newpage
\begin{figure}[h!]
{CH$_3$OCH$_2$OH {(towards IRAS 16293A)}}\\
\includegraphics[scale=0.54]{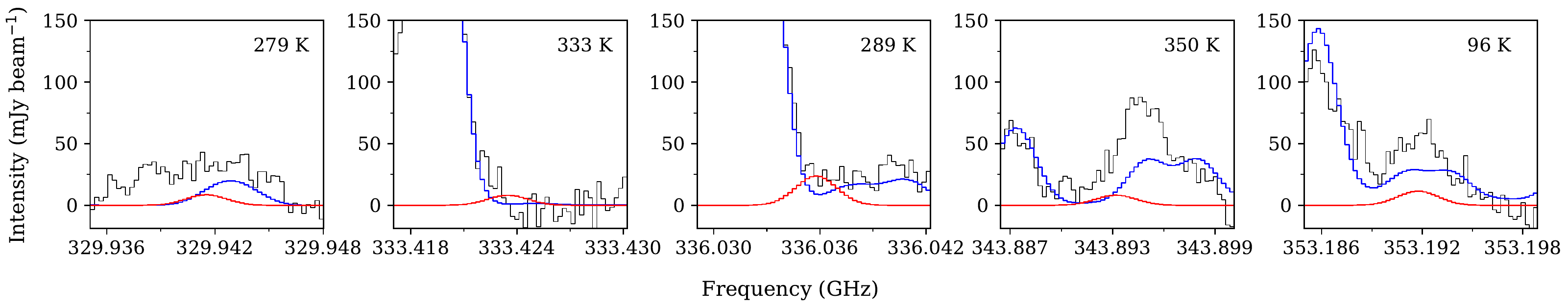}

{CH$_3$OCH$_2$OH {(towards IRAS 16293B)}}\\
\includegraphics[scale=0.54]{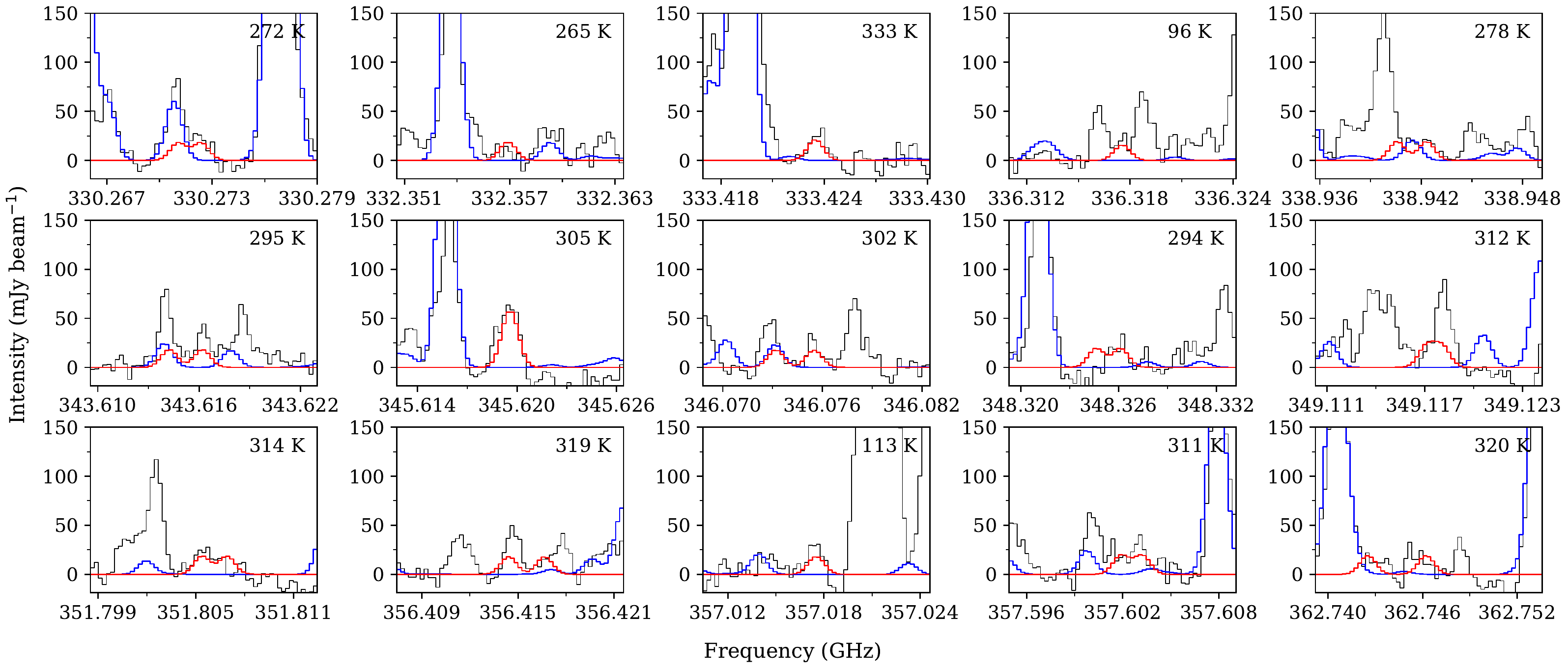}

{NH$_2$CHO}\\
\includegraphics[scale=0.54]{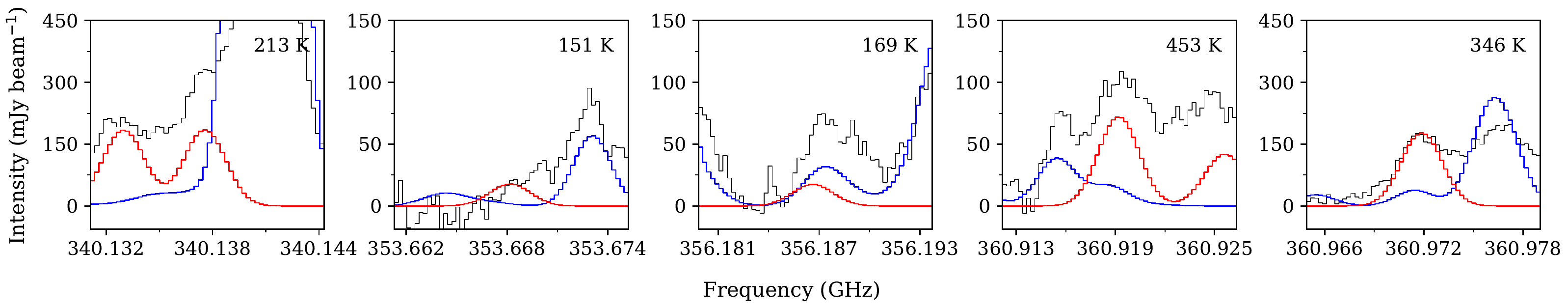}

{C$_2$H$_5$CHO}\\
\includegraphics[scale=0.54]{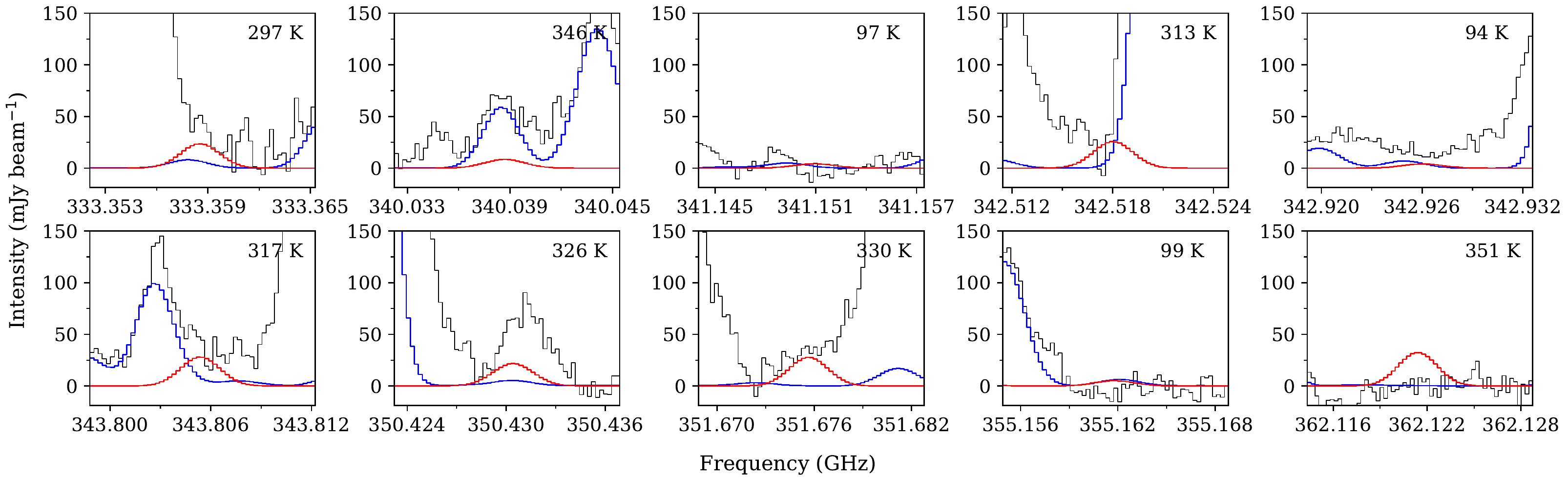}
\caption{Representative selection of transitions of CH$_3$OCH$_2$OH towards IRAS 16293 A and B and NH$_2$CHO and C$_2$H$_5$CHO towards IRAS 16293A only. The synthetic spectra is over-plotted in red, the reference spectrum in blue, and the data in black. The upper energy level of the transition is indicated in the top-left corner.}
\end{figure}

\newpage
\begin{figure}[h!]
{CH$_2$DCHO {(towards IRAS 16293A)}}\\
\includegraphics[scale=0.54]{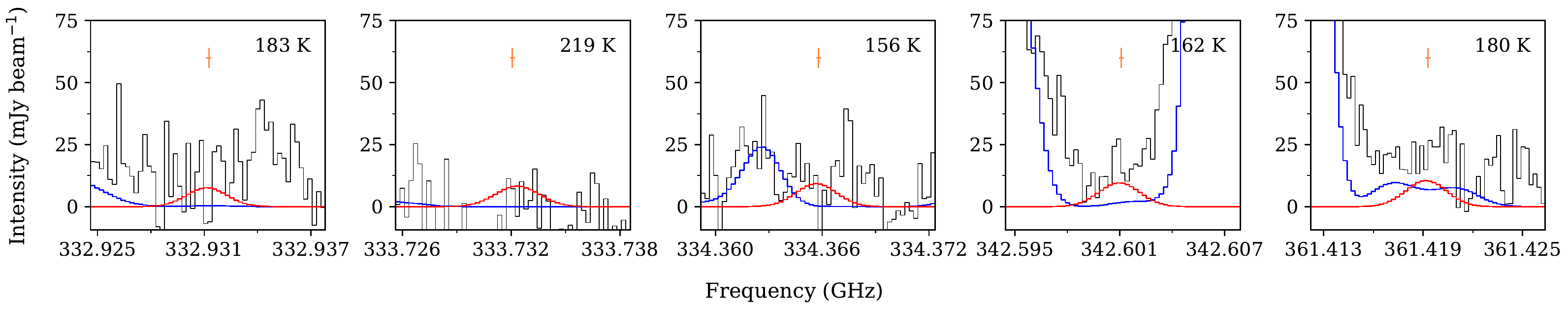}

{CH$_2$DCHO {(towards IRAS 16293B)}}\\
\includegraphics[scale=0.54]{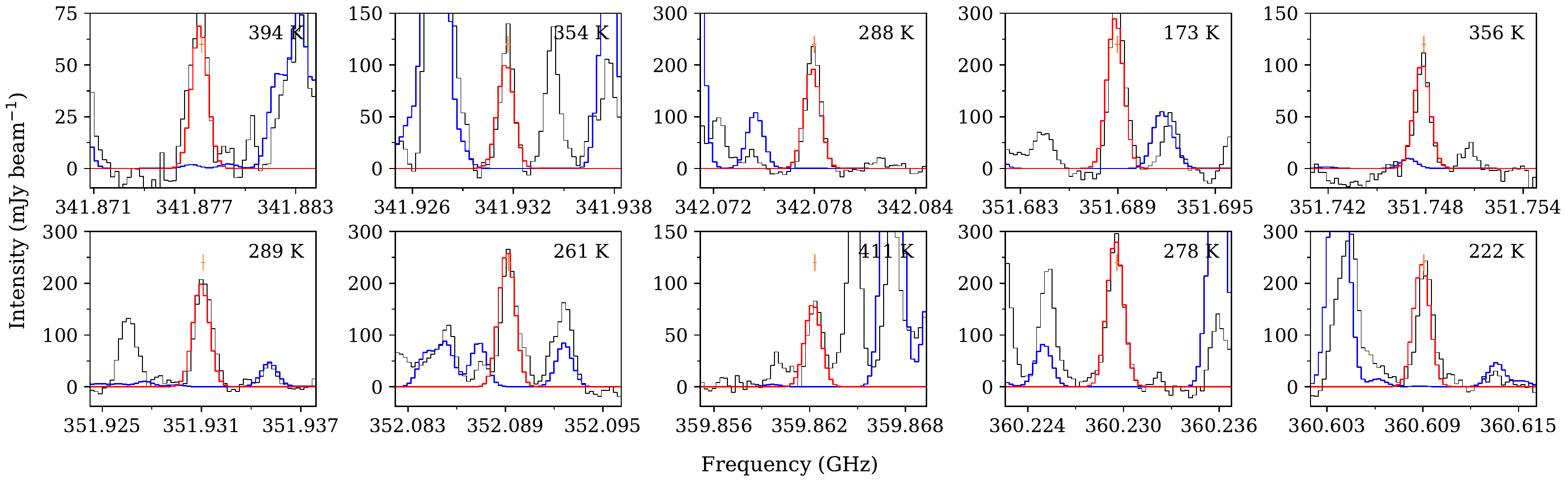}

{CH$_3$CDO {(towards IRAS 16293A)}}\\
\includegraphics[scale=0.54]{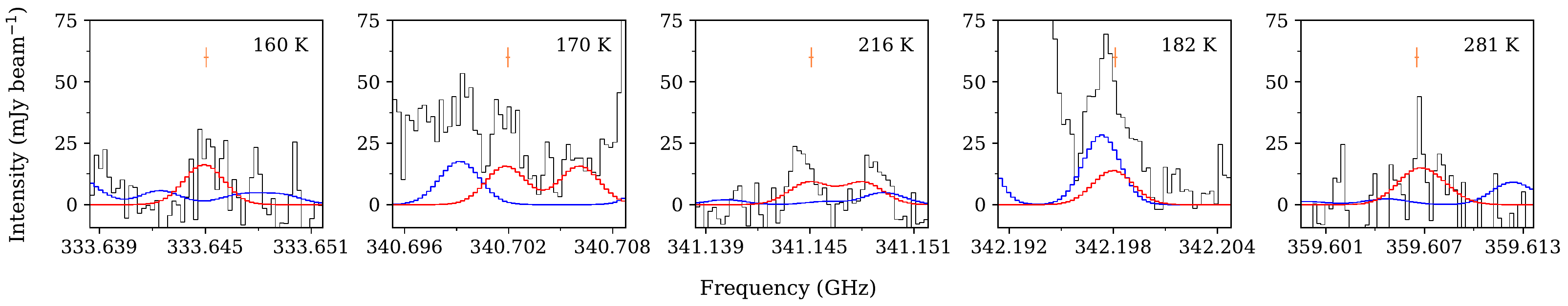}

{CH$_3$CDO {(towards IRAS 16293B)}}\\
\includegraphics[scale=0.54]{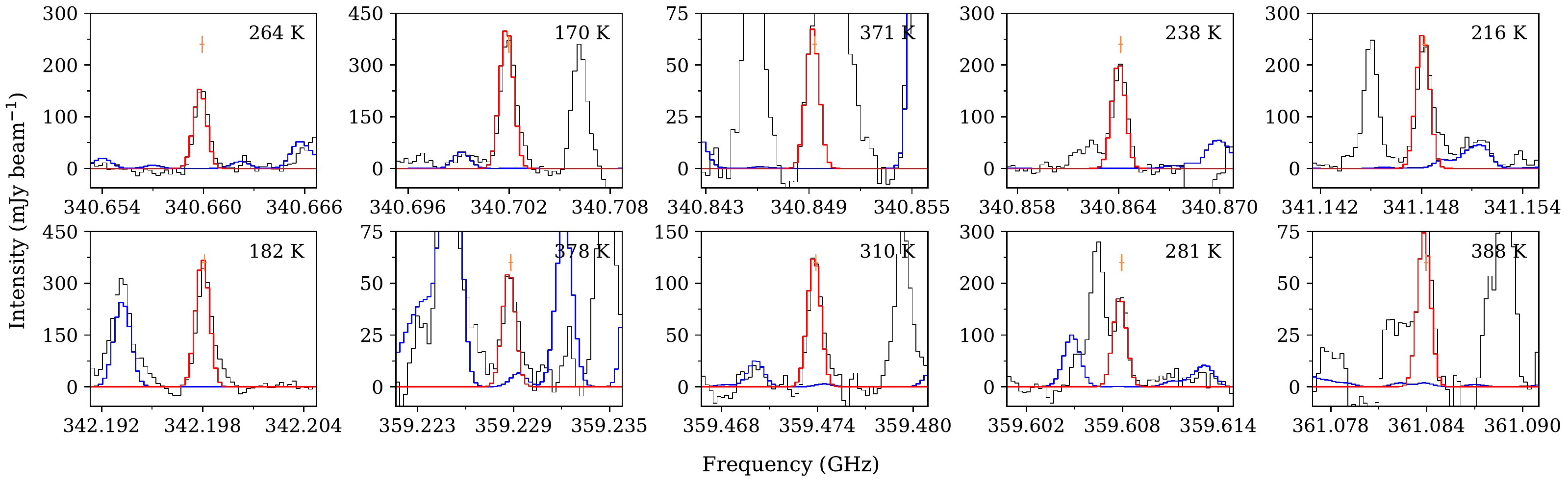}
\caption{Representative selection of transitions of CH$_2$DCHO and CH$_3$CDO towards IRAS 16293 A and B. The synthetic spectra is over-plotted in red, the reference spectrum in blue, and the data in black. The upper energy level of the transition is indicated in the top-left corner.}
\end{figure}

\end{document}